\documentclass[preprint]{aastex}

\newcommand{\kms}{km~s$^{-1}$}
\newcommand{\subsun}{\mbox{$_{\odot}$}}
\newcommand{\etal}{{\it et al.\/}}
\newcommand{\teff}{$T_{eff}$}
\newcommand{\grav}{log($g$)}

\newcommand{\eqw}{$W_{\lambda}$}

\begin{document}

\title{The Chemical Evolution of the Draco Dwarf Spheroidal Galaxy\altaffilmark{1} }

\author{Judith G. Cohen\altaffilmark{2} and Wenjin Huang\altaffilmark{3} }

\altaffiltext{1}{Based on observations obtained at the
W.M. Keck Observatory, which is operated jointly by the California
Institute of Technology, the University of California and the National
Aeronautics and Space Administration}

\altaffiltext{2}{Palomar Observatory, Mail Stop 105-24,
California Institute of Technology, Pasadena, California 91125
(jlc@astro.caltech.edu)}

\altaffiltext{2}{Palomar Observatory, current
address: University of Washington, Astronomy, Box 351580, Seattle,
Washington, 98195 (hwenjin@astro.washington.edu)}

\begin{abstract}

We present an abundance analysis based on high resolution
spectra of 8 stars selected to span the full range in metallicity
 in the Draco dwarf spheroidal galaxy.  We find
[Fe/H] for the sample stars ranges from $-1.5$ to $-3.0$~dex.
Combining our sample with previously published work for a total
of 14 luminous Draco giants, we show that
the abundance ratios [Na/Fe], [Mg/Fe] and [Si/Fe] for the Draco giants
overlap those of
Galactic halo giants at the lowest [Fe/H] probed, but are
significantly lower
for the higher Fe-metallicity Draco stars.  
For the explosive $\alpha$-elements
Ca and Ti, the abundance ratios for Draco giants with [Fe/H] $> -2.4$~dex
are approximately constant and slightly sub-solar, well below
values characteristic of Galactic halo stars.
The $s$-process contribution to the production of heavy elements
begins at significantly lower Fe-metallicity than in the Galactic halo.

Using a toy model we compare the behavior of the abundance ratios within
the sample of Draco giants
with those from the literature of Galactic globular clusters, and the Carina
and Sgr dSph galaxies.  The differences appear to be related to the
timescale for buildup of the heavy elements, with Draco having
the slowest rate.

We note the presence of a Draco giant with [Fe/H] $< -3.0$~dex in our sample,
and reaffirm that the inner Galactic halo could have been formed
by early accretion of Galactic satellite galaxies and dissolution
of young globular clusters, while the outer halo could have formed
from those satellite galaxies accreted later.

\end{abstract}

\keywords{galaxies: individual (Draco), galaxies: abundances, galaxies: dwarf}

\section{Introduction\label{intro}}

The Draco dwarf spheroidal (dSph) galaxy is a nearby 
satellite of the Milky Way.  It was extensively studied
by \cite{baade61}, who constructed a CMD and searched
for variable stars.  HST imaging by \cite{grillmair98}
suggested that the bulk of the stars are old, with a luminosity
function similar to that of old, metal-poor Galactic globular
clusters.  Although \cite{hernandez00} find
a median age for Draco stars of 7~Gyr from HST archival images,
the recent HST study by \cite{orban08}, which 
provides coarse age distributions for Draco
and several other Galactic satellites, yields a somewhat older
mean age
for this stellar population.
There is no evidence for multiple main sequences nor any sign
of ongoing star formation.  A very low
upper limit for neutral hydrogen gas in Draco was established
by \cite{young99}.

Many radial velocity surveys
of large samples of stars have been carried out to
confirm membership and 
measure the stellar velocity dispersion as a function
of radius in Draco as well as in the other  Milky Way satellites; 
\cite{armandroff95} is an early example of such for Draco,
while \cite{walker07} highlights the very large
$v_r$ datasets that
can be assembled today.
The observed $\sigma_v$ in Draco  is 
unexpectedly high, given the
low luminosity of the system, and remains flat to large radii,
which according to \cite{munoz} (see also the references therein)
requires the presence of large
amounts of dark matter.
These studies generally ignore the issue of potential ongoing 
tidal disruption
affecting the internal kinematics of the Milky Way satellite,
which does not appear to be a concern in the case of Draco
\citep{megacam}.

There is great current interest in the detailed
properties of satellites of the Galaxy in light of our
greatly improved hierarchical cold dark matter cosmological models, which
gave rise to the the missing
satellite  problem \citep{klypin99}.  This is only enhanced
by the  discovery that the dSph galaxies appear to be dark matter dominated systems,
unlike globular clusters of similar total stellar mass.
With the advent of 
efficient high dispersion
spectrographs, large area CCD detectors, and 10-m class telescopes,
studying stars in at least the nearer Galactic satellites
at high spectral dispersion has become feasible. 

Early moderate resolution spectroscopic studies of Draco by \cite{zinn78} and by
\cite{lehnert92} established the presence of an abundance spread
of $\sim$1~dex within Draco.  
In this paper we present detailed
abundance analyses of a sample of 8 luminous Draco stars near
the RGB tip, which more than doubles the sample of Draco stars from
the earlier work of \cite{shetrone1} and of \cite{shetrone2}. 
Our goal is understanding the chemical evolution
of Draco, and how this and other dSph galaxies may
be related to the population of
Galactic halo field stars and to Galactic globular
clusters.   The sample is presented in \S\ref{section_param}
where the procedure for determining their stellar parameters is described.
The next section describes the observations, while \S\ref{section_analysis}
gives the details of the abundance analysis.
We compare
our results to those for Galactic
halo field stars in \S\ref{section_field} and introduce our
toy model for abundance ratios in \S\ref{section_toy_model}.  A comparison
with the predictions of nucleosynthesis is given in 
\S\ref{section_nuc} and with chemical evolution models
in \S\ref{section_chemev}. This is followed by a discussion of 
role the dSph satellite galaxies might have played the
formation of the Galactic halo in \S\ref{section_discuss}.  
After a brief summary we discuss in an appendix
the Fe-metallicity calibration for the infrared Ca triplet,
and
compare our derived metallicities for Draco giants with those inferred from 
CaT measurements
and from Stromgren photometry.

\section{Stellar Sample and Stellar Parameters \label{section_param}}

Our sample in the Draco dSph galaxy contains 8 stars; details
are given in Table~\ref{table_sample}.
It was selected from Table~2.9 of \cite{winnick03} to include
stars which are known radial velocity members of this satellite
to the Galaxy
at or near the RGB tip spanning the full range in color 
and in metallicity and not previously observed at high spectral
resolution.
Winnick measured the infrared Ca triplet in moderate
resolution spectra of Draco stars obtained using the multi-fiber Hydra
spectrograph at the WIYN telescope.  Her sample was chosen from
the photometric survey of \cite{piatek01} and previous
radial velocity surveys \citep*[see, e.g.][]{armandroff95}.   
Only confirmed radial
velocity members are listed in her table; carbon stars
known to be members of Draco were excluded.  She developed
her own calibration of the relationship between Ca triplet
indices and Fe or Ca metallicity based on data for
Galactic globular clusters.  Details of the calibration and related
issues are discussed in the appendix.

We adopt the procedures described in \cite{cohen02} and used in all
subsequent work by the first author 
published to date to determine the stellar parameters
for our sample of luminous Draco giants.
Our \teff\ determinations are based on the broad band colors
$V-J$ and $V-K$.
The optical photometry (we use $V$ only, the $V-I$ color
supports the deduced \teff\ in most cases, but is not used) is 
from the SDSS \citep{york00} using the transformation equations of
\cite{smith02}.
The IR photometry is taken from 2MASS \citep{2mass1,2mass2}, and is
transformed from the 2MASS system to the Johnson-Bessell system
using the results of \cite{carpenter}.
The galactic extinction is from the map of 
\cite{extinct98}; E($B-V$) does not exceed $0.03$~mag for any star in the 
Draco sample.

We derive surface gravities by combining these \teff\ 
with an appropriate theoretical
isochrone from the grid of \cite{yi01} assuming a fixed age of 12 Gyr. 
We use the [Fe/H]
values of \cite{winnick03} from the infrared Ca triplet as an 
initial guess.
We iterate as necessary given the metallicity we derive here through
analysis of our high resolution spectra.  

We calculate the dereddened $M_V$ for each sample star
(ignoring the thickness of Draco itself along the line
of sight), adopting 76~kpc
as its distance based on the RR~Lyrae
study in this dSph by \cite{distance}.  We use the $M_V$, which
are independent of the adopted stellar parameters, to check
for overall consistency with the adopted isochrones.  We find the position 
along the RGB in the adopted isochrone appropriate 
for the
initial estimate of metallicity for each star
that has that $M_V$ and
compare the corresponding \teff\ with our adopted value inferred
from the photometry for each star.
The two independently determined values of \teff\
are reasonably consistent  given the uncertainty in 
our adopted \teff\ of 100~K.  Achieving
this agreement is made easier since
the slope of the upper RGB in the CMD
is quite steep;  we find for $\Delta$\teff\ ~ = ~ 100~K 
that $\Delta M_V~ \sim 0.45$~ mag
and $\Delta${\grav}$~ \sim 0.25$~dex 
from the adopted isochrones.


The resulting stellar parameters, which have been
derived with no reference to the spectra themselves,
are given in the second and third columns of
Table~\ref{table_param}, as are their
heliocentric radial velocities.  The random uncertainties
in the adopted \teff\ from photometric errors
are 100~K.  This ignores systematic errors
which may be present.  The adopted uncertainties in \grav\ 
based on the evolutionary tracks of \cite{yi01}
for stars on the upper RGB  are 0.25~dex.

\begin{figure}
\epsscale{0.8}
\plotone{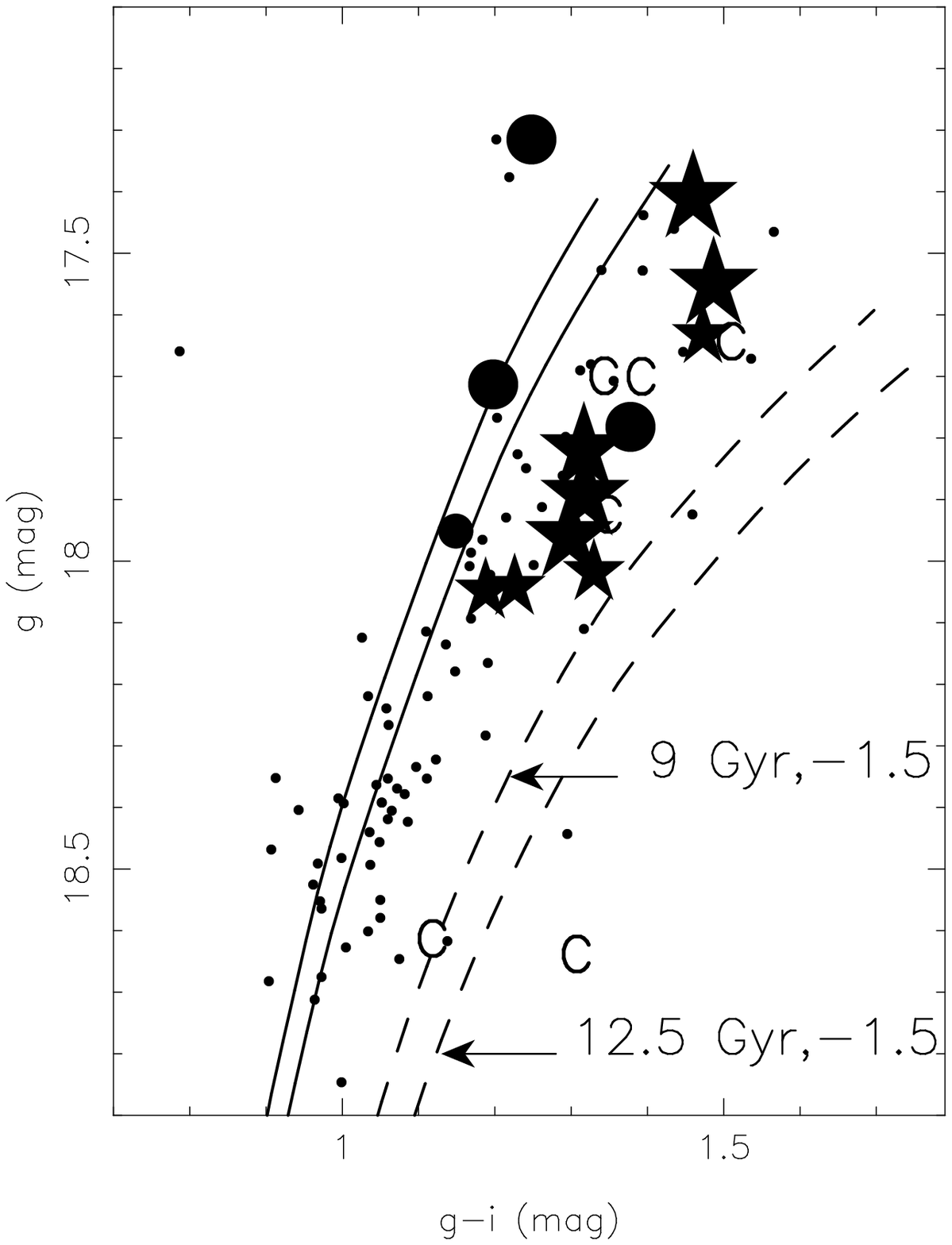}
\caption[]{Our Draco HIRES sample is shown in a plot of
$g'-i$
versus $g'$  (large symbols),
where filled circles indicate giants with [Fe/H] $< -2.5$~dex, and
star symbols have higher Fe-metallicity.
The sample of Draco stars studied by \cite{shetrone1} and by \cite{shetrone2}
are indicated by the smaller symbols.
The dots indicate Draco members from \cite{winnick03} with  photometry
from \cite{megacam} or, for the brightest stars, from the SDSS.  
Carbon stars that are confirmed members of Draco \citep[see, e.g.][]{cioni}
are indicated by the letter C.  All observerational data are  corrected for 
interstellar reddening.
Isochrones from the Dartmouth Stellar Evolution Database
\citep{dotter08} for [Fe/H] $-2.5$~dex with [$\alpha$/Fe] = +0.2~dex
(solid lines)
and for [Fe/H] $-1.5$~dex with [$\alpha$/Fe] Solar (dashed lines)
for ages 9 and 12.5 Gyr are shown.
\label{figure_isochrone}}
\end{figure}

Fig.~\ref{figure_isochrone} shows 
our sample of 8 luminous giants in Draco in a plot of
$g'-i'$
versus $g'$  corrected for interstellar reddening;
the previously studied sample of 6 giants
from \cite{shetrone1} and by \cite{shetrone2} is shown as well.
Members of Draco from the list of \cite{winnick03},
which excludes the known carbon stars, were cross indexed
with the CFHT Megacam photometry of \cite{megacam}\footnote{The
four brightest Draco giants do not appear in their Megacam database 
(presumably they were saturated); their photometry is
from the SDSS \citep{york00} DR5 release
\citep{sdss_dr5}, as is the $i'$~mag of the
brightest Draco giant with a HIRES spectrum.}  and are also displayed.
The positional tolerance was 1.0~arcsec; a few spurious
identifications may have occurred.
Superposed in this figure are isochrones from the Dartmouth
Stellar Evolution Database \citep{dotter08} for [Fe/H] 
 $-2.5$~dex with [$\alpha$/Fe] = +0.2~dex
(solid lines) and for [Fe/H] $-1.5$~dex with [$\alpha$/Fe] Solar (dashed lines)
for ages 9 and 12.5 Gyr.

Our HIRES sample of luminous Draco giants was selected to span
the full range in metallicity as inferred from 
the Ca triplet indices \citep{winnick03}.  Fig.~\ref{figure_isochrone} shows
it does cover
the full range in $g'-i'$ color of the upper RGB of Draco members.
The luminosity of the brightest Draco giants is in good agreement
with that predicted from the isochrones for the RGB tip as a
function of metallicity in the $g',i'$ colors.  The
deduced ages will be described later in \S\ref{section_age}.

\section{Observations \label{section_obs} }

The Draco stars in our sample were observed with HIRES \citep{vogt94} at 
the Keck~I telescope
during two runs, in June 2005 and in Sep. 2006.  
Sky conditions during the 2006 run were better than in the 2005 run; one star
from the earlier run was re-observed at that time.
Total exposure times were two hours per star,
broken up into 1800 or 2400 sec segments to expedite
removal of cosmic rays.
The instrument configuration yielded complete spectral coverage
in a single exposure from 3810 to 6700~\AA, and extends
to 8350~\AA\ with small gaps between orders.
The slit width was  1.1 arcsec, which corresponds to
a spectral resolution of 35,000 for all exposures.
  The signal-to-noise ratio
per spectral resolution element ranges from 40 to 85
(with 5 of the sample stars having at least 65) in the
continuum in the central order of the spectrum (at $\sim$6050~\AA),
depending on the brightness of the star and the sky conditions
during the exposure, most importantly on the seeing.  
The SNR at the bluest end of these spectra is poor, only 30 to 40
per spectral resolution element.
This SNR calculation utilizes only
Poisson statistics, ignoring issues of cosmic ray removal,
night sky subtraction, flattening, etc.   The observations
were carried out with the slit length aligned to
the parallactic angle.  

The processing of the spectra was done with 
MAKEE\footnote{MAKEE was developed
by T.A. Barlow specifically for reduction of Keck HIRES data.  It is
freely available on the world wide web at the
Keck Observatory home page, 
http:\\www2.keck.hawaii.edu/inst/hires/data\_reduction.html.} and Figaro 
\citep{shortridge93} 
scripts, and follows closely that described by \cite{cohen06}.
The equivalent widths were measured as described in \cite{cohen04}.
Due to the lower SNR in the blue, lines bluer than 4400~\AA\
were ignored if the species had sufficient other 
detected lines.
Lines with \eqw $> 175$~m\AA\ were
discarded except for two lines from the Mg triplet,
the Na~D lines and Ba~II lines in some of the stars; for these
key elements no or only a few weaker features could be detected
in most of the stars.  Sometimes only one line  of the Na~D doublet could
be measured; the $v_r$ of some of the Draco stars was such that the other
component of the doublet was mangled by NaD  line
emission from the Earth's atmosphere.
Table~\ref{table_eqw} lists the atomic parameters adopted
for each line and their equivalent widths measured in the spectra of
each of the Draco dSph stars. 

The very luminous giant Draco~XI-2 shows
strong emission in the red wing
of H$\alpha$ in its spectrum, with even stronger emission
seen in the blue wing.  No other star
in the sample shows any detectable emission in H$\alpha$.

\section{Analysis \label{section_analysis} }

The analysis is identical to that of \cite{cohen08}
and earlier references therein.
In particular we use the model
stellar atmosphere grid of \cite{kurucz93}
and a current
version of the LTE spectral synthesis program MOOG \citep{moog},
which treats scattering as LTE absorption.

Our analysis assumes classical plane
parallel stellar atmospheres and LTE, both for atomic
and for molecular features.
We adopt a Solar Fe abundance of log[$\epsilon$(Fe)] = 7.45~dex,
which is somewhat lower (by up to 0.10~dex) than that used by 
many groups.  
This leads directly to our [Fe/H] values for a given star being 
somewhat higher  and to our abundance 
ratios [X/Fe] being somewhat lower than those which would be inferred by 
most other teams.  
Our $gf$ values are generally taken from
Version 3.1.0 of the NIST Atomic Spectra Database 
(phsics.nist.gov/PhysRefData/ASD/index.html, NIST Standard Reference
Database 78).
A comparison of $gf$ values for Fe we adopt 
with those of the First Stars Project at the VLT \citep{cayrel04} shows
that the mean difference in  Fe
transition probabilities between these us is 
$\mid \Delta \mid ~ < ~ 0.01$~dex for Fe~I and 0.04 dex for Fe~II.
Corrections for hyperfine structure for
Sc~II, V~I, Mn~I, Co~I, Cu~I, Ba~II, and Eu~II, when necessary,
were used; the majority of the HFS patterns were adopted from
\cite{prochaska}.

Our abundances for C are from the 4320~\AA\ region of the
G band of CH, where the absorption is less than in
the main part of the G band at 4300~\AA.  O abundances
are from the forbidden line at 6300~\AA; sometimes 
the weaker 6363~\AA\ line is also detected. 
Our nominal Solar C and O abundances are 8.59 and 8.83~dex respectively.
We adopt a dissociation potential of  3.47~eV \citep{huber79} for CH.
We have checked that with our adopted molecular parameters
we can  reproduce the Solar spectrum of the G band of CH,
taken from \cite{wallace98}.
The adopted Solar abundances of C and of O are close to those of \cite{grevesse98},
but somewhat larger than  the values obtained using 3D model atmospheres 
by \cite{asplund04} and \cite{asplund05}. 
We use 1D model atmospheres to synthesize the molecular features,
ignoring any 3D effects, although \cite{collet07} suggest that
these may be very large at [Fe/H] $\sim -3$~dex, becoming less important
at higher metallicities.
They claim that  
CNO abundances may be overestimated by $\sim$0.8 dex as
compared to a 1D analysis when molecular bands are used in 
extremely metal-poor stars.  However,
we prefer not to attempt 3D corrections until a full grid of 3D model 
atmospheres or of corrections to CNO abundances derived from the
molecular bands from 3D to 1D models becomes
available.

Since the Draco stars are rather faint for 2MASS, the uncertainties
in the $J$ and particularly in the $K_s$ magnitudes are large.  We therefore
feel free to
slightly adjust \teff\ and \grav\  after the first pass
through the analysis to improve the ionization equilibrium and
slope of the abundances determined from the set of Fe~I lines
as a function of $\chi$ (the excitation potential of the lower level).
These spectroscopic stellar parameters are given in the
fourth and fifth column of Table~\ref{table_param} and
are the ones used subsequently.  With these values we were
able to achieve good ionization equilibrium for Ti and Fe as
well reasonable  excitation equilibrium of Fe~I.
Table~\ref{table_slopes} gives the slope of a linear fit
to the abundances determined from the set of Fe~I lines
as a function of $\chi$,
\eqw, and $\lambda$, which are most sensitive to \teff, $v_t$,
and the wavelength dependence of any problems in establishing the correct
location of the continuum (perhaps arising from the more severe
crowding towards bluer wavelengths) or of a
missing major source
of continuous opacity, respectively.  There are $\sim$ 55 to 150 Fe~I lines
detected in
each star, with $\chi$ ranging from 0 to $\sim$4.5~eV.  The only
slope which is large enough to be of concern
and which tends to have the most significant correlation coefficient 
is that with $\chi$ ($\mid cc(\chi) \mid ~ > ~ 0.4$ for some of the sample giants),
which depends largely on \teff.   In our final adopted
solutions, the Fe~I slope as a function of $\chi$
tends to be slightly negative, with values ranging from
$-0.03$ to $-0.10$~dex/eV.  However, since this slope
decreases by $\sim$0.2~dex/eV/($\Delta$\teff = 250~K), 
a decrease in \teff\
of a maximum of 125~K, consistent with our adopted \teff\ uncertainty,
would make all these slopes zero, and would decrease
the [Fe/H] derived from Fe~I lines by $\sim$0.2~dex.

One potential concern is the possibility of non-LTE in Fe affecting
the determination of stellar parameters using ionization equilibrium.
In particular
Fe overionization has been observed in M dwarfs in the Hyades
open cluster,
where the true Fe abundance can be established from a detailed
abundance analysis of the hotter  members, by 
\cite{schuler}.  They find that the overionization, in the
sense of the deduced abundance from Fe~II lines being higher than
that from Fe~I lines, is considerably smaller in the Hyades giants.
A similar problem in the same open cluster was found earlier by
\cite{yong}, who demonstrate that the Fe~I abundance in the coolest
Hyades dwarfs they studied is much
closer to the cluster mean Fe-metallicity.
Similarly, in resonance scattering 
\citep*[e.g. see][]{asplund_araa} 
the source function, S$_{\nu}$, is reduced to below the local 
Planck function,
thus leading to a stronger line in non-LTE.  Resonance scattering is
seen in the Na~D lines of metal-poor stars \citep*[e.g.][]{andrievsky07},
the OI triplet at
7774~\AA\ in the Sun and may have affected the abundances from the 
Ca~I 4226~\AA\ resonance line 
of \cite{mcwilliam95a}. 

The slightly negative Fe~I slope with excitation
potential mentioned above
may be a sign that overionization of Fe is occurring.  If this were the
case, to compensate  we would then have been driven to adopt a higher \teff\
than the actual value;
our derived [Fe/H] values would be too high as indicated above,
but the deduced abundance ratios would not be significantly
affected by such a decrease in \teff. With this in mind, we adopt 
asymmetrical uncertainties for \teff\ of +100~K, $-150$~K.
Since we have been
able to achieve satisfactory ionization equilibrium for Fe and for Ti
and at the same time
reasonably good excitation equilibrium for Fe 
with a single value of \teff\ which 
differs from that set solely from broad band photometry by 50~K or
less for 6 of the 8 Draco giants, we regard our choices for stellar
parameters  as satisfactory.  However, we could not find
a consistent set of stellar parameters for one of the 
most luminous and coolest Draco giants, Draco~XI-2, that would
simultaneously satisfy the above conditions. In particular
we could not achieve
Ti and Fe ionization equilibrium simultaneously, and had to adopt
compromise values.
Ideally, of course, one would like to have a full non-LTE 3D
analysis including both convection and spherical (as distinct
from plane parallel) layers for all species,  but at the present time this is 
not practical.

\section{Comparison with Galactic Halo Field Stars \label{section_field}}

In this section we study the behavior of abundance ratios
within Draco and compare them to those of Galactic halo field stars
in detail. The sample of Draco stars with detailed abundance analyses
based on high dispersion spectra is now 14, including the 8
we present here.  \cite{shetrone1} and \cite{shetrone2}
presented an analysis for 6 Draco members.
They used 1 hour exposures per star
and observed with HIRES prior to the major detector upgrade in 2004.  
The most metal-poor star they observed, Draco~119, was reobserved
with much longer HIRES exposures by \cite{fulbright} as the earlier
study only found an upper limit for Ba~II; they too could only
set upper limits for all elements heavier than 
Ni\footnote{\cite{koch_hercules} recently
found two stars with very low upper limits
for [Ba/Fe] yet very high [Mg/Fe] in the faint Hercules
dSph galaxy.  These are reminiscent of Draco~119, which
until now was believed to be unique in that no element
heavier than Ni could be detected in its spectrum in spite
of a major effort at Keck with HIRES by \cite{fulbright}.}.
Both studies adopted
essentially the same \teff\ for this star, but the recent work has \grav\ 0.6~dex
larger. Although they derive essentially the same [Fe/H] for
this star, there are substantial
differences in abundance ratios between these two works,
presumably largely due
to the much higher SNR of the \cite{fulbright} spectra.  In the following 
we always replace the \cite{shetrone2} abundances for Draco~119 with 
those of \cite{fulbright} and we refer to this collective data
set for six Draco giants as \cite{shetrone2}.
Our spectra of luminous stars in the Draco dSph 
are not as deep as that of the single star observed by \cite{fulbright},
but are substantially better than those of \cite{shetrone1} and
\cite{shetrone2}.

\begin{figure}
\epsscale{0.9}
\plotone{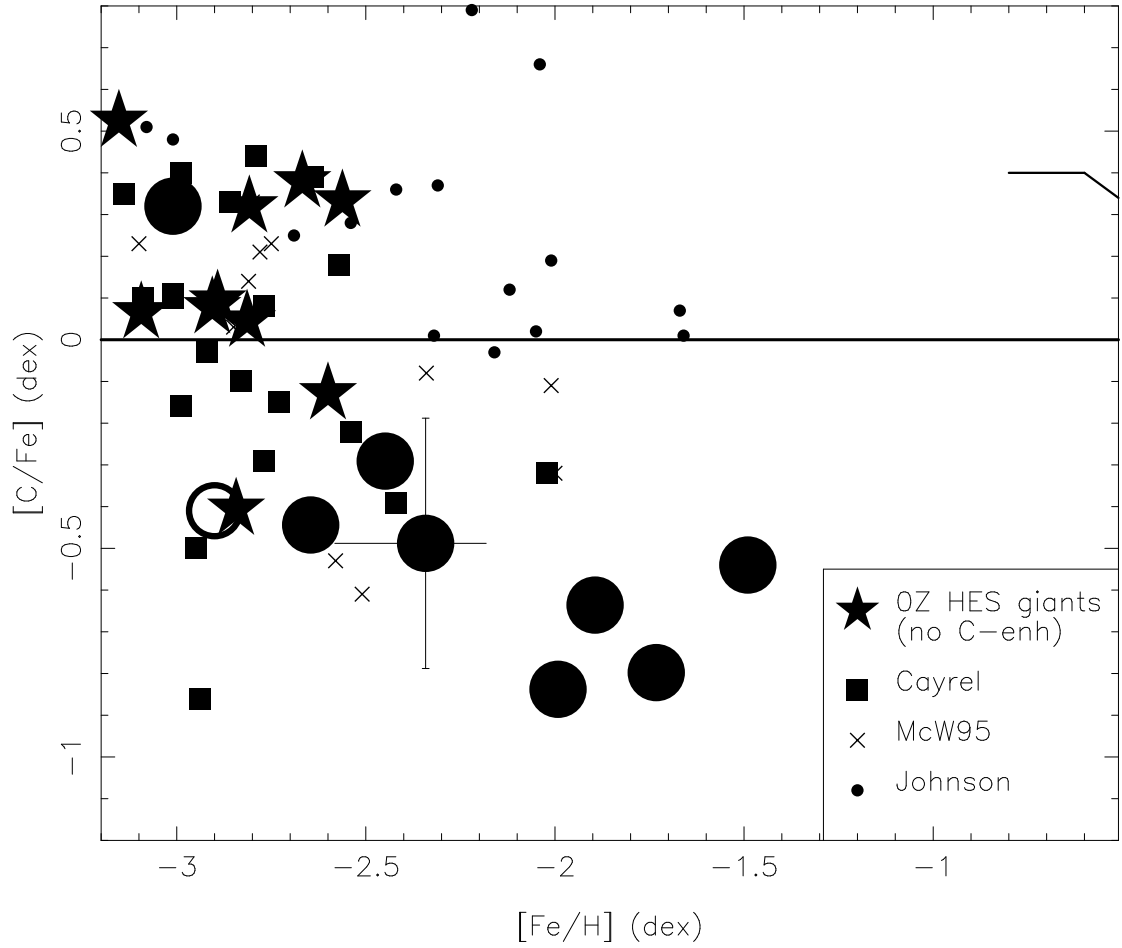}
\caption[]{[C/Fe] from the G band of CH vs [Fe/H] for Draco stars from our 
sample (large filled circles) with that of Draco~119 from \cite{fulbright}
(large open circle); these symbols are used for the remaining figures.
The symbol key for the other sources is given on each figure.
Typical uncertainties are shown for one star. 
The line represents the behavior of halo dwarfs from \cite{reddy06}.
\label{figure_cfeh}}
\end{figure}

\begin{figure}
\epsscale{0.9}
\plotone{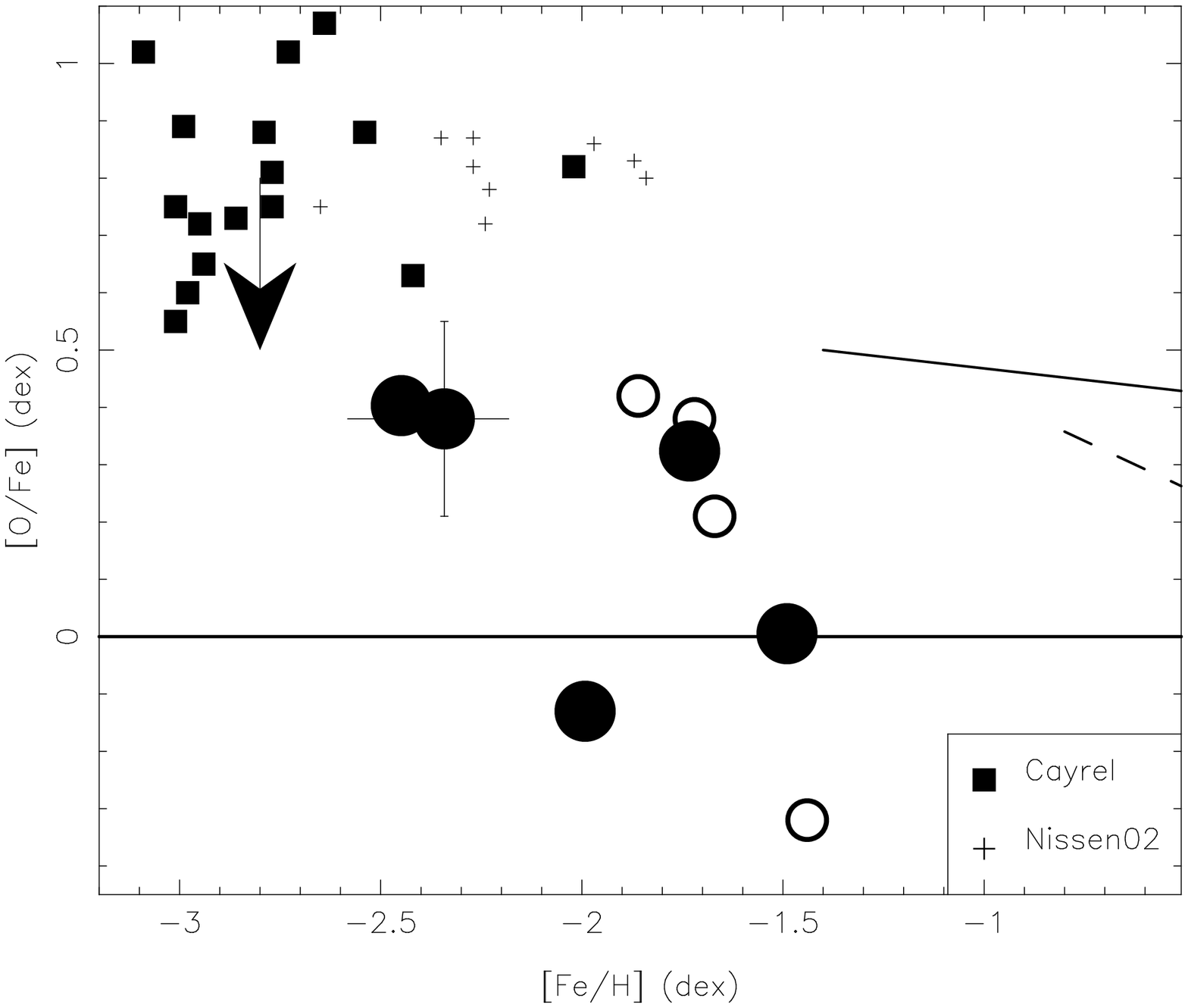}
\caption[]{[O/Fe] vs [Fe/H].  Draco stars from our 
sample (large filled circles) with those of \cite{shetrone2}
indicated by the somewhat smaller open circles.   
Typical uncertainties are shown for one star.
The small crosses are from \cite{nissen02}.  Linear fits to the
thick disk and halo stars (solid line) and thin disk (dashed line)
relations of \cite{ramirez07} are shown.  The arrow indicates
the probable magnitude of 1D to 3D model corrections required for
the \cite{cayrel04} and the Draco [O/Fe] values.
\label{figure_ofeh}}
\end{figure}

\begin{figure}
\epsscale{0.9}
\plotone{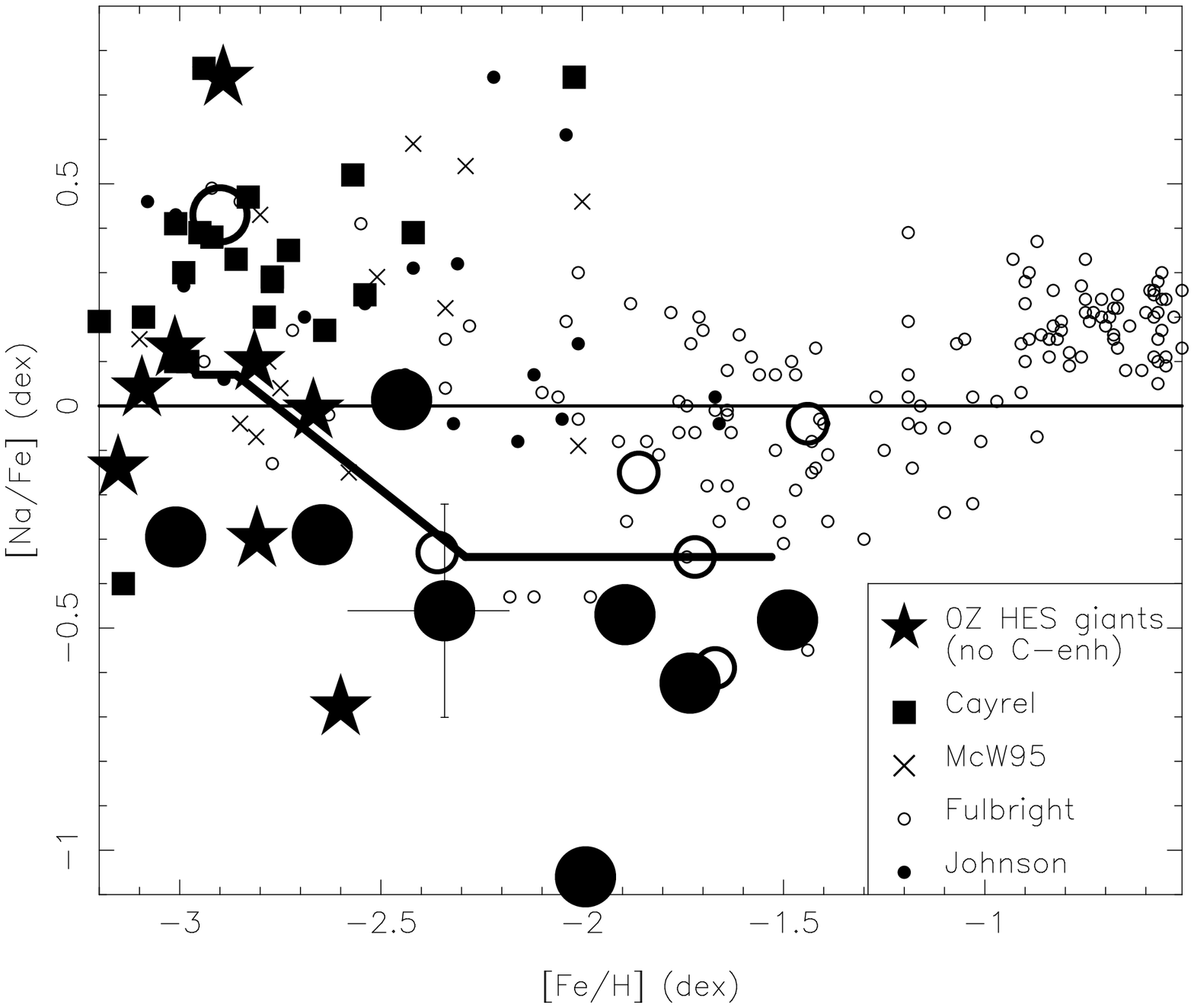}
\caption[]{[Na/Fe] vs [Fe/H] for Draco stars.  
Draco stars from our 
sample (large filled circles) with those of \cite{shetrone2}
indicated by the somewhat smaller open circles and the
updated value for Draco~119 from \cite{fulbright} as
the large open circle.  Typical uncertainties are shown for one star.
The symbol key for other sources is shown on the figure.  The thick line
indicates the fit of the toy model described in \S\ref{section_toy_model}
(see also Table~\ref{table_fit})   to the Draco data.
\label{figure_nafeh}}
\end{figure}

\begin{figure}
\epsscale{0.9}
%
%
\plotone{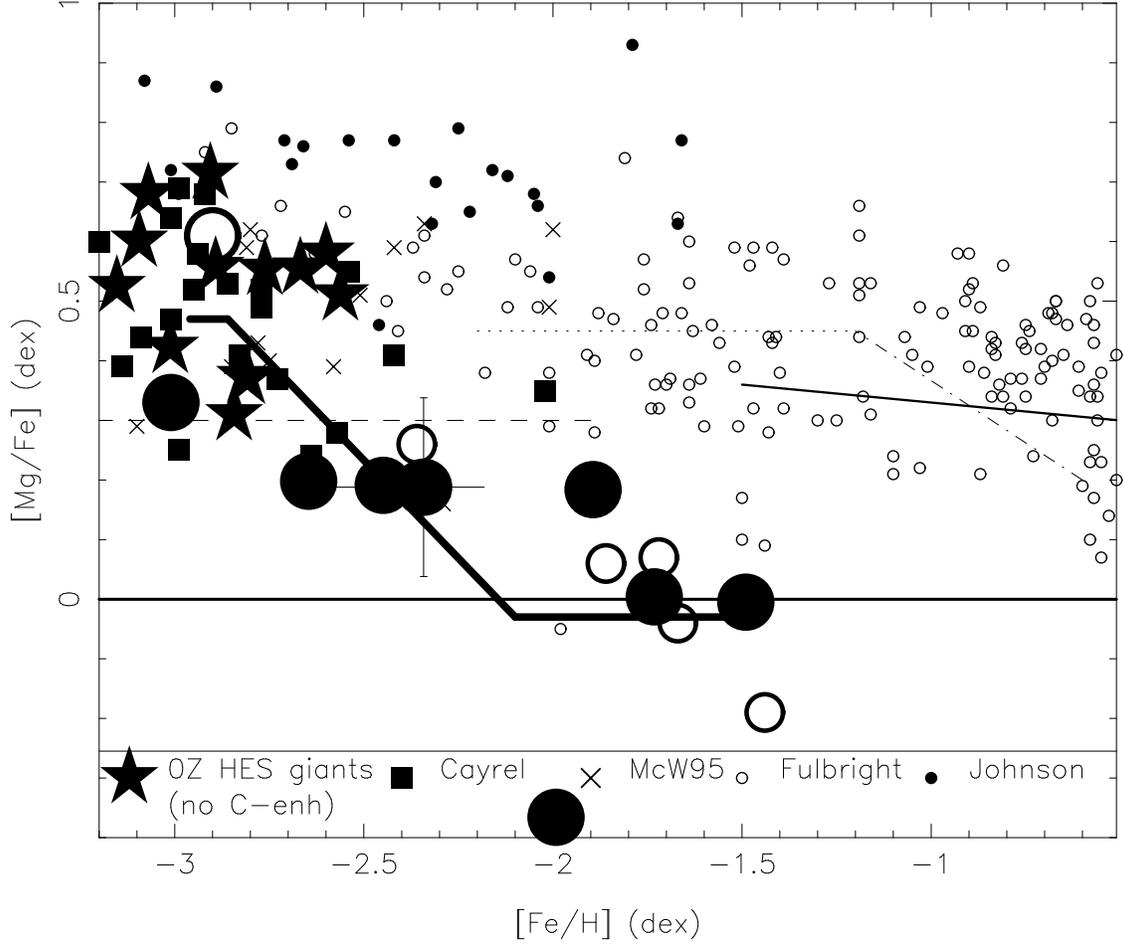}
\caption[]{[Mg/Fe] vs [Fe/H] for Draco stars.
See Fig.~\ref{figure_nafeh}
for details regarding the symbols for the Draco stars and uncertainties.
The symbol key for sources of data for Galactic halo field
stars is shown on the figure.  Note that 0.15~dex has been added to
the [Mg/Fe] values from \cite{cayrel04}; see the text for details.
The thick line
indicates the fit of the toy model described in \S\ref{section_toy_model}
(see also Table~\ref{table_fit})   to the Draco data.
The solid line is the mean relation for thick disk stars
from \cite{reddy06}.  The dotted line is the mean relation for 
inner halo stars from \cite{roederer08}, while his outer halo mean
is shown as the dashed line. 
\label{figure_mgfeh}}
\end{figure}

\begin{figure}
\epsscale{0.9}
\plotone{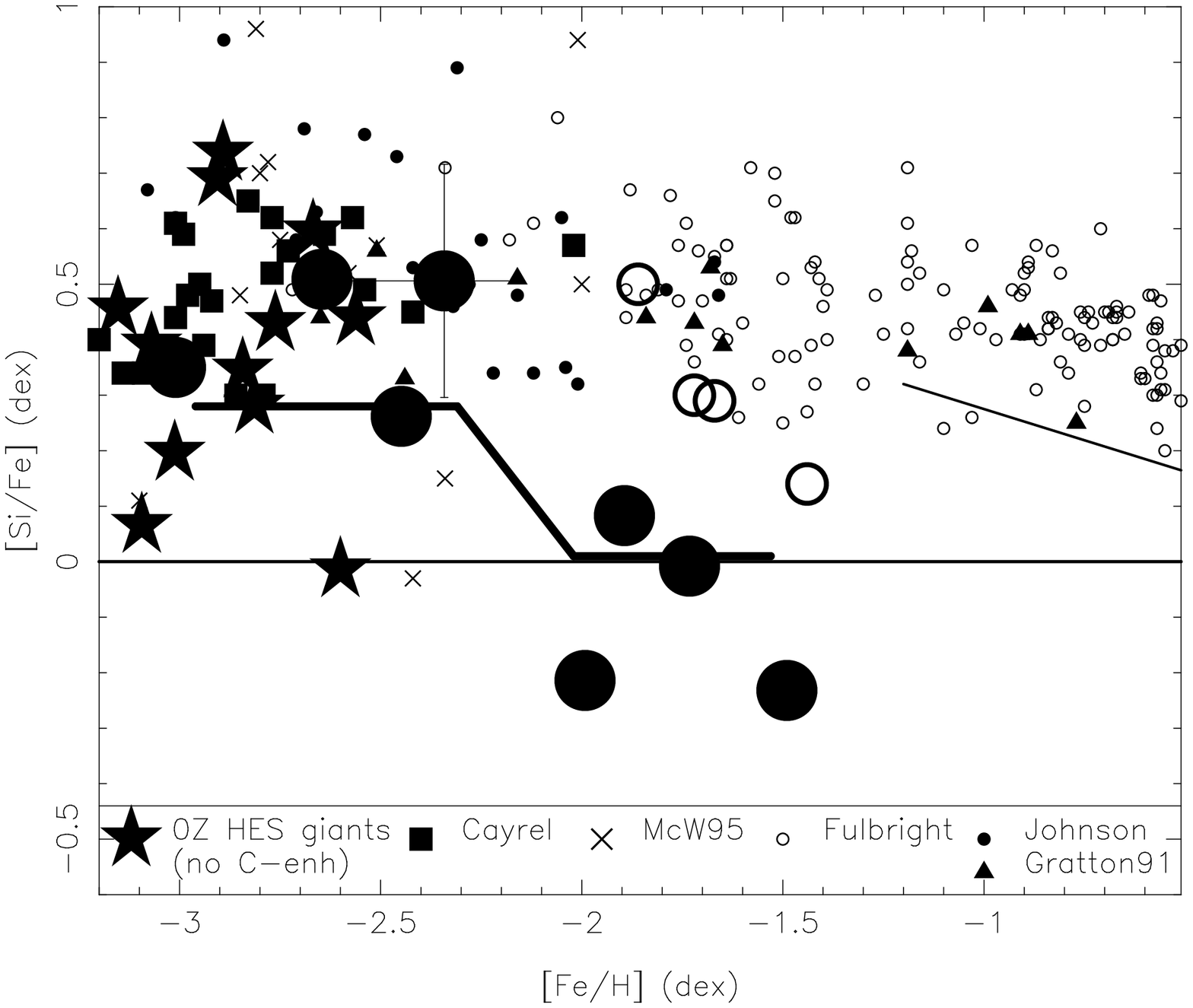}
\caption[]{[Si/Fe] vs [Fe/H] for Draco stars.
See Fig.~\ref{figure_nafeh}
for details regarding the symbols for the Draco stars and uncertainties.
The symbol key for sources of data for Galactic halo field
stars is shown on the figure. 
The thick line
indicates the fit of the toy model described in \S\ref{section_toy_model}
(see also Table~\ref{table_fit})   to the Draco data.  
The solid line is the mean 
relation for thick disk stars from \cite{reddy06}.
\label{figure_sifeh}}
\end{figure}

\begin{figure}
\epsscale{0.9}
\plotone{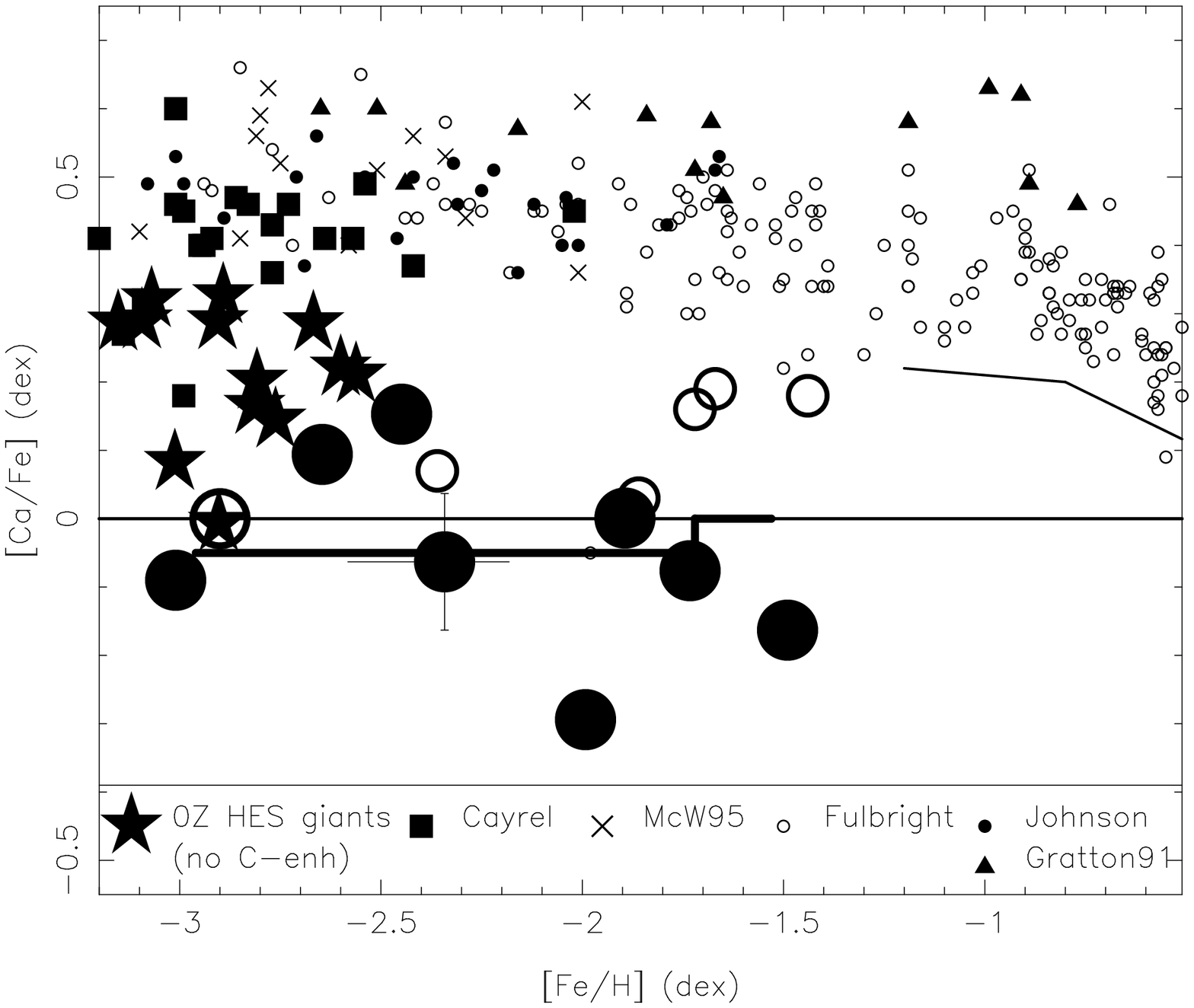}
\caption[]{[Ca/Fe] vs [Fe/H] for Draco stars.
See Fig.~\ref{figure_nafeh}
for details regarding the symbols for the Draco stars and uncertainties.
The symbol key for sources of data for Galactic halo field
stars is shown on the figure. 
The thick line
indicates the fit of the toy model described in \S\ref{section_toy_model}
(see also Table~\ref{table_fit})   to the Draco data. 
The solid line is the mean 
relation for thick disk stars from \cite{reddy06}.
\label{figure_cafeh}}
\end{figure}

\begin{figure}
\epsscale{0.9}
\plotone{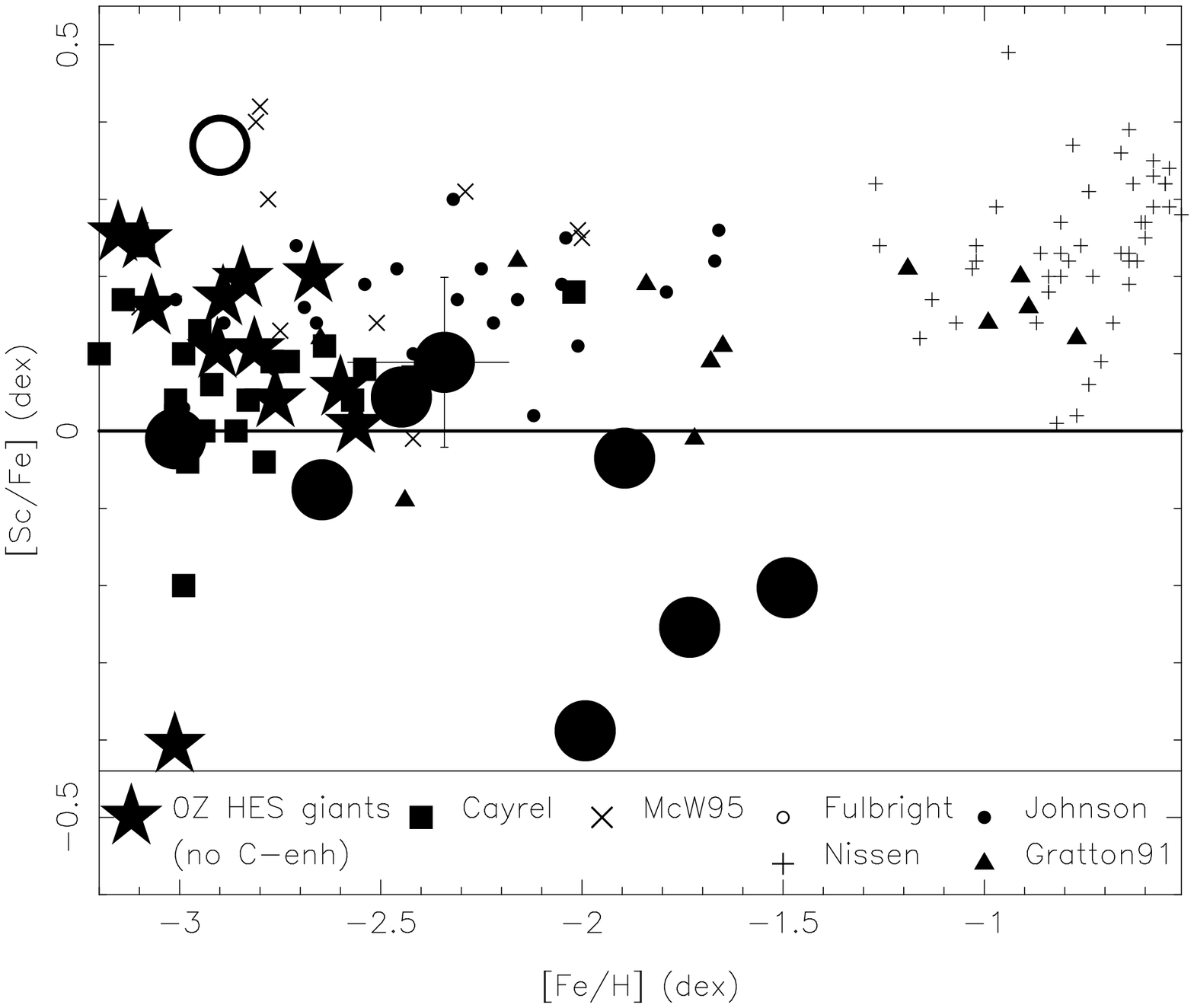}
\caption[]{[Sc/Fe] vs [Fe/H] for Draco giants.  See Fig.~\ref{figure_nafeh}
for details regarding the symbols for the Draco stars and uncertainties.
The symbol key for sources of data for Galactic halo field
stars is shown on the figure.
\label{figure_scfeh}}
\end{figure}

\begin{figure}
\epsscale{0.9}
\plotone{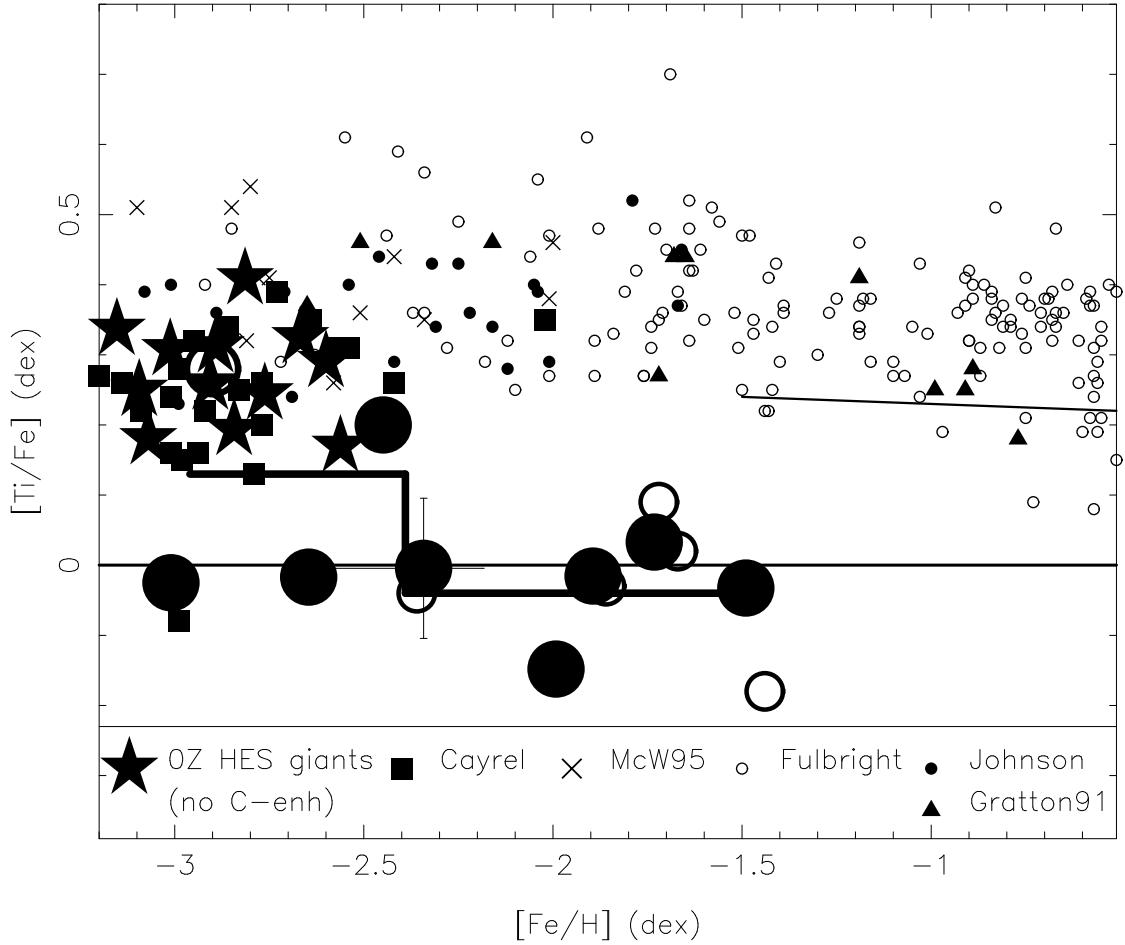}
\caption[]{[Ti/Fe] vs [Fe/H] for Draco stars.
[Ti12/Fe12], which relates ionized Ti to ionized Fe, and neutral
Ti to Fe~I, is shown for our Draco stars.
See Fig.~\ref{figure_nafeh}
for details regarding the symbols for the Draco stars and uncertainties.
The symbol key for sources of data for Galactic halo field
stars is shown on the figure.  The thick line
indicates the fit of the toy model described in \S\ref{section_toy_model}
(see also Table~\ref{table_fit})   to the Draco data.
The solid line denotes
the mean relation for the thick disk stars from \cite{reddy06}.
\label{figure_tifeh}}
\end{figure}

\begin{figure}
\epsscale{0.9}
\plotone{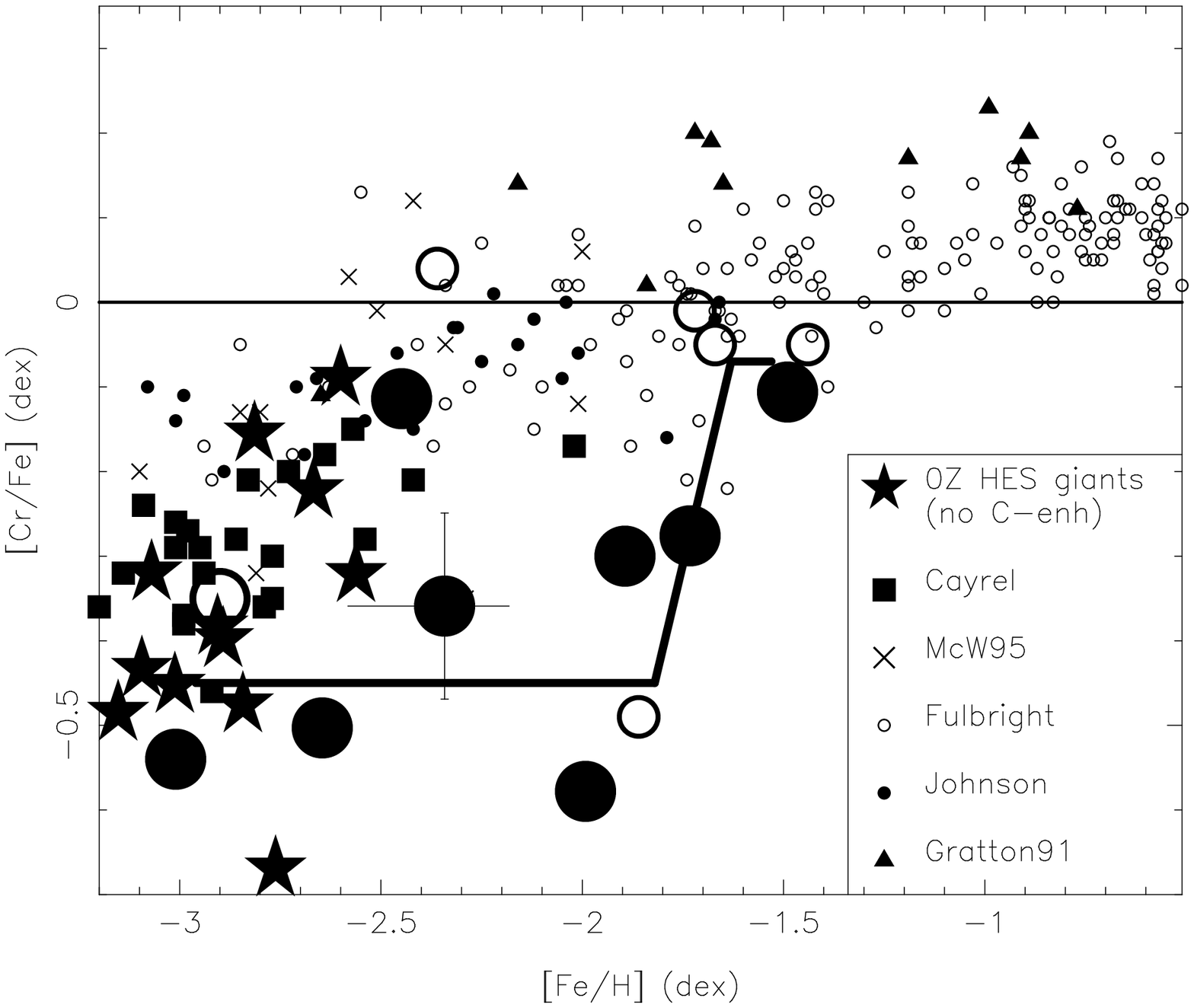}
\caption[]{[Cr/Fe]  vs [Fe/H] for Draco stars.
See Fig.~\ref{figure_nafeh}
for details regarding the symbols for the Draco stars and uncertainties.
The symbol key for sources of data for Galactic halo field
stars is shown on the figure.  The thick line
indicates the fit of the toy model described in \S\ref{section_toy_model}
(see also Table~\ref{table_fit})   to the Draco data.
\label{figure_crfeh}}
\end{figure}

\begin{figure}
\epsscale{0.9}
\plotone{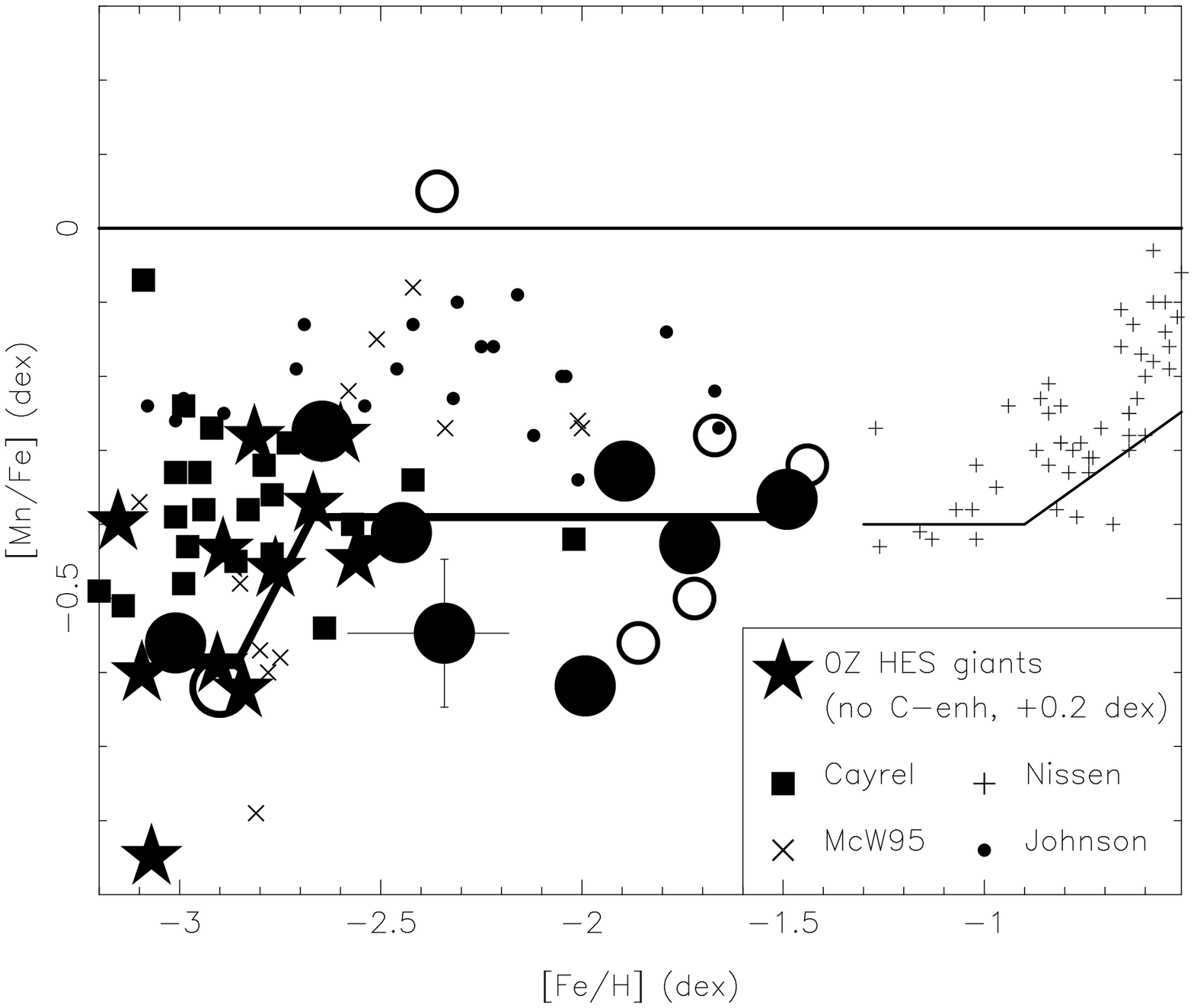}
\caption[]{[Mn/Fe] vs. [Fe/H] for Draco giants.
See Fig.~\ref{figure_nafeh}
for details regarding the symbols for the Draco stars and uncertainties.
The symbol key for sources of data for Galactic halo field
stars is shown on the figure.  The thick line
indicates the fit of the toy model described in \S\ref{section_toy_model}
(see also Table~\ref{table_fit})   to the Draco data.
The solid line denotes
the mean relation for the thick disk stars from \cite{reddy06}.
\label{figure_mnfeh}}
\end{figure}

\begin{figure}
\epsscale{0.9}
\plotone{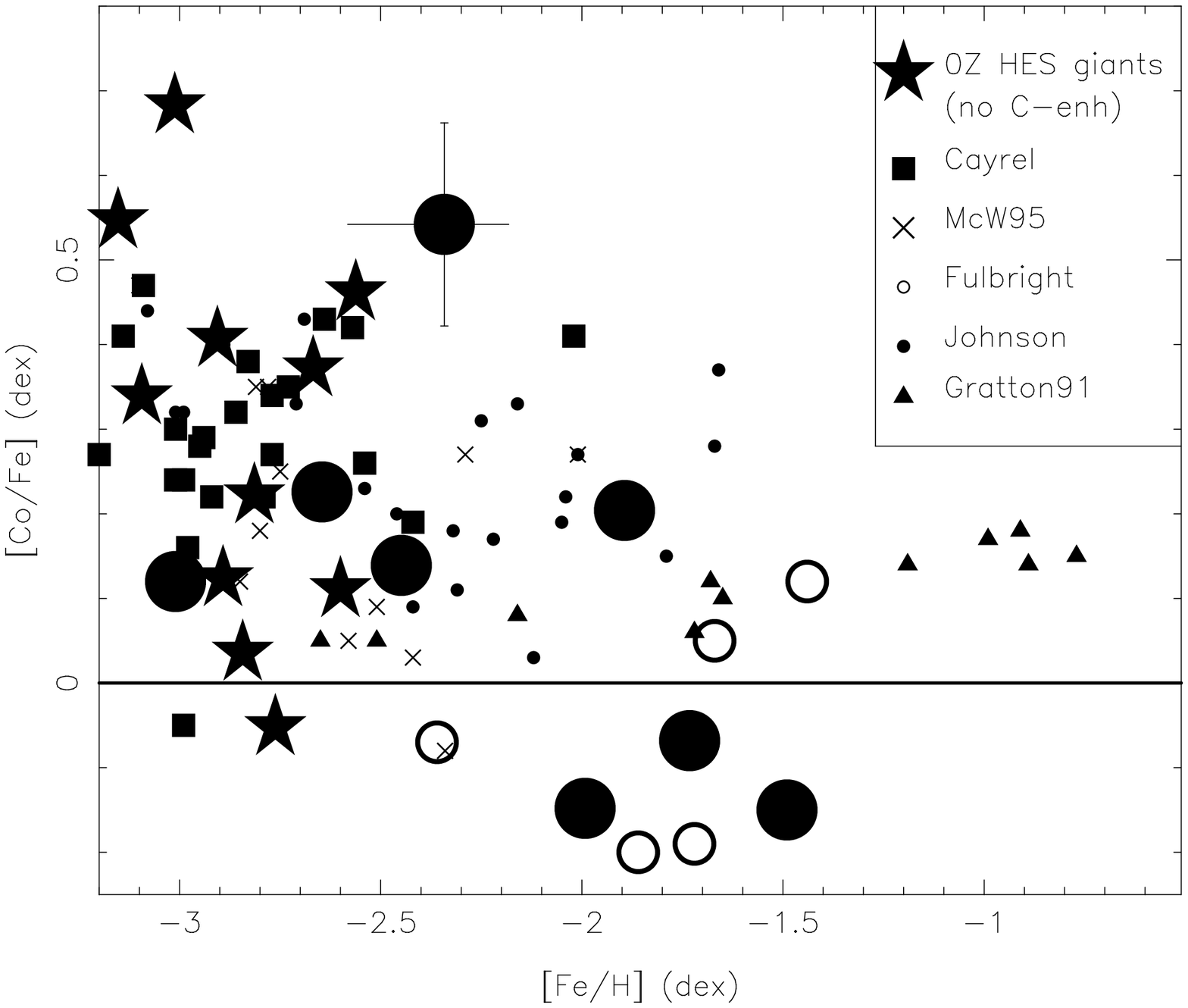}
\caption[]{[Co/Fe]  vs [Fe/H] for Draco stars.
See Fig.~\ref{figure_nafeh}
for details regarding the symbols for the Draco stars and uncertainties.
The symbol key for sources of data for Galactic halo field
stars is shown on the figure.
\label{figure_cofeh}}
\end{figure}

\begin{figure}
\epsscale{0.9}
\plotone{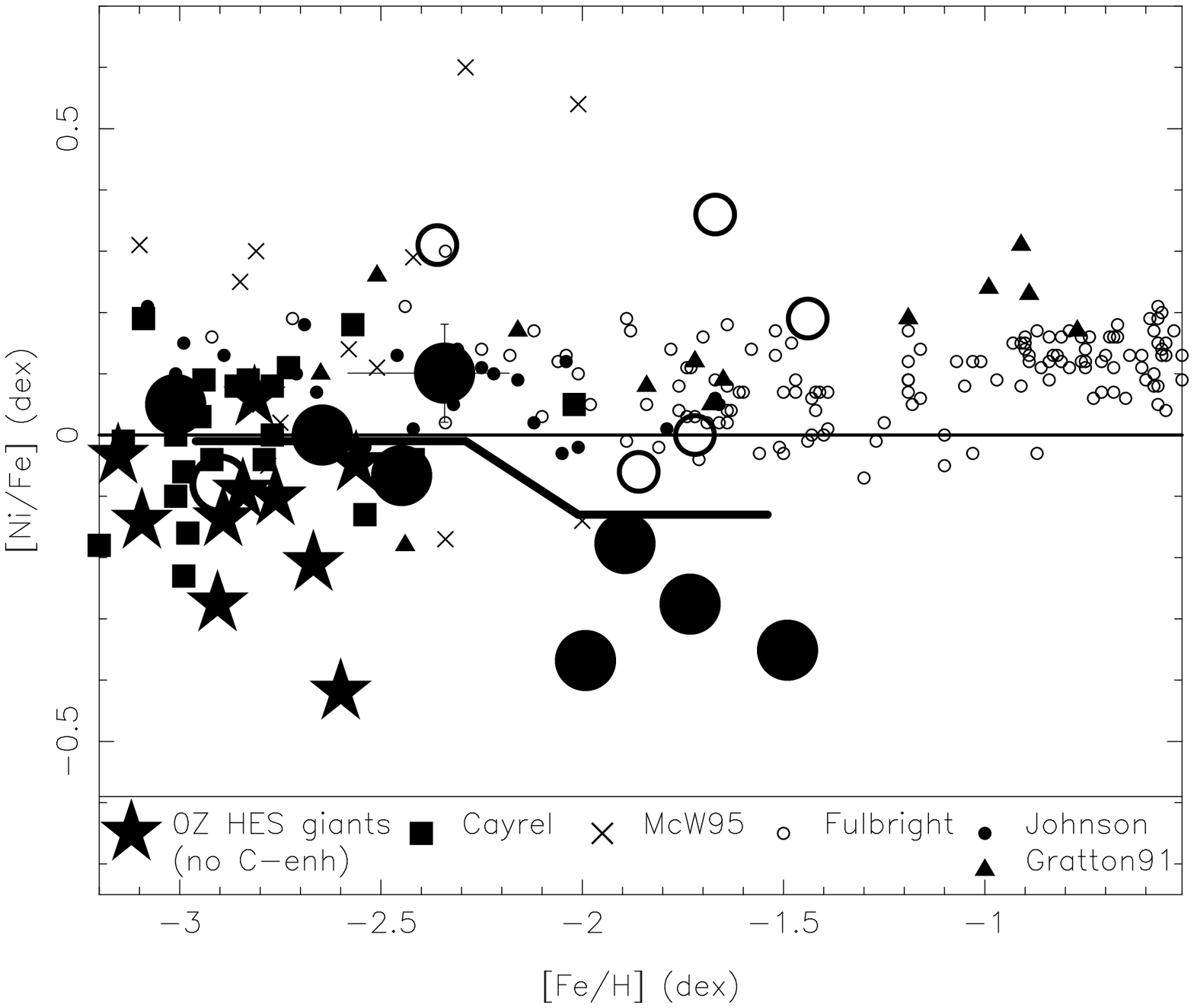}
\caption[]{[Ni/Fe]  vs [Fe/H] for Draco stars.
See Fig.~\ref{figure_nafeh}
for details regarding the symbols for the Draco stars and uncertainties.
The symbol key for sources of data for Galactic halo field
stars is shown on the figure.  The thick line
indicates the fit of the toy model described in \S\ref{section_toy_model}
(see also Table~\ref{table_fit})  to the Draco data.
\label{figure_nifeh}}
\end{figure}

\begin{figure}
\epsscale{0.9}
\plotone{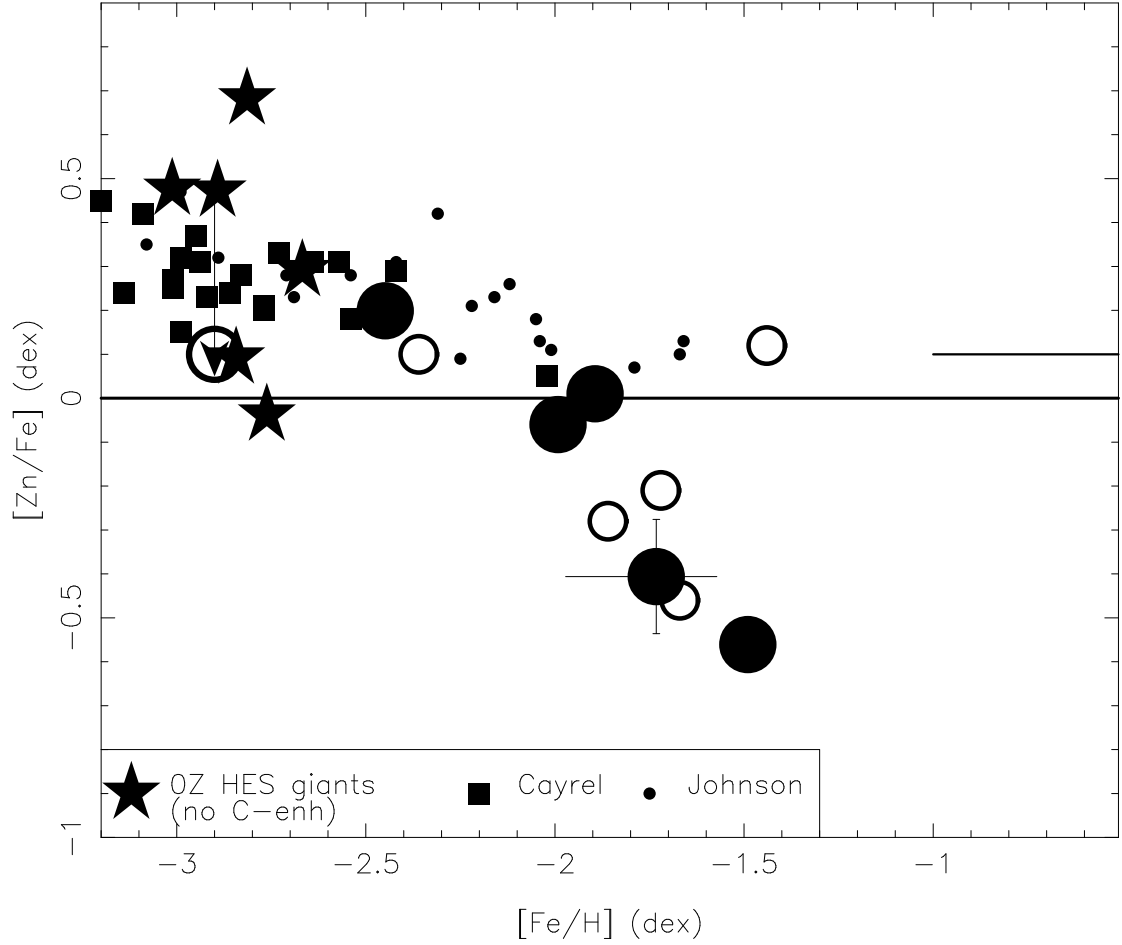}
\caption[]{[Zn/Fe]  vs [Fe/H] for Draco stars.
See Fig.~\ref{figure_nafeh}
for details regarding the symbols for the Draco stars and uncertainties.
The upper limit for Draco~119 from \cite{fulbright} is indicated by
 an arrow.
The symbol key for sources of data for Galactic halo field
stars is shown on the figure.  The behavior
of this abundance ratio in thick disk dwarfs from \cite{reddy06}
is indicated as a solid line.
\label{figure_znfeh}}
\end{figure}

\begin{figure}
\epsscale{0.9}
\plotone{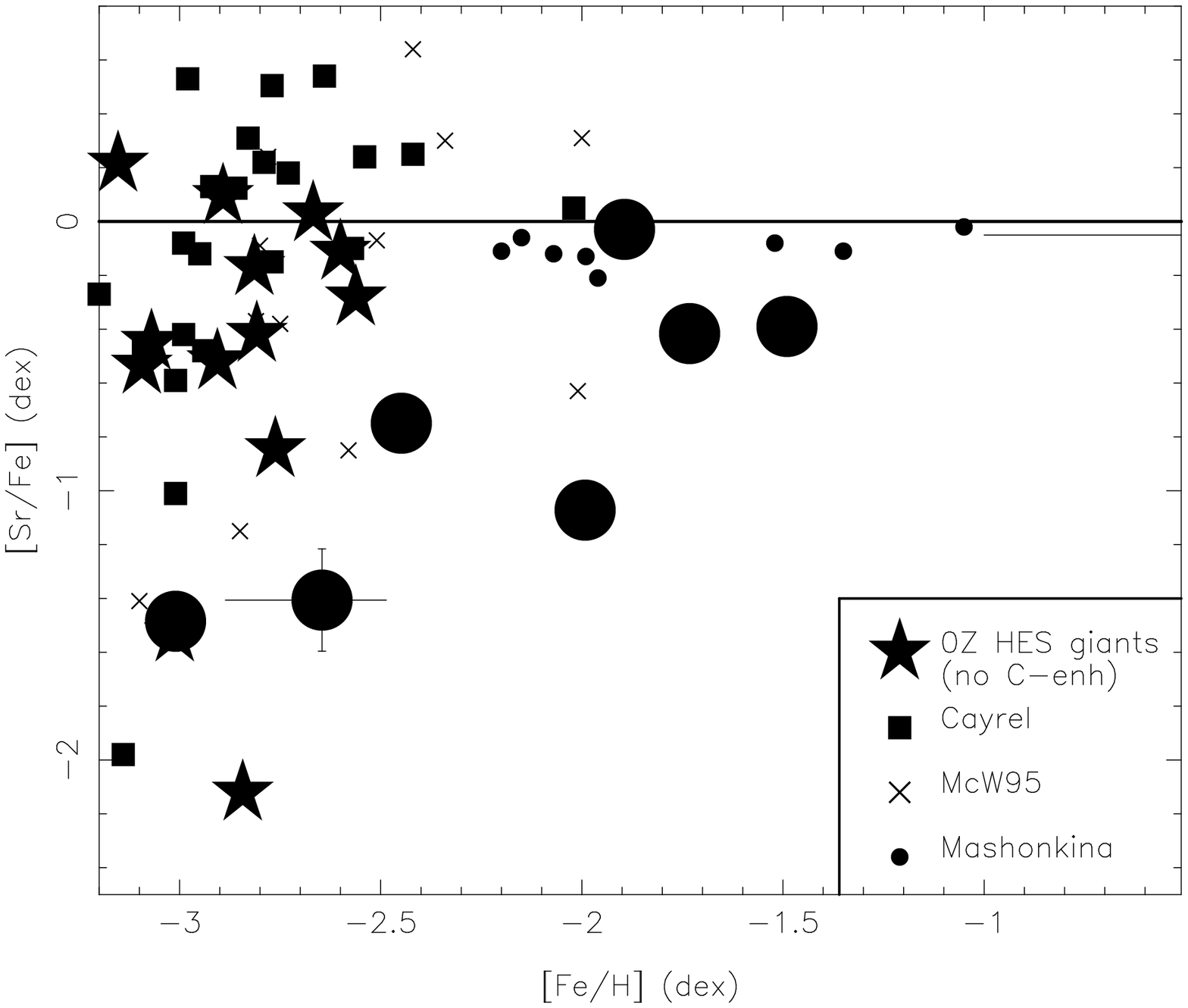}
\caption[]{[Sr/Fe]  vs [Fe/H] for Draco stars.
See Fig.~\ref{figure_nafeh}
for details regarding the symbols for the Draco stars and uncertainties.
The upper limit for Draco~119 from \cite{fulbright} is indicated by
 an arrow.  The First Stars data is from \cite{francois}.
The symbol key for sources of data for Galactic halo field
stars is shown on the figure. 
The thick line
indicates the fit of the toy model described in \S\ref{section_toy_model}
(see also Table~\ref{table_fit})  to the Draco data.
The behavior
of this abundance ratio in thick disk dwarfs from 
\cite{halo_sr} is shown as the solid line.
\label{figure_srfeh}}
\end{figure}

\begin{figure}
\epsscale{0.9}
\plotone{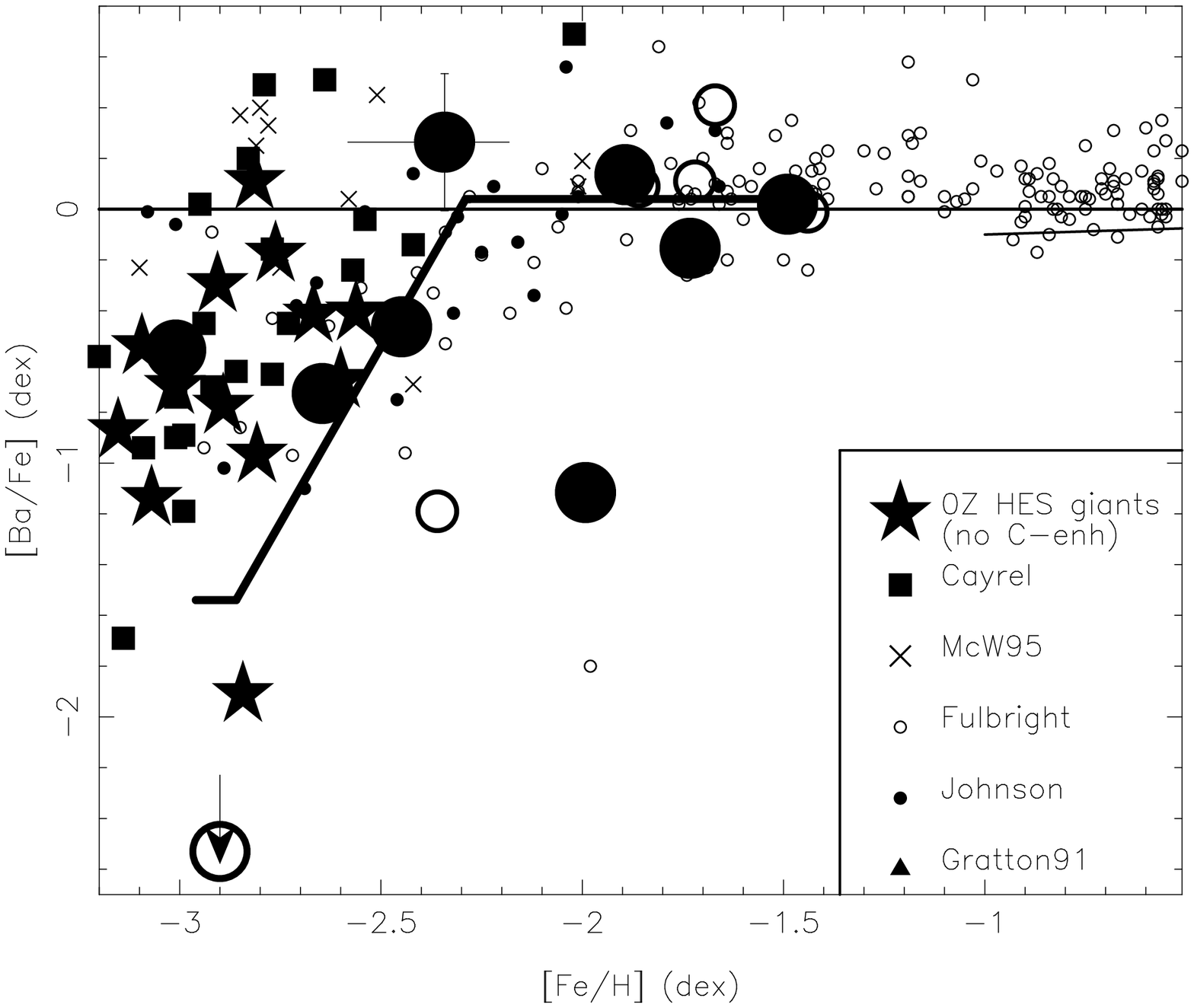}
\caption[]{[Ba/Fe]  vs [Fe/H] for Draco stars.
See Fig.~\ref{figure_nafeh}
for details regarding the symbols for the Draco stars and uncertainties.
The upper limit for Draco~119 from \cite{fulbright}  is indicated by an arrow.
The First Stars data is from \cite{francois}.
The symbol key for sources of data for Galactic halo field
stars is shown on the figure.  The thick line
indicates the fit of the toy model described in \S\ref{section_toy_model}
(see also Table~\ref{table_fit})  to the Draco data.
\label{figure_bafeh}}
\end{figure}

We proceed  by examining a series of plots 
(Figs.~\ref{figure_cfeh} to \ref{figure_bafeh}) in which we show
the Draco sample, both our 8 stars (indicated by large
filled circles) and the 6 observed
previously from  \cite{shetrone2} (denoted by
open circles, and less accurate than subsequent Draco studies), with
the more accurate revisiting of the star Draco~119 by \cite{fulbright} (the 
large open circle) used in preference to the earlier less accurate
values.  These figures
also display results from the 0Z project led by J.~Cohen
to datamine the Hamburg/ESO Survey for extremely metal-poor
stars.  Many of the most metal-poor candidates from this
work have been observed with HIRES at the Keck Observatory
and analyzed in a manner very similar to the present study
as described in \cite{cohen04} and \cite{cohen08},
with the difference that most of the spectra for the 0Z project
were taken further towards the blue than those of the Draco stars,
a move necessary because of the low density of lines in the red
in the spectra such low
metallicity stars.   Only the metal-rich end of
the 0Z project database, much of which is not yet published
(J.~Cohen, N.~Christlieb et al, in preparation),
is shown in these figures. 
A number of other halo field star surveys,
the most important of which at the lower metallicities probed here
is the First Stars Project  \citep{cayrel04}, are shown as well as
those stars from \cite{mcwilliam95a}
not re-observed by \cite{cayrel04}. \cite{fulbright02},
\cite{johnson02} and other sources noted in the symbol key
shown on each figure provide coverage at the higher metallicities
reached by the Draco stars.  \cite{nissen00} add higher
metallicity information for Sc and Mn.

We examine plots of [X/Fe] vs [Fe/H] for each element
and use them
to look for differences
in the chemical inventory
between Galactic halo field stars and the Draco sample,
which may be a function of Fe-metallicity.  
We will see that the differences are small, not larger than $\sim$0.3~dex
in most cases.  This means that some care is required
to ensure that all the abundances from the various sources
are homogeneous.  While we have not done a full check of this,
we have taken a few steps the first of which is
to adjust each survey to our set of Solar abundances,
particularly to our adopted value of [Fe/H], whenever
possible. Specific
cases where there are clear problems related to issues
of homogeneity between the various analyses are discussed individually below.
It should be noted that the analyses for our Draco sample, including the
methods used to
determine stellar parameters, are essentially identical
to those of the 0Z Project.

The trend of [C/Fe] versus [Fe/H] is shown in Fig.~\ref{figure_cfeh}, based
for the majority of these stars on the strength of the G band of CH.
The solid lines represent the mean behavior of thick disk dwarfs 
from the survey by \cite{reddy06}.
The C abundance in luminous giants is lowered substantially
from an initial [C/Fe] $\approx$ 0.0~dex 
due to intrinsic nucleosynthesis (the CN cycle of H burning)
followed by dredge-up to the stellar surface of processed material
within which C has burned to N
\citep[see e.g.][]{cohen_m15}.  
Thus  the initial
C abundance of the Draco giants cannot be determined.
Note that in Fig.~\ref{figure_cfeh} and those that follow the asymmetric
uncertainties we adopted in \S\ref{section_param}
are shown for [Fe/H], but not for abundance ratios
[X/Fe]; for the latter the larger uncertainty is plotted.

It is quite difficult to measure  O abundances in metal-poor giants. 
The set of features that can be used
is very limited and each has problems.
This has resulted in considerable
controversy about O abundances in metal-poor stars in recent years,
see e.g. the discussion in \cite{melendez06}.
The forbidden OI lines at 6300 and 6363~\AA\ line are very weak, and
the 7770~\AA\ triplet, which has substantial non-LTE effects, 
is not detectable.   [O/Fe] ratios for the Draco giants
and for a compilation of surveys in the literature are shown
in Fig.~\ref{figure_ofeh}.

The arrow in Fig.~\ref{figure_ofeh} indicates the probable
correction for 1D to 3D effects required for luminous
giants given by \cite{cayrel04}, which has not been implemented,
but which would bring the plateau in [O/Fe] down to a mean
level of $\sim$+0.5~dex. 
The lines are linear fits from \cite{ramirez07} to their
samples of
thick disk and halo dwarfs (solid line) and to thin
disk dwarfs (dashed line).  They use only 
the 7770~\AA\ triplet, with appropriate non-LTE corrections;
these lines become detectable in dwarf stars.
The net result is that the
Draco giants appear low in [O/Fe] when compared to 
samples of field halo giants which rely on the same 6300~\AA\
forbidden line. The more metal-rich
thick disk and halo dwarfs can be made consistent in [O/Fe] with
the metal-poor luminous giants in the halo and in the Draco
giants once the
3D to 1D correction suggested by \cite{cayrel04} is implemented.

Fig.~\ref{figure_nafeh} suggests
that there is a separation
of $\sim$0.2~dex for [Na/Fe]  at a
fixed [Fe/H] between the two large surveys of very metal-poor
halo field stars, i.e. the First Stars
Survey led by R.~Cayrel and the 0Z Survey led by J.~Cohen. 
\cite{cayrel04} use only the two NaD lines to determine the
Na abundance, as was the case for most of the stars
from the 0Z project;  only few of the more metal-rich
giants in the 0Z database had detectable 5682,~5688 Na~I lines.  
The origin of this offset
is not clear.  It could result from differences in the samples.
The 0Z sample is $\sim$3~mag fainter on average than is the First
Stars sample, and so is picking up many relatively nearby lower 
luminosity RGB giants,
as well as a few very distant giants near the RGB tip; the distribution
in distance and in luminosity
might be somewhat different within the much brighter 
First Stars sample.   [Na/Fe] ratios vary enormously from star
to star among very metal-poor stars, perhaps as a function
of Galactocentric radius, or  
perhaps as a result
of deep mixing, symptoms of which have been found among
very metal-poor halo giants by \cite{spite}, or contamination from a former
AGB companion.

Non-LTE corrections for the NaD doublet
for metal-poor giants calculated by \cite{takeda_na} are $\sim -0.2$~dex, but
no non-LTE corrections for Na were applied to 
any of the Draco stars nor to
those from the 0Z or First Stars projects.
In spite of these concerns, the
Draco giants clearly have [Na/Fe] somewhat lower than the
Galactic halo field stars, with Draco~XI-2, a high luminosity
giant in our sample,
having an extremely low [Na/Fe].   At the lowest metallicities,
the Draco sample appears to overlap the Galactic halo field stars.
 

Fig.~\ref{figure_mgfeh} shows the
important $\alpha$-element Mg, another element
with only a few accessible features in our Draco spectra.
The published values from \cite{cayrel04} for
the First Stars project have been increased by 0.15~dex.
This was done 
in accordance with \cite{bonifacio09}, who state
in {\S}6.1 of their paper that an error was made by the First Stars group in the
measurement of \eqw\ for the strong Mg triplet
lines in the more metal-rich giants in their sample
such that [Mg/Fe] was underestimated by this amount
for these stars in the tables of \cite{cayrel04}. 
The $gf$ values for Mg~I lines are uncertain, and
systematic differences of $\sim$0.05~dex may arise
between surveys from the choice of $gf$ values for the Mg lines,
depending on the specific Mg lines measured in a particular star,
as discussed in {\S}7.4 of \cite{cohen04}.
We also note that
the sample from \cite{johnson02} is systematically high
in [Mg/Fe] across more than 1~dex in [Fe/H] compared
to  both the 0Z project and the First Stars Project. 
She took her Mg~I $gf$ values from three sources; only those
taken from the most recent of the three \citep{fuhrman_mg} agree well with
the values we adopt.  We find
that [Mg/Fe] definitely is low in the Draco stars at
[Fe/H] $> -2$~dex, but at the lowest metallicities
the Draco giants overlap with the 0Z Survey and First Stars Survey
giants, once the correction suggested by \cite{bonifacio09} is implemented.

The  behavior of the $\alpha$-element Si is
shown in Fig.~\ref{figure_sifeh} with the mean relation for thick
disk stars from \cite{reddy06} indicated.  The figure shows
good agreement  between the 0Z and First Stars Project
abundance ratios for this element;
[Si/Fe] from \cite{johnson02} is again slightly high.  The behavior
of [Si/Fe]  in the Draco giants is similar to that described above
for [Mg/Fe], i.e.  the lowest metallicity
Draco giants overlap
the Galactic halo stars, while at somewhat
higher metallicities, where the Galactic halo field stars
still have elevated [Si/Fe], $\sim$~+0.5~dex, the Draco
stars have fallen to the Solar value.

The explosive $\alpha$-element Ca also has problems with inconsistencies
between the two large surveys of very metal-poor halo stars,
the First Stars Project and the 0Z Project.  An offset of $\sim$0.2~dex
is required for agreement between them.  
This problem appears to originate in our 0Z database,
as the mean [Ca/Fe] for dwarfs given in \cite{cohen04}, which
is $\sim$0.15~dex higher than that found here, agrees
well with that of the First Stars Survey 
(Cayrel et al 2004 for halo giants, Bonifacio et al 2009 for halo
dwarfs, their two means for [Ca/Fe] in Galactic halo field stars
are identical).
The $gf$ values adopted
are essentially identical for the two large extremely metal-poor halo field
star surveys, and are those of
NIST.  Many of the older analyses use those from \cite{cagf_old},
which scatter widely around the recent values.  Thus, as is also the case for Mg~I,
ideally one must correct each star individually, checking
which specific Mg~I and Ca~I lines were used and what $gf$ values were adopted.
Irrespective of
this issue, the luminous Draco giants have [Ca/Fe] lower
than that of the Galactic halo stars over the full
range of [Fe/H] found in Draco.  At the lowest Fe-metallicities
probed, they come close to the abundance ratios for Ca from
the 0Z project, but don't quite reach as high in [Ca/Fe].


Figs.~\ref{figure_scfeh} and \ref{figure_tifeh}
show the behavior for [Sc/Fe] and [Ti/Fe] respectively.
The mean relation for thick
disk stars from \cite{reddy06} is shown for the latter.
In both cases there is good agreement between the
abundance ratios deduced by
the 0Z Project and the First Stars Project.
For the explosive $\alpha$-element
Ti as well as for Sc the
metal-rich Draco stars fall well below the
halo field. [Ti/Fe] is slightly sub-solar
in this regime; [Sc/Fe] is definitely sub-solar. 
Among the lowest metallicity Draco giants [Sc/Fe]
overlaps the halo field star samples, while
for [Ti/Fe] the Draco stars still appear $\sim$0.2~dex  low
compared to the Galactic halo field stars.

At high metallicities, [Cr/Fe] (Fig.~\ref{figure_crfeh}) appears lower among
the Draco stars than among the Galactic halo field stars.
The difference, if it exists at all, is  smaller
for the lowest metallicity Draco stars.
[Mn/Fe] (Fig.~\ref{figure_mnfeh})
declines rapidly from Solar to about $-0.4$~dex
as [Fe/H] decreases in Galactic halo field stars.  A known problem discussed in
\cite{cohen04} requires that the Mn abundance be increased
by 0.2~dex for stars where the 4030 resonance triplet lines are used.
These are the strongest Mn~I lines in the optical and the only
ones accessible for extremely metal-poor stars, but are too blue
to be included in our Draco analyses.  This offset has been applied
to the 0Z Project analyses; 
the First Stars Project values
already contain this offset.  The Draco giants overlap
the Galactic halo field stars at all Fe-metallicities.

Fig.~\ref{figure_cofeh} displays the [Co/Fe] ratios which
for Galactic halo stars
rise rapidly from near the Solar ratio as [Fe/H] decreases below $-2$~dex.
It appears that the lowest Fe-metallicity Draco giants follow
this trend, with the higher metallicity Draco giants 
perhaps lying
slightly lower than do Galactic halo field stars.  
However, there is only one Co~I line
detected in most of these stars; only a few have measurements for
the stronger line
at 4121~\AA, which is uncomfortably far in the blue, as
confirmation of the deduced abundance ratio. Thus any 
conclusion regarding the behavior of [Co/Fe] in Draco is 
still uncertain.

The nickel abundance relative to Fe (Fig.~\ref{figure_nifeh}) 
appears to fall below that of the
halo field (which has [Ni/Fe] at the Solar ratio over the
entire range of Fe-metallicity) among the higher metallicity
Draco stars.  The [Ni/Fe] ratios
for the lowest metallicity Draco stars overlap
in Fig.~\ref{figure_nifeh} those of  Galactic field halo stars.

The Galactic halo field samples from the 0Z Project and the
First Stars Project overlap well for the abundance ratio [Zn/Fe].
Among field halo stars, [Zn/Fe] is close to the Solar ratio
but rises rapidly below [Fe/H] $\sim -2$~dex, as shown
most recently for halo dwarfs by \cite{nissen08}.  In the Draco giants,
[Zn/Fe] behaves similarly to the Co and Ni abundance ratios
for intermediate metallicities, where the Draco stars
fall below those in the Galactic halo.  At the lowest
metallicities, [Zn/Fe] may rise above the Solar value, but there are not
enough detections among the lowest metallicity stars in
the Draco sample to be certain.

The Galactic halo field samples from the 0Z Project and the
First Stars Project \citep[data for Sr and Ba is from][]{francois} 
overlap well for the abundance ratio [Sr/Fe]
versus Fe-metallicity shown in Fig.~\ref{figure_srfeh}. 
The Draco stars over the full range of Fe-metallicity have
[Sr/Fe] ratios that
fall below the bulk of the Galactic sample,
although, except for Draco~119, which only has an upper limit
to [Sr/Fe] from \cite{fulbright}, they are not outside the range 
of the low outliers in
the much larger  Galactic sample. 
The Sr~II lines used are
the resonance lines at 4077 and 4215~\AA; they are
uncomfortably far
in the blue, where the SNR of our Draco spectra is rather low.
In addition, the
Draco stars are right at the tip of the RGB, and hence are cooler
and more luminous than almost all of the stars in the Galactic halo
surveys of the 0Z Survey and most of those
in First Stars Project, as well as essentially all
in the surveys of somewhat higher metallicity field halo stars used here.
One might be concerned that some differential error, for
example, the magnitude of the non-LTE corrections, might
affect our comparisons shown in Fig.~\ref{figure_ofeh} to
\ref{figure_bafeh}. This might be an issue
for [Sr/Fe] in view of the large positive non-LTE corrections
for low metallicity stars
calculated by \cite{halo_sr} and by \cite{short06}
which appear from the limited number
of cases studied to be
highly variable with \teff\ and \grav.  However,  
the sample of luminous giants in the Galactic
globular cluster NGC~7099 of J.~Cohen (unpublished)
are closer in \teff\ to the Draco giants.  These GC
giants with [Fe/H] $\sim -2.4$~dex
show Solar [Sr/Fe] and [Co/Fe] as well, suggesting that any such differential
effect is probably not very large.

Fig.~\ref{figure_bafeh} shows the abundance ratios [Ba/Fe],
with the mean for the Galactic thick disk from \cite{reddy06} indicated as a solid
line.
The Draco stars as well as the Galactic halo field stars display
a very large range in [Ba/Fe] at low metallicities
(see Fig.~\ref{figure_bafeh}).  Most of the Draco stars
lie within the regime occupied by the majority of the
Galactic halo stars from the 0Z Project and the First Stars project.
There are however two Draco giants that are moderately low,
though still within the range of the Galactic giants, while
Draco~119 \citep{fulbright02}, which only has an upper limit
for [Ba/Fe], is an extremely low outlier
and lies below any known Galactic star.

\subsection{The Outer Galactic Halo \label{section_outer} }

The SDSS \citep{york00,sdss_dr5} has obtained moderate resolution
spectra and multi-color photometry of
very large samples of
Galactic halo stars.  Their database is  
large enough and uniform enough
to determine the properties (spatial, kinematic, and chemical)
of such stars as a function of their Galactocentric distance,
which program was carried out by 
\cite{carollo}.  They demonstrate in a definitive way that
the outer halo with $R_{GC} > 15$~kpc contains
in the mean stars 
which have a more spherical spatial distribution with 
low retrograde net
rotation about the center of the Galaxy and with lower metallicity
in the mean than does the stellar population of the inner halo. 
However, their discussion is restricted to Fe-metallicities. 

Much smaller samples of outer halo stars in the local neighborhood
have been isolated and their chemical inventory analyzed in
detail in several previous studies, in particular
by \cite{nissen97} and by \cite{stephens99}, who searched for local
dwarfs with retrograde motions.  
They comment on some peculiarities unearthed in their abundance
analyses of such retrograde dwarfs, which they consider
as outer
halo stars, as compared to metal-poor dwarfs with normal orbital
characteristics.  Such anomalous dwarfs tend
to have low $\alpha$-ratios and perhaps have [Ni/Fe] somewhat below the
Solar value;
\cite{stephens99} found $<$[Ni/Fe]$>~ = ~ -0.09~{\pm}0.07$~dex.
These are reminiscent of some of the differences noted earlier between
the Galactic halo field stars and the giants from the Draco dSph galaxy.

\cite{roederer08} has recently compiled a sample of halo stars
with published detailed abundance analyses and with parallaxes
so that their orbits within the Galaxy can be derived.  
He classifies his sample on the basis of kinematics
into inner halo and outer halo stars, and has again found
a small deficit in [Mg/Fe] in outer halo stars as compared to inner
halo ones shown by the dotted and dashed lines in Fig.~\ref{figure_mgfeh},
but he does not find any difference for the other $\alpha$-elements
he could study, Ca and Ti.  He also finds
a somewhat larger
scatter in abundance ratios in outer halo stars than in
inner halo stars for [Ni/Fe] and for [Ba/Fe].  Much more work
remains to be done along these lines, but these tantalizing
hints suggest that to find stars with the chemical inventory
of the Draco dSph one must look far out in the halo rather
than at the closer in samples that have been well studied
over the past decade by the First Stars Project of \cite{cayrel04}.
A fraction of the 0Z Survey resides in the outer halo (in situ,
not stars moving with retrograde velocities through the Solar
neighborhood, see Sch\"orck et al 2009)
but the necessary parallaxes for most of these stars are not
available, or are so uncertain and close to zero to be useless.

The many abundance analyses for giants in Galactic globular clusters,
reviewed by \cite{gratton_araa}, reveal abundance ratios that
follow those seen in the inner halo, so that the
stellar population in the inner halo
could easily be composed of globular clusters that dissolved
before the correlations among the light elements 
C,N,O, Na, Mg and Al (widely believed to
result from contributions from intermediate mass AGB stars)
were imprinted.
However, there are a few exceptions such as Pal~12, which according
to \cite{cohen04} has a chemical inventory similar to that of intermediate
metallicity dSph stars. Specifically [Ca/Fe] is almost Solar, 
as is [Ti/Fe], and Na is very deficient.  This provides
evidence, also supported by kinematic data and calculation
of Galactic orbits, that Pal~12 originally was a cluster in
the Sgr dSph galaxy, which is  currently being accreted by the Milky Way.
During the process of accretion Pal~12 was presumably
stripped from the Sgr dSph to become part of the extended Sgr stream;
\cite{layden00} suggest several other Milky Way globular clusters 
were also stripped from the Sgr dSph, one of which is Terzan~7.
An abundance analysis of this globular cluster by \cite{sbordone05}
shows the same anomalies as noted above for Pal~12.  
The globular cluster M54 is at the location of the nucleus
of the Sgr dSph galaxy, but \cite{siegel07} suggest that it is not the
nucleus itself, but rather was pulled in very close to it by dynamical
friction.  Many of the most distant Galactic GCs, which tend
to show somewhat younger ages than the inner halo GCs
\citep[see, e.g.][]{acs_ages}, are now
believed to have been accreted by the Milky Way.

\section{A Toy Model for the Draco Abundances \label{section_toy_model} }

\begin{figure}
\epsscale{0.9}
\plotone{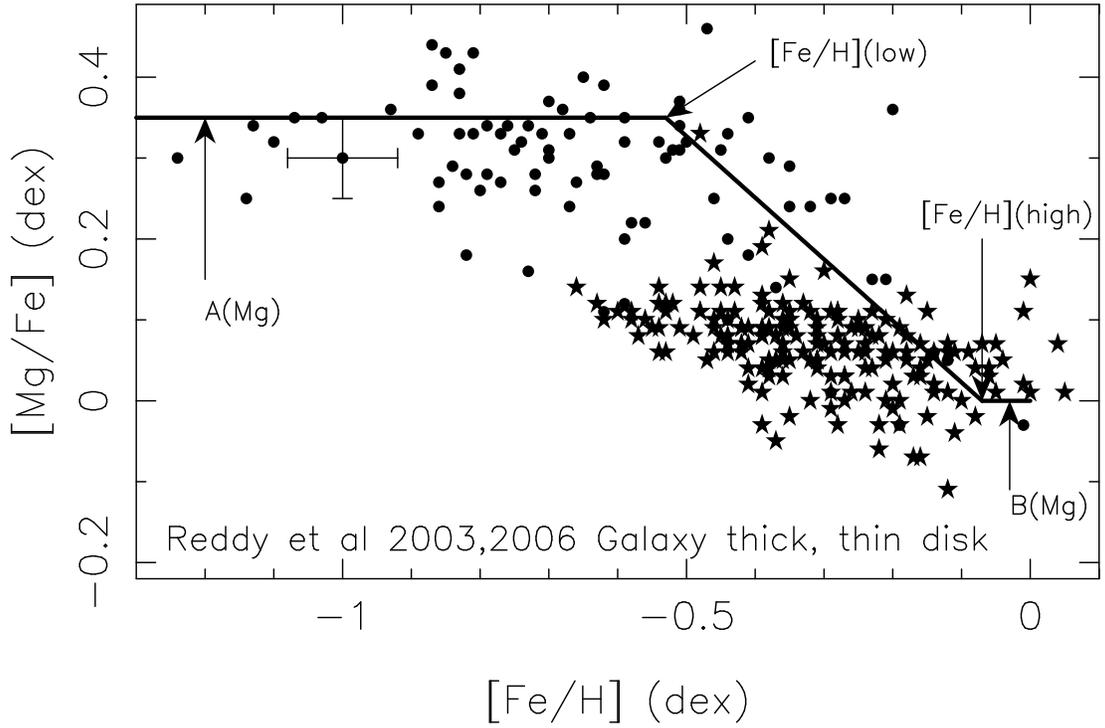}
\caption[]{Abundance ratios [Mg/Fe] are shown as a function of [Fe/H] for a 
large sample
of Galactic thick disk stars (filled circles) from \cite{reddy06} and thin disk stars
(star symbols)
from \cite{reddy03}.  Typical errors are shown for a single star. 
The best fit using our toy model for the thick
disk stars is indicated by the solid lines.  The values of the key
parameters are marked.
\label{figure_reddy}}
\end{figure}

The 8 Draco giants in our sample have [Fe/H] between $-1.45$
and $-3.05$~dex. To provide a context for the understanding
of our results we develop a toy model for the behavior
of abundance ratios  [X/Fe] vs [Fe/H] to apply to the Draco sample.
We look for guidance to the behavior of abundance ratios in Galactic
populations, the thin disk, thick disk, and Galactic halo field stars
since the same nucleosynthetic processes are involved, although they
may contribute different relative fractions to the chemical  inventory
in different environments.
Fig.~\ref{figure_reddy} shows a plot of [Mg/Fe] vs [Fe/H] for thin
and thick disk F and G dwarfs based on the detailed abundance
analyses of \cite{reddy03} and of \cite{reddy06}
respectively.  We note, as have many others, that at low metallicities,
the [Mg/Fe] ratios for thick disk stars
appear to reach a plateau level.  Both thin and thick disk stars
reach the Solar ratio at approximately the Solar Fe-metallicity.
Galactic halo field stars show a similar behavior  with
[Mg/Fe] reaching a comparable low Fe-metallicity plateau level
\citep[see, e.g.][]{cohen04}.

We characterize the data for the abundance ratios
for each element X detected by first finding A(X),
the value of [X/Fe] at the lowest [Fe/H] found in the sample  
and B(X), a value for
stars at the high end of the Fe-metallicity of the sample. 
We assume that  the set of values of [X/Fe] for the  sample
can be modeled by
constant [X/Fe] $\equiv$  A(X) over
the metallicity range [Fe/H](A) $<$ [Fe/H] $<$ [Fe/H](low,X), and B(X)
over the range [Fe/H](high,X) $<$ [Fe/H] $<$  [Fe/H](B).  
Between  [Fe/H](low,X) (the knee of the distribution)
and [Fe/H](high,X) the values of [X/Fe] are assumed to change linearly
with [Fe/H].  We thus have a plateau in [X/Fe] over
the range [Fe/H](A) to [Fe/H](low,X) and another plateau
at a value of [X/Fe] which may be different at the
high metallicity end from [Fe/H](high,X) to [Fe/H](B),
with a straight line connecting the two plateaus.
We solve for the two parameters in this toy model
[Fe/H](low,X) and [Fe/H](high,X)
by minimizing the variance around the fit.   Thus our
model has four variables whose values are determined directly from the
dataset of [X/Fe] as a function of [Fe/H], with two additional
fit parameters.

Our toy model is based on the behavior of abundance
ratios in the Galactic disk and halo as exemplified by Fig~\ref{figure_reddy}.
Our model fits to the [Mg/Fe] ratios for thin and thick disk Solar neighborhood
F and G dwarfs from \cite{reddy03} and \cite{reddy06} are given in Table~\ref{table_fit};
the latter is shown in this figure.  These fits are a good representation
of the behavior of the data for abundance ratios of Mg and of
other elements as a function of [Fe/H] for Galactic stellar
populations.

Our toy model fits offer important clues for the importance
of various nucleosynthesis processes in Draco as compared to
in the Galactic thick disk and halo stellar populations.
The parameters of the toy model depend on the nucleosynthetic
yields for the production channels for
each of the elements X and Fe,
the IMF, the rate of star formation,
accretion, loss of gas via galactic winds, interaction between the dSph
and the Milky Way via tides, ram pressure stripping, etc. as will
be discussed in \S\ref{section_nuc}.

We apply this toy model to
the sample of 14 Draco giants.
We use
the two lowest metallicity and three highest metallicity stars in
the Draco sample to determine the plateau values
for [X/Fe], A(X) and B(X), as well as the minimum and maximum
[Fe/H] values over which the model is fit,
[Fe/H](A) and [Fe/H](B).    Weights
are halved for the 5 stars with lower accuracy spectra.  Note
that these four values are calculated directly from the data.
We solve for the two parameters in this toy model
[Fe/H](low,X) and [Fe/H](high,X)
by minimizing the variance around the fit
using our sample of Draco giants described above.

We have applied this model to 10 elements for which sufficient
accurate data is available for Draco members.
The  resulting parameters 
for each element
are listed in Table~\ref{table_fit} and the fits are shown
when available in Figs.~\ref{figure_nafeh} to \ref{figure_bafeh}.
The uncertainties in A(X) and in B(X) are approximately
those of $\sigma$[X/Fe]
for a single Draco star from our sample.  The latter values are given in
Table~\ref{table_sens_rel}.  Thus, for example, for [Mg/Fe] they are
$\pm$0.14~dex, while A(Mg) is 0.50~dex larger than B(Mg). 
The decline in [Mg/Fe] as [Fe/H] increases is highly significant,
as is the increase in [Cr/Fe], [Sr/Fe]
and [Ba/Fe] as [Fe/H] increases, while [Ca/Fe] and [Ti/Fe] are
constant at approximately the Solar value to within the uncertainties.

Even when the change between the low and high metallicity abundance
ratio is clearly statistically significant, as is the case for Mg,
the values for the knees of the distribution, [Fe/H](low)
and [Fe/H](high) are quite uncertain due to the small sample
of Draco giants coupled with the uncertainty of the individual
[Mg/Fe] determinations for each Draco giant.  
However, the possibility of a straight line
connecting  A(X) to B(X) without any plateau at either low
or high metallicities is included as part of the search for the
best knee values.  This straight line without any
plateau 
has a variance which is $\sim$50\% larger than the
minimum found here for the knee values (listed in Table~\ref{table_fit})
for the [Mg/Fe] abundance ratios.

We consider whether our simple toy model provides an adequate
fit to the Draco abundance ratios.
We assume that the minimum $\sigma$[X/Fe] for all elements is 0.15~dex,
which is for some elements somewhat 
larger than the value given in Table~\ref{table_sens_rel}.
We then calculate $\chi^2$ for the 
the 14 Draco giants between their
observed abundance ratios and the values from the toy model fits.
We find that six of the 10 elements with such fits
have $\chi^2 \leq 22$ for all 14 stars.
The same is true for two additional cases of the 10
when one highly discrepant giant (Draco XI-2), which
is a low outlier, is ignored.
This suggests that the uncertainties in [X/Fe] have been
slightly underestimated by perhaps 20\% or that the adopted
toy model is somewhat too simple to describe the full 
range of behavior seen in Figs.~\ref{figure_nafeh}
to \ref{figure_bafeh}.  However, as an initial approach
towards a simple model, our toy model seems to represent the data
regarding abundance ratios for the Draco giants
reasonably well.

There are a number of elements for which the toy model could not
be applied due to the low number of stars with measured [X/Fe]
or to our judgement, based on issues discussed in
\S\ref{section_field},  that the abundance ratios are not reliable enough.
Table~\ref{table_range}
provides means, extreme values and dispersions for all
species detected in the Draco sample.

\subsection{Summary of the Comparison of the Toy Model Parameters Between Draco and 
Galactic Populations \label{section_toy_gal} }

Using the toy model described above
we compare the behavior of the abundance ratios within
the sample of Draco giants
with those of Galactic stellar populations in the thin disk,
the thick disk, and the halo.  The [X/Fe] values at low
metallicity, A(X), for the elements Na, Mg, Si, Cr, Mn, and Ni
and perhaps Sc and Ti are identical to those of
the thick disk to within the
uncertainties.  [Ca/Fe] is approximately constant at a slightly
sub-solar values over the full range of [Fe/H] spanned by the
Draco giants.  It falls somewhat below the
halo values even at the lowest Fe-metallicities probed.
The uncertainties are still large, and for Ca the
systematic differences between the two large scale surveys
of Galactic halo field stars (our 0Z survey and the First Stars
survey at the VLT led by R.~Cayrel) remain to be resolved.
[Cr/Fe] and [Mn/Fe] both increase
as [Fe/H] increases in Draco and in Milky Way halo stars.
For the heavy neutron capture elements Sr and Ba, the
abundance ratios at the lowest metallicity are at or below
the most extreme stars in the much larger halo samples;
[Sr/Fe] and [Ba/Fe] both also increase towards the Solar ratio
as [Fe/H] increases in Draco and in the halo field.  Intermediate
[Fe/H] Draco stars in general have lower abundance ratios
than do the Galactic halo field stars.

The Galactic populations all approach [X/Fe] = 0 close to 
or at the
Solar Fe-metallicity, i.e. B(X) and [Fe/H](high,X) $\sim 0$. 
A comparison of Fig.~\ref{figure_mgfeh}, [Mg/Fe] vs.
[Fe/H] for the Draco stars,
with Fig.~\ref{figure_reddy}, the same for the thick and thin disk
stars in the Milky Way, suggests that the toy model can
adequately represent both the Galactic and Draco data, and that
that the shape of the mean relation is similar.  But in the
Galactic populations, as is shown in Table~\ref{table_fit}, 
the approach toward Solar ratios for many elements begins
at a considerably higher [Fe/H].  For example. 
[Fe/H](low,Mg) is $-0.53$~dex for
the Milky Way thick disk and less than $-2$~dex for Draco.

\subsection{Comparison with Nucleosynthesis Predictions \label{section_nuc}}

The parameters of our toy model 
have close ties with nucleosynthesis.
The dependencies on star formation rates, yields of various nuclear reactions, etc.
of the various parameters in the toy model are most clear 
in the case of elements which are produced largely by sources
different from those of Fe.  The plateau value A(X) 
at low [Fe/H] presumably
represents the abundance ratio [X/Fe] from  only those sources
of material processed through the nucleosynthesis channels
that were active at very early times, i.e.  SNII from
core collapse of massive stars, events
characteristic
of the earliest epochs of star formation.  
The plateau level
B(X) represents the abundance ratio 
for element X at late times when all relevant sources,
SNIa, SNII, AGB, novae, etc are contributing.  These abundance
ratios are independent of accretion of primordial material and,
for a well mixed ISM, of galactic winds.

The abundance ratios for the $\alpha$-elements, produced strongly
in SNII, but not much in SNIa, and for the heavy neutron capture elements
fall into this category.   
On the other hand, when the nucleosynthesis channels for element X
are approximately the same as those for Fe, we expect
that the abundance ratio will be approximately constant, with
A(X) $\approx$ B(X) $\approx 0$, ignoring issues of metallicity
dependent yields.

There is overall agreement of the plateau values at both high and low [Fe/H]
for the $\alpha$-elements between the Draco giants and 
field halo stars in the Milky Way.  [Mg/Fe] has the largest
range, as might be expected since Mg, unlike Ca or Si, is
produced only in SNII, while Ca and Si are produced
in both SNII and SNIa \citep{woosley}.  A(Ca) is 2$\sigma$ low
compared to the Galactic halo field stars; all other $\alpha$-element
plateau values are close to those of the halo field stars.
This suggests 
that the relative contributions to [X/Fe] of SNII at early times and of
a mixture of various sources, dominated by SNIa, at later times
are similar in the cases being compared.   The yields
depend both on the rates of various nuclear reactions (which
are mass dependent) and on the initial mass function for star
formation.  The former is fixed by physics, so this implies
that there cannot be much variation in the latter, i.e. in the
initial mass function (and in the mass loss during the course of
evolution of massive stars).  The agreement at high [Fe/H] 
suggests that the relative contributions of at least the dominant
sources must also be the same in the two environments, constraining
SNIa production.

The major differences appear to be confined to the fit parameters
[Fe/H](low) and [Fe/H](high) which define the point at which
[X/Fe] drops off the low-metallicity plateau and where it approaches the Solar
value respectively.  These parameters are measures of the
delay between when the contribution of SNII is dominant and
when SNIa contribute at full strength. 
SNIa are believed to arise from thermonuclear explosion of
a white dwarf whose mass increases, presumably through mass
transfer in a binary system, until explosive carbon
burning is ignited.  Thus [Fe/H](low)  measures
how much [Fe/H] builds up
over the timescale required for the binaries responsible for
SNIa to form, to explode, and to begin to contribute significantly to the
chemical inventory of the ISM in the dSph, while [Fe/H](high,X)
refers to the time when the SNIa reach the full level of
their contribution.  

\cite{greggio08} offer
a recent determination of the distribution of delay times
between an episode of star formation and the resulting
SNIa events. 
 This delay pattern for a specific burst of star formation
is to first order fixed by the IMF. 
The Fe-metallicity built up
in Draco over this fixed timescale, corresponding to
[Fe/H](low), is lower in Draco than it is in the
Galactic halo.  Given that we have already offered some evidence
that the IMFs are similar, this must constrain some combination
of the star formation efficiency, the rate of loss of gas processed through
stars and hence enhanced in both X and Fe, and the rate
of accretion of primordial gas consisting mostly of H, each
of which affect the relationship between $<$[Fe/H]$>$ in the 
stellar system 
and time.  More detailed models for chemical evolution incorporating
the full dependence of these factors are required to further disentagle
and evaluate the contributions of each of these.

The light elements C, O, Na and Mg show abundances relative to Fe
in luminous Draco giants at similar Fe-metallicity
which have a range far larger than can be ascribed to observational
error.  These elements are believed to have been formed largely
in the interiors of massive stars through nuclear fusion reactions,
then dispersed into the interstellar medium by SNII explosions,
all of which happens quite rapidly
after the initiation of star formation.  The rates of production of 
the various elements depend
on the  initial mass of the SNII progenitor 
and its subsequent mass loss history,
and these dependencies differ not just in amplitude
but even in sign among the light elements \citep{woosley}.
Deep mixing similar to that seen in luminous Galactic
globular cluster giants may also be important for
these elements.

Draco has $M_V \sim -10$~mag, brighter than all 
Galactic globular clusters except $\omega$ Cen and NGC~6715. 
Most of Draco's stars are old, but we cannot
be sure that the entire population formed in a single burst.
The stellar population  may be small enough
that stochastic effects of the formation of the very rare most
massive stars 
discussed by \cite{carigi08}
may become significant especially among the most metal-poor
stars in Draco.    This may be the source of some or all of the variations
among the light elements, although why such stochastic effects
are not seen in even low mass globular clusters is an interesting
question. 

\cite{arnett} \citep[see also][]{clayton03} discusses the production
of Na, which is believed to occur in the interiors of massive stars
and to depend on the
neutron excess, which in turn depends on the initial heavy
element abundance in the star.  Na thus has both a primary and
a secondary nucleosynthesis channel.  This presumably gives rise to
the very low [Na/Fe] abundances seen among the Draco giants.
Sc production in SNII is highly dependent on the
details of the explosion, including both the metallicity
and mass of the progenitor \citep{woosley,limongi03}.
Stronger odd-even effects are found for lower metallicity
and, in the case of Sc, for lower mass progenitors \citep{limongi03}.
Thus a relative absence of the higher mass SNII
with $M > 35M$\subsun\ might
give rise to the low [Sc/Fe] in the Draco sample.

The Fe-peak elements Cr through Zn are the highly stable end products
of nucleosynthesis via fusion.  In the Draco stars, these elements 
mimic the behavior
of the Galactic halo field stars, except that [Ni/Fe] is somewhat
lower than expected in the range $-2 <$ ~Fe/H]~$< -1.5$~dex.
In the Solar neighborhood and inner Galactic halo, Ni closely
follows Fe so that [Ni/Fe] = 0.0~dex, the Solar ratio.
The calculations of \cite{woosley} suggested that the nucleosynthetic
yield for Ni in SNII is metallicity dependent, increasing at higher
metallicity. 
\cite{ohkubo} recently showed that the same for SNIa, 
which may explain the somewhat low [Ni/Fe]
in intermediate metallicity Draco giants, but their result is
sensitive to the details of the model for the SNIa explosion.
[Cr/Fe], [Zn/Fe] and perhaps [Co/Fe] are also
lower in intermediate metallicity Draco giants than in Galactic halo
field stars. \cite{timmes} found that the production of Zn in SNII
depends on the neutron excess, hence depends on the initial
metallicity, and thus should mimic the behavior
of Na and Ni.

\subsection{The Heavy Neutron Capture Elements \label{section_ncapture} }

Neutron-capture reactions require both a supply of neutrons
and the presence of seeds, presumed
to be Fe-peak nuclei.  It is widely
believed that the $s$-process occurs in the interiors (but well outside
the cores) of intermediate
mass AGB stars, dredge up to the stellar surface follows. 
Processed material is then disseminated
into the interstellar medium through slow stellar winds. 
Models of the $s$-process, including the
effect of the initial stellar metallicity  on the resulting products,
are summarized by \cite{busso99}.
Among the expected metallicity effects is a shift in the relative
abundance of elements among the various ``peaks'' where the 
neutron-capture cross sections are small due to filled
shells in the nucleus; the abundance
of these specific elements at the end of neutron-capture are relatively high
compared to their close neighbors in the periodic table, thus producing
``peaks''.
Metallicity effects
in $r$-process nucleosynthesis, whose site
is not yet fully identified, are not as clearly understood at present.
Suggested sites include SNII, hypernovae \citep{qian08}, 
accretion-induced collapse of a white dwarf into a neutron star
in a binary system \citep{qian03}, and
neutrino-driven winds from a nascent neutron star as reviewed
by \cite{qian07}.

We compare the predicted behavior of the heavy neutron capture elements
with that observed in Draco to determine which neutron capture process
dominates.  In very metal-poor stars in the
Galactic halo the chemical inventory of the heavy elements is
from the $r$-process only.  There was not enough time for 
the formation and evolution of any intermediate mass AGB stars, hence
no $s$-process
contribution to the ISM prior to the formation of these stars,
although subsequent  mass transfer across a binary system can alter
the surface abundances in the former secondary
star and produce large s-process enhancements \citep{cohen06}.
Such cases can be identified by the accompanying
large enhancements of C as well. The best diagnostic abundance
ratios are  Eu/Ba or Eu/La,
where Eu is formed almost entirely by the $r$-process, while
Ba and La are formed mostly via the $s$-process, at least in the Sun.
(Ba, La, and Eu all belong to the same neutron-capture peak.)
No La features have been detected in any Draco dSph star.

\begin{figure}
\epsscale{0.6}
\plotone{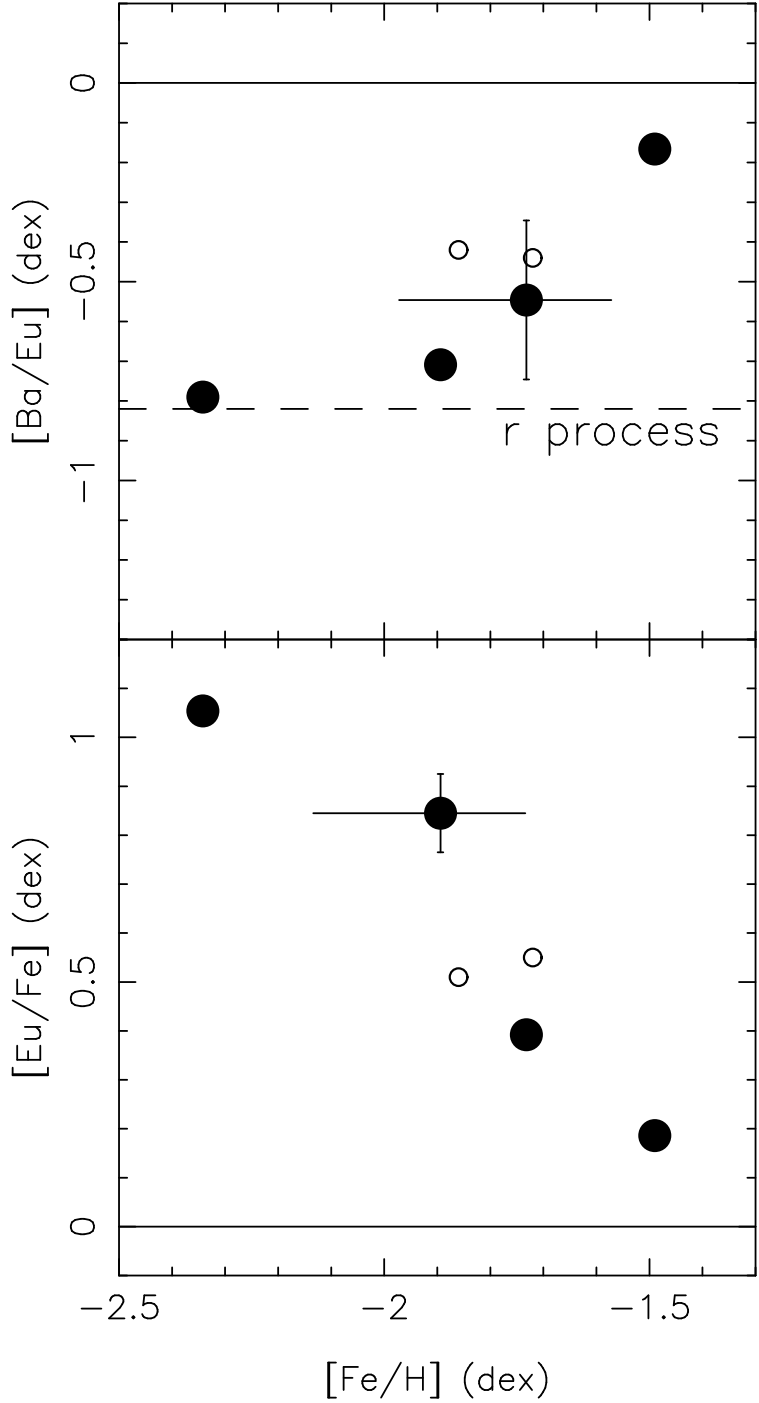}
\caption[]{[Ba/Eu] (upper panel) and [Eu/Fe] (lower panel) vs [Fe/H] 
is shown for the four giants our Draco sample
and the two from \cite{shetrone2} with detected Eu.  
The Solar ratio is the solid horizontal
line, while the dashed horizontal line is the $r$-process
ratio from \cite{simmerer}.
Typical uncertainties are shown for one 
Draco star.
\label{figure_baeu}}
\end{figure}

The upper panel of Fig.~\ref{figure_baeu} shows [Ba/Eu]  
as a function of [Fe/H] for the four Draco giants 
from the present sample and the two from \cite{shetrone2}
with detectable Eu~II lines.  The $r$-process ratio shown in the top
panel is taken from
\cite{simmerer}; the solar ratio is a mixture of $r$ and $s$-process
material, while the pure $s$-process ratio for [Ba/Eu]
lies higher than the top
of the figure.  They suggest that 
in the Galactic halo, signs of the $s$-process begin only at
[Fe/H] $> -2.6$~dex, and a mean [Eu/La] ratio halfway between
the pure $r$-process value and the Solar ratio is reached only
at [Fe/H] $\sim -1.4$~dex. The survey of cool metal-poor local dwarfs
of \cite{mashonkina} reaches the halfway
point in [Eu/Ba] from pure $r$-process to the Solar mixture only 
at [Fe/H] $\sim -0.5$~dex.  Thus the
result from Fig.~\ref{figure_baeu} 
is clear;
the lowest [Fe/H] Draco stars have $r$-process dominated heavy
neutron capture elements, but the $s$-process becomes important at 
a Fe-metallicity significantly lower than is characteristic
of the Galactic halo.

\begin{figure}
\epsscale{0.6}
\plotone{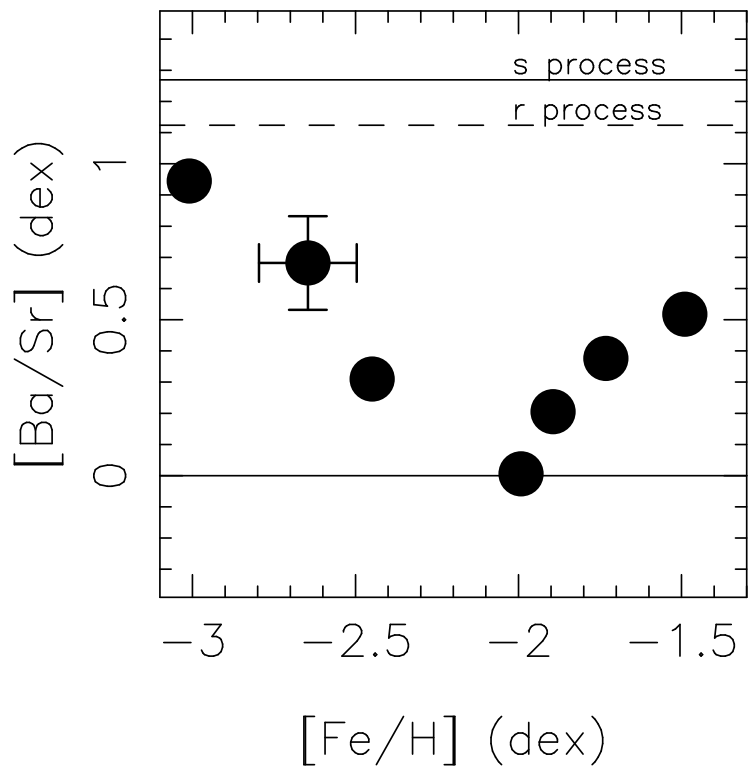}
\caption[]{[Ba/Sr] is shown as a function of [Fe/H] for our Draco sample.
The Solar ratio is the lower solid horizontal
line.  The pure $s$ and pure $r$-process
ratios from \cite{simmerer} are indicated.  A typical error bar is shown
for one star.
\label{figure_basr}}
\end{figure}

Fig.~\ref{figure_basr} shows [Ba/Sr] as a function of [Fe/H] for our
Draco sample, together with the $s$ and $r$-process ratios 
from \cite{simmerer}.  Only the lowest metallicity star lies
close to the pure $r$-process ratio.  The others deviate from
both the pure $r$ and the pure $s$-process predictions in the sense that
production of additional Sr by some other mechanism is required.
\cite{travaglio04}
discuss the origin of Sr (and Y and Zr, all part of the same peak) in
very metal-poor stars and suggest a new additional process of neutron capture,
which they call the ``lighter element primary process'', while 
\cite{qian08} suggest a different origin for additional
Sr in very metal poor stars.


\section{Chemical Evolution of the Draco dSph Galaxy \label{section_chemev} }

Recent models of chemical evolution for the disk, bulge, and halo of the Milky
Way based on the precepts first established by \cite{tinsley}
have been presented by several groups, including
\cite{timmes}, \cite{kobayashi}, \cite{prantzos}, and \cite{matteucci08}.
These models assume an initial mass function, a star formation history, 
a mass infall
history, outflow from Galactic winds, all as functions of time.
They generally assume complete and uniform
mixing of the gas over the total volume considered at all times with
the exception of the sophisticated model of 
\cite{marcolini06}, which includes 3D hydrodynamic
simulations of the ISM and the spatial variation of its properties.
Such models combine networks of nuclear reactions from SNII with yields
from \cite{woosley} or \cite{limongi03}, from SNIa with
yields generally from \cite{iwamoto}; some also add in
the contributions
to the light elements from AGB winds and novae.  They 
have been reasonably successful in reproducing the chemical
evolution of the major components of the Milky Way overall, although
failing in some (minor) details.

The evolution of the dSph galaxies differs in principle from
that of the Milky Way or its halo.  Their binding energies
are lower, so 
the importance of gas loss may be higher,
particularly material from SNII, for which the ejection velocity is 
significantly larger than the escape velocity so that a considerable fraction
of the enriched gas may be lost from the galaxy;
evidence for this happening today in small starburst galaxies
is given by \cite{martin02}. 
\cite{marcolini06} suggest that if the dark matter halo is 
sufficiently massive, SNII ejecta may be pushed far out into the halo,
away from the central region where stars form, but remain
confined within the halo.
Furthermore since
Draco at present shows no evidence for the presence of gas,
gas loss via a galactic wind or through interactions between the dSph satellite
and its host, the Milky Way, must have been important in the past. 
These galaxies also
show the consequences of lower star formation efficiency
which leads to slower star formation
overall without
the large initial burst that dominates nucleosynthesis in 
most of the Milky Way
components.  In a system where the star formation rate is more
constant with time, SNIa ejecta can become important contributors
before [Fe/H] just from SNII production builds up to 
high values near $\sim -1$~dex.  It is this time delay
between the SNII and SNIa contributions that dominates discussion
of the chemical evolution of dSph galaxies.  Note that once substantial gas loss 
begins, either through a wind or an interaction with the Milky Way,
the star formation rate drops, the SNII rate drops precipitously,
and thus the injection of $\alpha$-elements into the ISM drops,
while SNIa continue unaffected, so the Fe content of the ISM
continues to rise.  This then causes the drop in [$\alpha$/Fe] ratios 
seen in Draco and in the other dSph galaxies.

\subsection{Age -- Metallicity Relation for Draco \label{section_age}}

A very useful diagnostic of the star formation rate as a function
of time is the age -- metallicity relationship.  We construct this
for Draco using the [Fe/H] available from detailed
abundance analyses for our sample and that of \cite{shetrone2}.
We take the remaining members of Draco listed by \cite{winnick03}
(the 6 known carbon stars  in Draco are excluded), adopt her
[Ca/H] values, and convert them to [Fe/H] values (see the appendix).
Photometry is taken from \cite{megacam}, who used the Megacam
camera on the CFHT
to image Draco in the $g$', $r$', and $i$' bands.
We use the isochrones of \cite{dotter08} which are available
for the SDSS colors.  We adopt a relation between
[$\alpha$/Fe] and [Fe/H] based on our results described above.  
Given [Fe/H], [$\alpha$/Fe],
the colors, the distance of Draco and the adopted reddening, we can
determine the age of each Draco giant. 

\begin{figure}
\epsscale{0.9}
\plotone{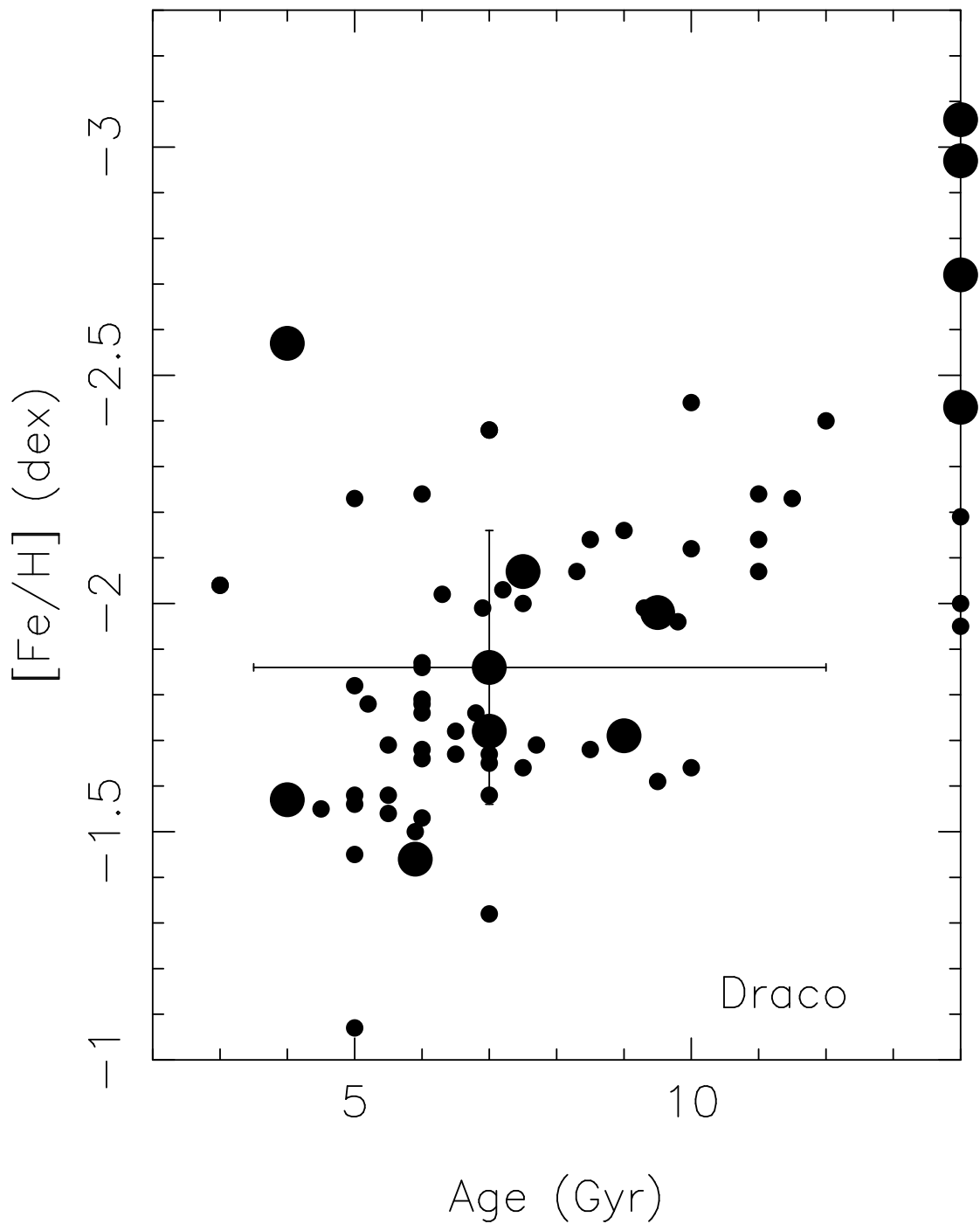}
\caption[]{The age of each giant with $M_i < -2.0$~mag
known to be a member of Draco  from Table~2.9 of \cite{winnick03}.
The Dartmouth isochrones \citep{dotter08} were used with
[Fe/H] set to [Ca/H] as derived by \cite{winnick03} or are
from HIRES spectra (large filled circles). Typical error
bars are shown for a single star.
\label{figure_age_metal}}
\end{figure}

We do this for each star with $M_{i'} < -2.0$~mag.  
The isochrones along the RGB converge too much in the ($g$' -- $i$') color
to attempt this
for stars less luminous than this.  The results are shown in 
Fig.~\ref{figure_age_metal}.  The median age for the
luminous Draco giants is
7~gyr. The uncertainty for each individual star
is large as an uncertainty in [Fe/H] of 0.3~dex, typical
of that values derived from the Ca triplet (see the appendix), introduces an
age error of $\sim 5$~Gyr.  At a fixed [Fe/H], the ($g$' -- $i$') color
changes by $\sim$0.015~mag/Gyr, thus requiring extremely accurate
and well calibrated
photometry. The random photometric
measurement errors are not important compared to those arising from
the uncertainty in [Fe/H],
but systematic errors of only a few hundredths of
a mag due to problems in the calibration of the
photometry could be serious.  Also 
systematic errors in the
transformation
between the theoretical \teff, $L$ plane
and the $g$', $i$' system adopted by the grid of isochrones,
if present, could seriously bias the derived ages.   
HST studies of Draco \citep[see, e.g.][]{orban08} reach the 
main sequence region, 
have much more accurate photometry for the giants, and perhaps more carefully
calibrated isochrones, but they lack metallicity information
for individual stars.  They too have difficulty distinguishing
small age differences at age $\sim$10~Gyr.

Fig.~\ref{figure_age_metal} suggests a rough age-metallicity relation, with
star formation extending over perhaps 5~Gyr beginning about 10~Gyr ago
over which the mean
[Fe/H] decreased by a factor of $\sim$5, although the large uncertainties
in age introduce considerable scatter.  The apparently youngest stars
could be AGB stars as they are bluer at a given luminosity than are
first ascent RGB giants of the same luminosity.

\subsection{Comparison With Other dSph Galaxies \label{section_comp_dsph} }

In recent years there has been a tremendous improvement in abundance
data for dSph satellites of the Milky Way due largely to technical
improvements and the construction of 8 to 10~m telescopes.
The situation prior to 2008 is reviewed by \cite{geisler08}.
As of today, although there is a large ongoing project
to study dwarf galaxies  \citep[the DART project, ][]{tolstoy03} at the VLT,
there are only two other dSph galaxies with published detailed
abundance analyses from high dispersion spectra
for 14 or more stars  to which we can compare
our Draco results.  These are the Sgr dSph (the main core, not the stream)
\citep{monaco05,sbordone07}
and the Carina dSph galaxy, for which
\cite{koch_carina}  combines his analysis of 10 giants
with 5 from the earlier study by \cite{shetrone03}.

\begin{figure}
\epsscale{0.85}
\plotone{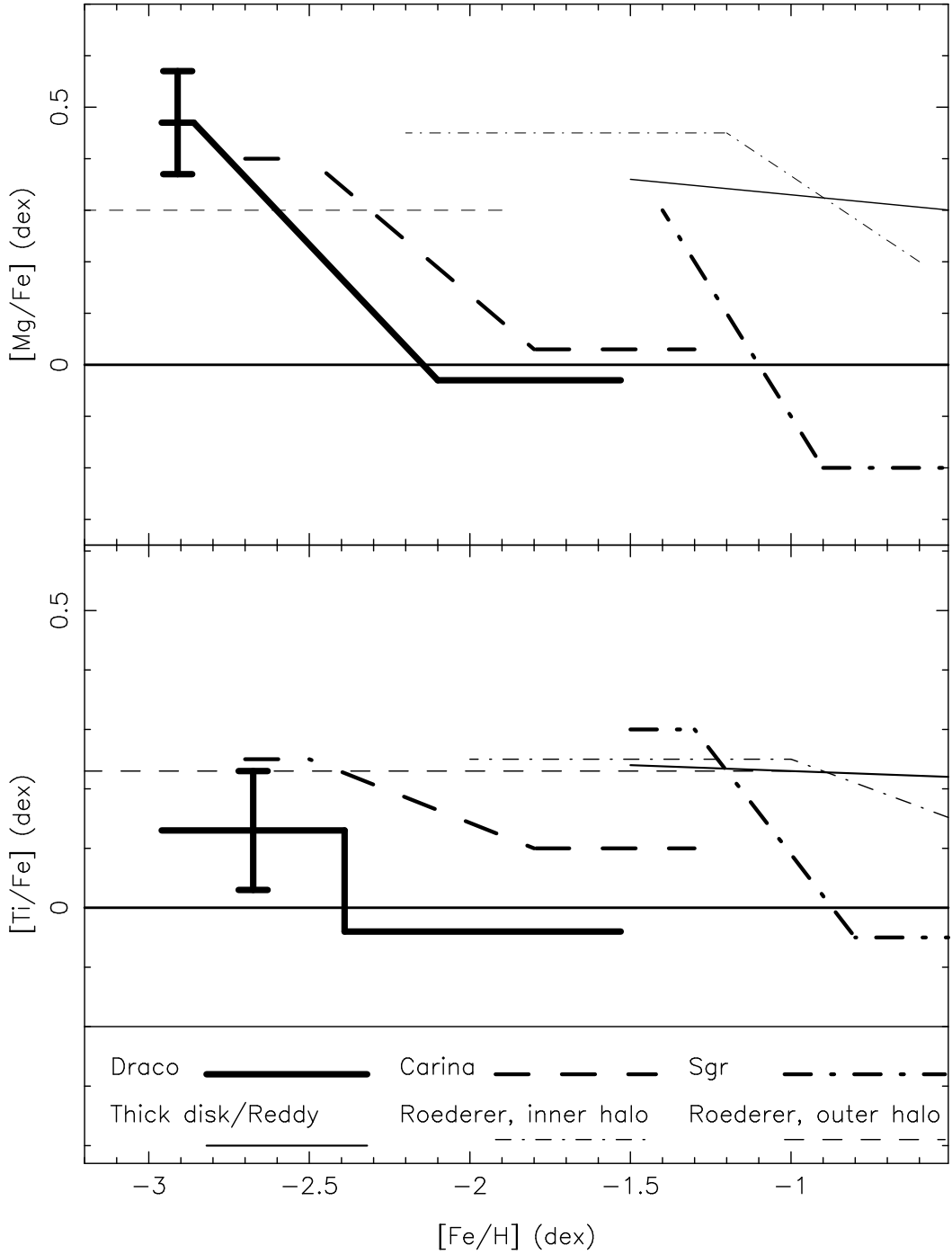}
\caption[]{The toy model fit for [Mg/Fe] vs [Fe/H] (top panel)
and for [Ti/Fe] (bottom panel) for the sample
of 14 Draco giants is shown together with that for the Sgr 
\citep{monaco05,sbordone07}
and Carina \citep{koch_carina} dSph galaxies.  Fits for
Galactic components to data from \cite{roederer08} and from
\cite{reddy06} are shown as well.  Typical errors in abundance
ratios for the average of two stars are shown.
\label{figure_dsph_2panel}}
\end{figure}

We apply our toy model to the recent data for the Carina and the Sgr dSph
galaxies.  Fig.~\ref{figure_dsph_2panel} for [Mg/Fe] and 
for [Ti/Fe] show the fits for these two galaxies, for Draco, and 
for the Milky Way thin and thick disk. 
This figure clearly demonstrates that to first order the form
of the [X/Fe] vs [Fe/H] relations, at least for Mg and for Ti,
are identical to within the uncertainties, 
but what is changing is the Fe-metallicity range,
and the knee values [Fe/H](X,low) and [Fe/H](X,high) for
these two elements. 

This reinforces our previous comparison between the Draco
abundance ratios and those of the stellar population of
the Milky Way.  While the initial low metallicity abundance ratios
and the final high metallicity ones are identical to
within the uncertainties for most elements, 
the key differences lie in the [Fe/H] values
corresponding to the knee values, i.e. in the timescale
at which the overall metallicity of the system increases.
The Draco system has the slowest evolution of metallicity in its
stars of the three, as well as the lowest mean [Fe/H] for its giants,
Carina is intermediate, and Sgr is closest 
to the Milky Way.  In addition, each dSph that
has sufficient data for the heavy neutron capture elements appears
to have contributions from the $s$-process that begin at somewhat
lower [Fe/H] than that characteristic of the Milky Way.

\cite{matteucci08} reviews models for the chemical
evolution of the dSph galactic satellites of the Milky Way
that reproduce the behavior
of the $\alpha$-elements.  Presumably the agreement at
the lowest [Fe/H] values probed here, where the Galactic halo
stars overlap the Draco giants, is a consequence of 
a chemical inventory to which only SNII contributed.
\cite{lanfranchi04} present
detailed models for the evolution of
6 of the dSph Milky Way satellites, including Draco,
which try to reproduce not only the chemical evolution
but also the total stellar mass and  their individual star
formation histories as derived from CMD studies.
They vary the star formation efficiency
and galactic wind rate to reproduce
the dSph characteristics inferred from available data as of that time.
Their model for Draco has
the lowest star formation efficiency 
and the weakest galactic wind of these 6 dSph galaxies, with a
a single burst lasting 4~Gyr
which occurred 6~Gyr ago.  To within the uncertainties
of the measurements and the models, they succeed in reproducing the
almost flat  [Ca/Fe] relation with [Fe/H] of Fig.~\ref{figure_cafeh} 
as well as the steep
decline in [Mg/Fe] vs [Fe/H] of Fig.~\ref{figure_mgfeh}.
\cite{carigi02} present
chemical evolution models for four dSph satellites of the
Milky Way, but Draco is not included in their study.
\cite{marcolini06} presents the most sophisticated model,
in that it includes a detailed study of the state of the ISM
as a function of time
within each of the N small volumes that together comprise
 the dSph galaxy.  Star formation, SN explosions, etc are
 considered in a statistical manner within each volume.  A dark
 matter halo is also included.
 \cite{marcolini06} apply their model to predict
 [O/Fe] vs [Fe/H] for Draco, but do not consider any other elements.
 Our relationships for the other $\alpha$-elements in Draco do not
 agree with their prediction for the behavior of [O/Fe].

As is the case for the Milky Way, conventional models 
similar to those reviewed by \cite{matteucci08} 
can explain in
general the abundance ratios seen in Draco.  However, there are some
problems when one looks in detail.  For example, \cite{lanfranchi08},
who address the production of heavy elements beyond the Fe peak
in dSph galaxies,
overpredict by a factor of more than 10
the ratio Ba/Eu in the most
metal-poor Draco stars.  The cause of the (small) difference 
in behavior of [Mg and Si/Fe] vs [Ca, and Ti/Fe] at the lowest
metallicities in Draco is not clear, particularly
since Si is  an explosive $\alpha$-element
while Mg is a hydrostatic one.   How this  behavior relates to the mass
distribution of the SNII progenitors, given that one also
needs to reproduce the odd-even effect at [Sc/Fe], is not obvious.  
Qualitatively similar differences in the behavior
of the $\alpha$-elements vs [Fe/H] are also seen in the Galactic
bulge \citep{fulbright_bulge}, but again there are differences
in detail as the separation between hydrostatic and explosive
$\alpha$-elements is cleaner there, i.e. [Si/Fe] behaves like 
[Ca/Fe] and [Ti/Fe]
in the Galactic bulge.
 
\cite{tsujimoto06} suggested  that the low $\alpha$-element
signature in the dSph  galaxies is a reflection of the
contribution of massive stars with smaller rotation compared
to solar neighborhood stars, instead of the
conventional interpretation that this is a consequence of
a low star formation rate.  Current data are significantly better
than what was available to him in 2006.  They show
that  the values of [$\alpha$/Fe] within the 
lowest Fe-metallicity Draco giants
approach and often reach those of
Galactic stars, providing additional
support for the conventional interpretation that this results from
a low star formation rate.

\section{dSph Galaxies and the
Formation of the Galactic Halo \label{section_discuss}}

Whether the Galactic halo could have been formed by accretion of
satellite dwarf galaxies
has become a question of great current interest; see e.g. \cite{tolstoy03},
\cite{shetrone2}, among others. The
cold dark matter cosmological model leads to hierarchical galaxy 
formation in which massive galaxies should be surrounded by
numerous lower mass satellite halos;  the stellar component
of the Galactic halo presumably
resulted at least in part from the dissolution of some
of these satellites.  But the small
number of satellite galaxies known around our Galaxy
was in conflict with this prediction, a situation often called the
``missing satellites problem'' \citep{klypin99}. 
Recently, however,
a substantial number of
new satellite galaxies of the Milky Way have been discovered
through searching the SDSS 
\citep[see, e.g.][]{belokurov06}.
If one multiplies the new discoveries by the
fraction of the sky not surveyed by the SDSS, taking into account
the detection efficiency for finding a low luminosity satellite
in the SDSS data
as a function of distance and total luminosity of the satellite
quantified by \cite{koposov08}, the ``missing satellites'' 
problem is well on
the way to solution.

From the point of view of the chemical inventory of the Galactic halo
versus that of the dSph satellites of the Galaxy,
the initial answer to the question poised above given by 
\cite{tolstoy03}
was negative.  \cite{shetrone03} and his earlier work
demonstrated that dSph galaxies clearly have
a chemical inventory which reveals signatures of a lower 
star formation rate and these low luminosity satellites
appear to have a smaller fractional contribution
of SNII in the total chemical inventory 
than in the Galactic halo itself.  
But since our work in Draco and that published for the Carina
and for the Sgr dSph galaxies show that
abundance ratios among stars in dSph galaxies tend
to overlap those of Galactic halo giants at the lowest
Fe-metallicities probed, 
it is possible that the satellites were accreted early in their
development.  Their properties as we observe them today
would then not be relevant to this issue.
\cite{helmi06} claim, however, that
early accretion is still ruled out because 
of the metallicity distribution
function (MDF) they deduce for four dSph galaxies.
Given the 
metallicity distribution
function (MDF) they use for the Galactic halo, they claim that
dSph galaxies would be expected
to contain at least a few stars with [Fe/H] $< -3.0$~dex, while
they have not to date  detected any such stars in the four dSph galaxies
in which they have extensive samples from the DART project.

However, we have contributed to refuting this argument 
by finding a giant in our Draco sample with [Fe/H] from a high
dispersion spectrum below $-3$~dex, the first such discovered.
\cite{kirby08} have found a few more such stars in the 
still lower luminosity satellite galaxies using moderate
resolution spectra, as have (after this paper was submitted)
\cite{frebel09} \footnote{We suspect that the stars with [Fe/H] $< -3$~dex
in the Bootes~I dSph galaxy found
by \cite{norris08} may not actually be so Fe-poor, as their
Fe-metallicities are from Ca lines assuming [Ca/Fe] = +0.3~dex, which
is probably too large for stars in a dSph galaxy, so they
require further verification.}.
Furthermore, as is discussed in the appendix, the calibration
adopted by the DART project
to convert measurements of the strength of the Ca infrared triplet into
Fe-metallicities is quite different at very low metallicities from
that we use here, and with their adopted calibration, it is quite unlikely
that they would have found any stars with [Fe/H] $< -3$~dex in
their dSph samples, even if they did exist, an issue also
discussed by \cite{kirby08}.  In this context, it is interesting to note
that the star in Sextans recently found to have [Fe/H] $-3.2$~dex by
\cite{aoki09} has [Fe/H](CaT) from the DART team
of $-2.5$~dex assigned in \cite{helmi06}, later revised to
$-2.7$~dex by \cite{battaglia08} \citep[see Fig.~6 of][]{aoki09}.
In addition \cite{hes_mdf} 
recently completed 
a determination of the halo MDF based
based on the Hamburg/ESO Survey  
which shows that completeness corrections are important in the
MDF derived from the HES. 

Collectively this very recent work
serves to help reestablish
the scenario for the formation of the Galactic halo via accretion
of satellite galaxies as viable.  The material
now  in the inner halo of the Galaxy had to have been
accreted early in the star formation history of the dSph galaxies,
giving time for orbital mixing to eliminate traces of discrete
stellar streams,
while satellite galaxies accreted somewhat later could contribute
to populating
the outer halo, which shares many of the abundance anomalies of
the dSph galaxies.  Dissolved globular clusters had to disperse
fairly quickly before the light element correlations
among Na, Mg and Al developed,
as these are not seen among halo field stars.

Finally we
comment on the possible presence of a significant
intermediate age population in the Draco dSph galaxy.
\cite{aaronson83} found several carbon stars in this
dwarf galaxy.  \cite{shetrone_cstar},
in an effort to determine the highest metallicity reached
in Draco, probed
stars redward of the RGB, but found that they were
either non-members or carbon stars, to establish an upper
bound for this dSph of [Fe/H] $\sim -1.45$~dex (The highest
metallicity star in our Draco sample has [Fe/H] very close to this
value.)
\cite{cioni} recently surveyed Draco in the near-IR
and found a few more carbon star candidates.  The presence
of these C stars has been used to argue for an intermediate
age population in Draco, and for a difference between the
stellar population of Draco vs that of the Galactic globular cluster
system, where such C-rich stars are very rare.

\cite{shetrone_cstar} point out that the carbon star specific frequency
in Draco is 25 -- 100 times higher than that of the Galactic
globular clusters.  However, study of the halo stellar population
found in the Hamburg/ESO Survey by the 0Z Project 
\citep{cohen_cstarfreq} reveals the presence of
numerous very carbon-rich stars at [Fe/H] $< -2.5$~dex.  While
the exact frequency of such stars is uncertain
\citep[see the discussion in][]{cohen_cstarfreq}, it is somewhere
between 14 and 25\%.  This is a metallicity dependent phenomenon,
with carbon rich stars becoming more frequent as [Fe/H] decreases
and the amount of C which must be dredged up to get to
a situation where $\epsilon$(C)$~ > ~ \epsilon$(O) consequently drops.  
It is for
this reason that such C-stars are rare in the Galactic globular
clusters, most of which have [Fe/H] substantially larger than that
of the mean of the Draco stellar population.
In addition, the IR photometry of
\cite{cioni} demonstrates that Draco's C-stars are at luminosities
near or well below the RGB tip; they are not the very luminous intermediate
age C-stars seen in the LMC.  We suggest instead that these 
are these just ordinary low metallicity C-stars similar to those
in the Galactic halo, and in themselves do not imply the presence
of an intermediate age population.

\section{Summary \label{section_summary}}

We present an abundance analysis based on high resolution
spectra obtained with HIRES on the Keck~I Telescope
of 8 stars in the Draco dwarf spheroidal galaxy.
The sample was selected to span the full range in metallicity
inferred from \cite{winnick03}, who used moderate resolution spectroscopy
for radial velocity members of this dSph galaxy found by earlier surveys,
e.g. \cite{armandroff95} and others.
Her CaT indices of the strength of the near-infrared Ca triplet correlate 
well with [Ca/H] we derive from our detailed abundance analyses
with differences
from a linear fit of only $\sigma = 0.17$~dex; $\sigma$ increases
to 0.28~dex for linear fit of differences when considering
how well the CaT indices predict [Fe/H] as determined from our
HIRES spectra.  Winnick's calibration differs significantly at 
low metallicities from that adopted by the DART project, an
important issue in the search for etremely metal-poor stars in
dSph galaxies discussed in the appendix.

We use classical plane-parallel (1D) LTE models from the Kurucz grid
\citep{kurucz93} with a recent version of the stellar abundance
code MOOG \citep{moog}.
[Fe/H] for our sample stars ranges from $-1.5$ to $-3.0$~dex.
Combining our sample with previously published work
of \cite{shetrone1} and \cite{shetrone2} for 6 Draco giants, 
and using the analysis of better spectra
for one of these 6 stars by \cite{fulbright}, gives a total
of 14 luminous Draco giants with detailed abundance analyses.
We find that
the abundance ratios [Na/Fe], [Mg/Fe], [Si/Fe], [Cr/Fe],
[Ni/Fe], [Zn/Fe], and perhaps [Co/Fe] for the Draco giants
overlap those of
Galactic halo giants at the lowest [Fe/H] probed, but 
for the higher Fe-metallicity Draco stars are
significantly lower than those of Milky Way halo giants.
For the explosive $\alpha$-elements
Ca and Ti, the abundance ratios are low over the full metallicity range of
the Draco dSph stars compared to Galactic halo giants, being
closer, but still slightly low, at the lowest Fe-metallicities, traits
shared by outer Galactic halo field stars.
Nucleosynthetic
yields sensitive to the neutron excess, hence to the initial metallicity
of the SN progenitor \citep{timmes}, may  be  important in explaining the
origin of differences between Draco giants and Galactic field stars
for several of the abundance
ratios studied here. 

The heavy neutron capture elements Sr and Ba have abundance ratios
with respect to Fe
comparable to or lower than those of the lowest halo giants selected
from much more extensive surveys.
The $s$-process contribution to the production of heavy elements
inferred from the ratio of Ba to Eu
begins at significantly lower Fe-metallicity than in the Galactic halo.

The dominant uncertainty in these results is the possibility of differential
non-LTE or 3D effects
between the very cool luminous giants in our sample from the
Draco dSph and the comparison halo field and globular cluster stars,
which are somewhat hotter.  With a 30-m telescope it will be possible
to reach lower luminosity and somewhat hotter Draco giants where these
issues will be less important. 

We develop a toy model which we use to illuminate these trends, and to
compare them with those of Galactic globular clusters and of
giants from the Carina and Sgr dSph galaxies.  The model has
consists of a plateau in [X/Fe] at low and at high [Fe/H]
connected by a straight line.  There are
four parameters determined directly from the data
defining the plateau characteristics and two additional
fit parameters which indicate the knee
values.  Since there is such good agreement in almost
all cases for the abundance ratios at the lowest metallicity within
a given sample and also the highest metallicities sampled, the fundamental
contributors to their chemical inventory 
(SNII at the lowest
 metallicity and SNIa plus other sources at the highest [Fe/H])
behave in very similar ways
 in all these environments.  We thus infer that the IMF for massive stars must be
 similar as well.
 
The key differences lie in the [Fe/H] 
corresponding to the knee values, i.e. in the timescale
at which the overall metallicity of the system increases.
The Draco system has the slowest evolution of metallicity in its
stars of the three, as well as the lowest mean [Fe/H] for its giants,
Sgr is intermediate, and  the Carina dSph is closest 
to the Milky Way  halo and the 
thick disk.  This timescale is
conventionally
interpreted as a measure of the star formation efficiency combined
with the rates of accretion of pristine materials and 
of gas lost via galactic winds.  Our new data will enable
much more sophisticated modelling of the chemical evolution
of Draco with more detail than our simple toy model can provide.

We note the
presence of a Draco giant with [Fe/H] $< -3.0$~dex in our sample.
This combined with other recent evidence for a small number
of extremely low metallicity stars in other dSph galaxies
reaffirms that the inner Galactic halo could have largely been formed
by early accretion and dissolution of Galactic satellite galaxies and by
globular clusters which dissolved prior to the imprinting of an AGB signature,
while the outer halo could have formed largely 
from those dSph galaxies  accreted later. 

\cite{shetrone2} point out that the fraction of Carbon stars in
Draco is is 25 -- 100 times higher than that of the Galactic
globular clusters, and this has been viewed as an indication
of the presence of an intermediate age stellar population.  However,
the 0Z Project \citep{cohen_cstarfreq} has established that the
frequency of carbon rich stars among very metal-poor Galactic
halo field stars is surprisingly high.  
We thus suggest that
the presence of carbon stars in the Draco dSph galaxy should
not be regarded as a sign of  an intermediate age
stellar population there; this is instead a signpost of the low
mean Fe-metallicity of the stellar component of this dSph galaxy. 

The age--metallicity relationship established by combining
photometry, spectroscopic metallicities, and isochrones
suggests a fairly extended period of star formation in
Draco with a duration of $\sim$5~Gyr beginning about 10~Gyr ago
over which the mean Fe-metallicity
increased by a factor of $\sim$5.  However, the uncertainty
in the age of any individual star is high, there may be large systematic
errors, and the 
apparently youngest stars
could actually be older AGB stars.

\acknowledgements

The entire Keck/HIRES and LRIS user communities owes a huge debt to
Jerry Nelson, Gerry Smith, Steve Vogt, and many other
people who have worked to make the
Keck Telescope and HIRES a reality and to operate and
maintain the Keck Observatory. We are grateful to the
W. M.  Keck Foundation for the vision to fund
the construction of the W. M. Keck Observatory.  The authors wish 
to extend
special thanks to those of Hawaiian ancestry on whose sacred mountain
we are privileged to be guests.  Without their generous hospitality,
none of the observations presented herein would
have been possible.

The authors are grateful to NSF grant AST-0507219  for partial support.
This publication makes use of data from the Two Micron All-Sky Survey,
which is a joint project of the University of Massachusetts and the 
Infrared Processing and Analysis Center, funded by the 
National Aeronautics and Space Administration and the
National Science Foundation.

\section{Appendix: The Validity of CaT and Stromgren Photometry for Draco 
\label{section_comp}}

We have relied on the rough abundances determined
from the infrared Ca triplet by \cite{winnick03} to
select our sample of stars in the Draco dSph galaxy.  Here
we examine the reliability of her metallicities, relevant
both for the present case and as an object lesson 
for the analysis of such data in other nearby galaxies.  We 
only discuss calibration issues here; we assume
that the actual measurements of the Ca triplet are correct, an issue
which is complicated by how the wings of these strong lines
are handled.

Winnick calibrates her results
with observations of stars from Galactic globular clusters
assuming [Ca/Fe] is the same
in Draco as it is in Galactic globular clusters.
It should be noted that the corresponding relationship
between  [Fe/H] and $W$\'~, the weighted sum of the equivalent widths of the three
lines of the Ca triplet, corrected to the level of the horizontal
branch, with the same assumption regarding [Ca/Fe], adopted
by the DART group \citep[equations 7 and 13 of][]{battaglia08}
gives [Fe/H] 0.44~dex higher than that of \cite{winnick03}
for   $W$\'~ = ~ 0.

\begin{figure}
\epsscale{0.9}
\plotone{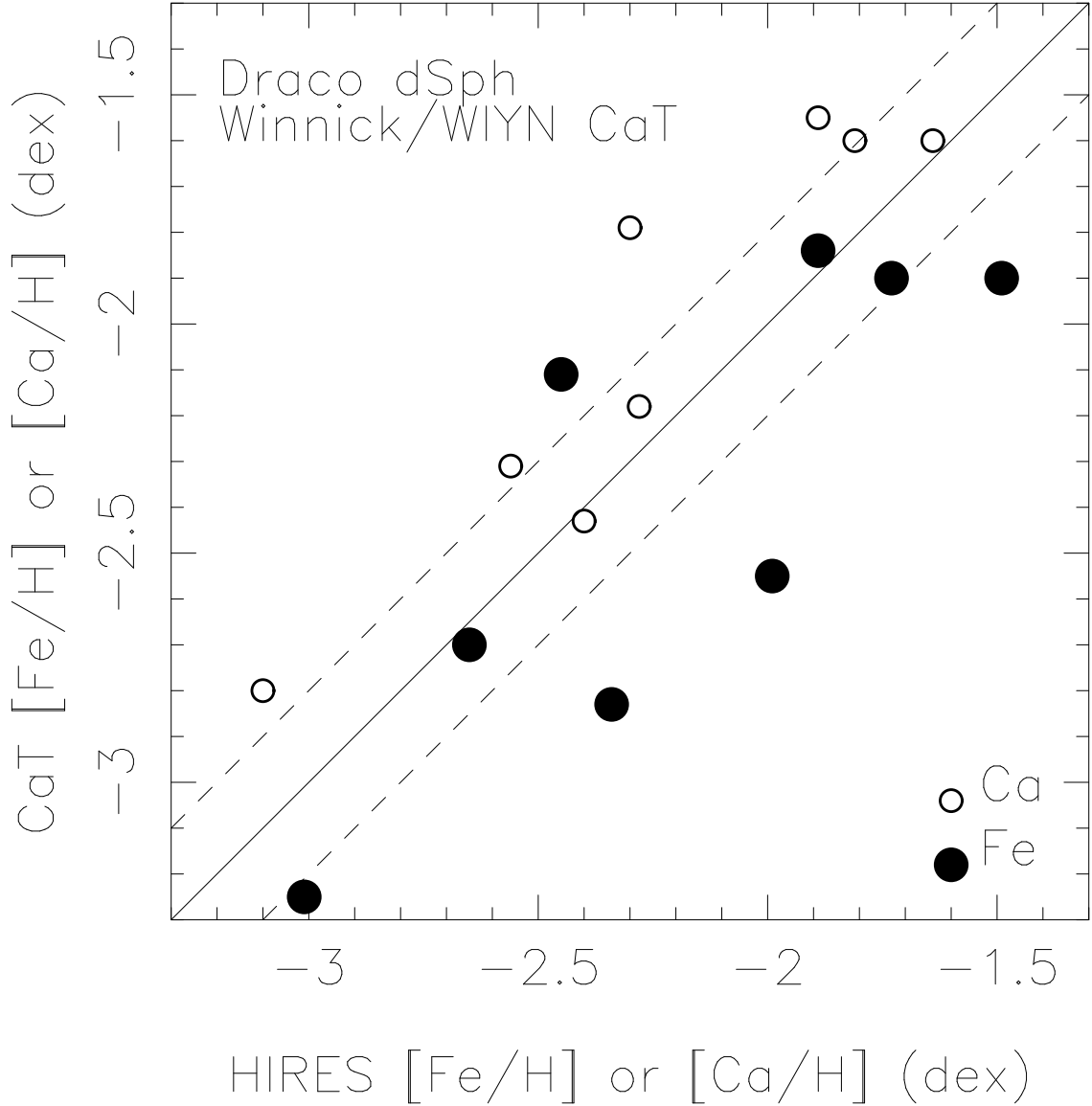}
\caption[]{The [Fe/H] (filled circles) and [Ca/H] values derived
from indices of the strength of
the infrared Ca triplet by \cite{winnick03} are compared to
the results of our high-resolution detailed abundance analyses
for our sample of 8 RGB stars in the Draco dSph galaxy.
The solid line
represents equality, while the dashed lines show offsets between
the two determinations of $\pm 0.2$~dex. 
\label{figure_winnick}}
\end{figure}

Fig.~\ref{figure_winnick} shows a comparison for
[Ca/H] and for [Fe/H] for the 8 stars in our sample in Draco
with the values derived by \cite{winnick03} from her
infrared Ca triplet survey.    For the entire metallicity range
probed in Draco [Ca/Fe] is essentially Solar, while in 
globular clusters it is $\sim$+0.3~dex.  So we would
expect Winnick's [Fe/H] values to be low by this amount
compared to our HIRES results, as is indeed seen in this figure.
It is not clear why her [Ca/H] values are somewhat higher than
the HIRES results, but her calibration for that only includes
one globular cluster below [Fe/H] $=~-1.4$~dex, i.e. over
the entire metallicity range of interest here.

Using our sample of 8 Draco luminous giants with HIRES detailed
abundance analyses, we find a linear fit

$$ [Ca/H] ~ = ~ -2.84 (\pm 0.12) ~ + ~ 0.434 (\pm 0.08)~W'(Ca) $$

\noindent where $W'(Ca)$ is Winnick's measured line strength for
the infrared Ca triplet, with a dispersion about the fit
of 0.17~dex.  This is in agreement to within the uncertainties
with the relation given by Winnick. A linear fit for [Fe/H] gives

$$ [Fe/H] ~ = ~ -2.77 (\pm 0.13) ~ + ~ 0.421 (\pm 0.08)~W'(Ca). $$

\noindent The constant here is 0.33~dex larger than that of
Winnick, not surprising given that her relation is calibrated to
globular clusters, which have [Ca/Fe] typically $\sim +0.3$~dex.
The dispersion about the linear fit, 0.28~dex, is 
somewhat larger than for [Ca/H], again
not surprising, since we are using a measurement of a Ca feature
to determine a Fe abundance.

Four of the stars in our sample are included in the recent 
catalog of Draco members with metallicities derived from Stromgren photometry by
\cite{faria07}.
The two most metal-poor of these (Draco~3157 and Draco~19219)
have [Fe/H](Stromgren) 0.5 and 1.0~dex higher respectively
than found from our detailed abundance analysis, presumably
due to the low sensitivity of photometry at very low metallicity.
The two higher metallicity stars in common show somewhat better agreement.

{}

\clearpage

\begin{deluxetable}{rccccc}
\tablewidth{0pc}
\tabletypesize{\scriptsize}
\tablecaption{The Sample of Stars in the Draco dSph
\label{table_sample}}
\tablehead{
\colhead{ ID\tablenotemark{a} } &
\colhead{Coords.\tablenotemark{a} } &
\colhead{V } &
\colhead{I } &
\colhead{Date Obs. } &
\colhead{Exp. Time } \\
\colhead{  } &
\colhead{(J2000) } &
\colhead{(mag, from SDSS) } &
\colhead{(mag, from SDSS) } &
\colhead{ } &
\colhead{(sec)} }
\startdata

361     &   17 20 34.19~~~$+$57 53 32.14  &  17.53 & 16.15  &  09/2006 &  7200 \\
3053    &   17 19 35.46~~~$+$57 58 46.78  &  17.48 & 16.09  &  09/2006 &  7200 \\
3150    &   17 19 36.37~~~$+$57 51 25.1\phn   &  17.38 & 15.98  &  06/2005 &  
   7200 \\ 
3157    &   17 19 41.85~~~$+$57 52 19.47  &  16.91 & 15.56  &  
    09/2006\tablenotemark{b} &  7200 \\
19219   &   17 21 43.69~~~$+$57 57 12.49  &  17.32 & 16.01  &  09/2006 &  7200 \\
19629   &   17 21 58.23~~~$+$57 56 04.31  &  17.33 & 15.97  &  09/2006 &  7200 \\
21456   &   17 18 10.43~~~$+$57 52 09.2\phn   &  17.08 & 15.63  & 06/2005 & 
    7200  \\
XI$-$2  &   17 19 17.53~~~$+$58 01 07.1\phn   &  16.96 & 15.38  &  06/2005
  & 7200  \\
\enddata
\tablenotetext{a}{The star names and coordinates are from 
\cite{winnick03}. Many of the names were originally 
assigned by  \cite{baade61}.}
\tablenotetext{b}{The 2 hour exposure from 06/2005 is not as good and
was not used.}
\end{deluxetable}

\begin{deluxetable}{rcc cc cr}
\tablewidth{0pc}
\tablecaption{Stellar Parameters for the Draco Giants \label{table_param}}
\tablehead{
\colhead{ ID } &
\colhead{$T_{\rm eff}$ } &
\colhead{$\log (g)$ } &
\colhead{$T_{\rm eff}$ } &
\colhead{$\log (g)$ } &
\colhead{$v_t$ } & \colhead{$v_r$} \\
\colhead{} &
\colhead{(K) (phot) } &
\colhead{(dex) (phot) } &
\colhead{(K) (spec) } &
\colhead{(dex) (spec) } &
\colhead{(km~s$^{-1}$)} & \colhead{(km~s$^{-1}$)} }
\startdata

361      &   4375 &  0.75 &   4420 &  0.75 & 2.0 & $-288.4$ \\
3053     &   4394 &  1.0 &    4350 &  1.0  & 2.1 & $-279.2$ \\
3150     &   4430 &  0.6 &    4380 &  0.6 &  2.3 & $-279.8$ \\
3157     &   4490 &  0.7 &    4510 &  0.7 & 2.4 & $-295.7$ \\
19219    &   4490 &  0.75 &   4500 &  0.75 & 2.9 & $-295.7$ \\
19629    &   4445 &  0.8 &    4460 &  0.8 & 2.9 & $-303.2$ \\
21456    &   4263 &  0.7 &    4430 &  0.7 & 2.3 & $-305.6$ \\
XI-2     &   4184 &  0.5 &    4350 &  0.6 &  2.4 & $-316.0$ \\
\enddata
\end{deluxetable}

\clearpage
\begin{deluxetable}{lccr rrrr rrrr}
\tabletypesize{\scriptsize}
\tablewidth{0pc}
\tablecaption{Equivalent Widths for Eight Stars in the Draco dSph
\label{table_eqw}}
\tablehead{
\colhead{$\lambda$} &
\colhead{Ion} &
\colhead{$\chi$ } &
\colhead{$\log (gf)$ } &
\colhead{361 } &
\colhead{3053 } &
\colhead{3150 } &
\colhead{3157} &
\colhead{19219} &
\colhead{19629} &
\colhead{21456} &
\colhead{XI$-$2} \\
\colhead{\AA} &
\colhead{ } &
\colhead{(eV) } &
\colhead{(dex) } &
\colhead{(m\AA) } &
\colhead{(m\AA) } &
\colhead{(m\AA) } &
\colhead{(m\AA) } &
\colhead{(m\AA) } &
\colhead{(m\AA) } &
\colhead{(m\AA) } &
\colhead{(m\AA) }}
\startdata
 6300.30 & [O~I] &   0.00 &  $-$9.780 &    17.5 &    36.0 &  \nodata &    10.5 &  
     \nodata &  \nodata &    14.7 &     9.0\tablenotemark{a}  \\
 6363.78 & [O~I] &   0.02 & $-$10.300 &    11.0 &    14.0 &  \nodata &  \nodata &  \nodata &  \nodata &  \nodata &  \nodata  \\
 5682.63 & Na~I  &   2.10 &  $-$0.700 &  \nodata &    12.8 &  \nodata &    10.0 &  \nodata &  \nodata &  \nodata &  \nodata  \\
 5688.19 & Na~I  &   2.10 &  $-$0.420 &    25.4 &    21.3 &  \nodata &    10.0 &  \nodata &  \nodata &  \nodata &  \nodata  \\
 5889.95 & Na~I  &   0.00 &   0.110 &   299.6 &   262.7 &   257.1 &   220.6 &   173.3 &   207.4 &   217.4 &   212.8  \\
 5895.92 & Na~I  &   0.00 &  $-$0.190 &   277.5 &   225.3 &   235.7 &   197.4 &   135.6 &   174.5 &  \nodata &  \nodata  \\
 4703.00 & Mg~I  &   4.34 &  $-$0.670 &   111.6 &   110.5 &   108.1 &    64.3 &    42.3 &    50.7 &    90.5 &    89.0  \\
 5172.70 & Mg~I  &   2.71 &  $-$0.380 &   353.6 &   326.6 &   294.5 &   234.6 &   192.7 &   222.4 &   246.9 &   243.0  \\
\enddata
\tablenotetext{a}{Very uncertain.  Could be an upper limit.}
\end{deluxetable}

\clearpage

\begin{deluxetable}{c rcr}
\tablewidth{0pc}
\tablecaption{Fit Fe~I Slopes With EP, Equivalent Width, and Wavelength 
\label{table_slopes}}
\tablehead{\colhead{Star ID} &
\colhead{$\Delta$[X/Fe]/$\Delta$(EP)\tablenotemark{a}} &
\colhead{$\Delta$[X/Fe]/${\Delta}[W_{\lambda}/\lambda]$} &
\colhead{$\Delta$[X/Fe]/${\Delta}\lambda$} \\ 
\colhead{} & \colhead{(dex/eV)} & \colhead{(dex)} & 
\colhead{($10^{-4}$dex/$\AA$)}
}
\startdata

361     & $-0.05$ & +0.02 & $-0.31$ \\
3053     &  $-0.04$ & $-0.07$ &  $+0.05$  \\
3150     & $-0.03$ &  $-0.01$ & $-0.18$  \\
3157     & $-0.10$ & 0.00 & $-0.09$ \\
19219    & $-0.08$ & 0.00 & $+0.18$   \\
19629    &  $-0.05$ & $-0.07$ & $+0.26$ \\
21456    &  $-0.08$ & $-0.01$ & $+0.08$ \\
XI-2     &  $-0.07$ & $-0.03$ & $+0.06$  \\
\enddata
\tablenotetext{a}{Typical range of EP is 4 eV.
This slope decreases by $\sim$0.2~dex/eV for an  increase
in \teff\ of 250~K.}
\end{deluxetable}

\clearpage

\begin{deluxetable}{l rrrr }
\tablewidth{0pc}
\tablecaption{Sensitivity of Absolute Abundances
\label{table_sens_abs}}
\tablehead{
\colhead{Species} & 
\colhead{$\Delta$log$({\epsilon})$}  &
\colhead{$\Delta$log$({\epsilon})$~($\Delta$ \grav} &
\colhead{$\Delta$log$({\epsilon})$~($\Delta$ Z(model)} &
\colhead{$\Delta$log$({\epsilon})$~($\Delta$ $v_t$}  
 \\
\colhead{} & \colhead{($\Delta~T_{\rm eff}~-250$~K)} &
\colhead{$-$0.5~dex)} & \colhead{$-$0.5~dex)} &
\colhead{+0.3~\kms)} \\
\colhead{} & \colhead{(dex)} & \colhead{(dex)}
& \colhead{(dex)} & \colhead{(dex)}  
}
\startdata
C(CH) &  $-0.15$ &   0.21 &   0.10 &   0.02 \\
O~I & $-$0.05 &  $-$0.18 &   0.06 &   0.00 \\
Na~I  & $-$0.66 &   0.17 &  $-$0.19 &   0.15 \\
Mg~I  & $-$0.32 &   0.19 &  $-$0.10 &   0.10 \\
Al~I  & $-0.69$ &   0.28 &  -0.34 &   0.15 \\
Si~I  & $-$0.11 &   0.03 &  $-$0.03 &   0.01 \\
K~I   & $-$0.39 &   0.06 &  $-$0.07 &   0.06 \\
Ca~I  & $-$0.27 &   0.07 &  $-$0.05 &   0.03 \\
Sc~II &  0.01 &  $-$0.14 &   0.03 &   0.03 \\
Ti~I  & $-$0.63 &   0.11 &  $-$0.13 &   0.06 \\
Ti~II &  0.03 &  $-$0.07 &  $-$0.01 &   0.12 \\
V~I   &  $-0.55$ &   0.08 &  -0.07 &   0.01 \\
Cr~I  & $-$0.53 &   0.13 &  $-$0.13 &   0.07 \\
Mn~I  & $-$0.37 &   0.11 &  $-$0.08 &   0.03 \\
Fe~I  & $-$0.41 &   0.09 &  $-$0.11 &   0.08 \\
Fe~II &  0.17 &  $-$0.09 &   0.00 &   0.12 \\
Co~I & $-$0.44 & 0.06 &  0.01 & 0.07 \\
Ni~I  & $-$0.38 &   0.05 &  $-$0.06 &   0.02 \\
Cu~I  & $-0.44$ &   0.08 &  $-0.07$ &   0.02  \\
Zn~I  & 0.05 &  $-0.03$ &  $-0.0$2 &   0.02 \\
Sr~II &  $-0.10$ &   0.00 &  $-0.03$ &   0.24 \\
Y~II  & $-$0.02 &  $-$0.13 &   0.03 &   0.03 \\
Zr~II &   0.04 &  $-0.02$ &  $-0.02$ &   0.11 \\
Ba~II & $-$0.14 &  $-$0.11 &   0.01 &   0.19 \\
Ce~II & $-$0.07 &  $-$0.11 &   0.02 &   0.03 \\
Nd~II & $-$0.09 &  $-$0.12 &   0.02 &   0.04 \\
Eu~II & $-$0.03 &  $-$0.17 &   0.05 &   0.01 \\
Dy~II & $-0.07$ &  $-0.11$ &   0.01 &   0.02  \\
\enddata
\end{deluxetable}

\begin{deluxetable}{l rrrrr c  r  }
\tablewidth{0pc}
\tablecaption{Sensitivity of Relative Abundances
\label{table_sens_rel}}
\tablehead{\colhead{Species} & 
\colhead{$\Delta$[X/Fe]} &
\colhead{$\Delta$[X/Fe]} &
\colhead{$\Delta$[X/Fe]} &
\colhead{$\Delta$[X/Fe]} &
\colhead{$\Delta$[X/Fe]} &
\colhead{Ion.} &
\colhead{Total Unc.\tablenotemark{c}}
\\
\colhead{ } & \colhead{($\Delta$ \teff} &
\colhead{($\Delta$ \grav} & \colhead{($\Delta$ Z(model)} &
\colhead{($\Delta$ $v_t$} & 
\colhead{$\Delta W_{\lambda}$\tablenotemark{a}} &
\colhead{Code\tablenotemark{b}}
\\
\colhead{} & \colhead{$-$150~K)} &
\colhead{$-$0.4~dex)} & \colhead{$-$0.5~dex)} &
\colhead{+0.3~\kms)} \\
\colhead{} & \colhead{(dex)} & \colhead{(dex)}
& \colhead{(dex)} & \colhead{(dex)} 
& \colhead{(dex)}   & \colhead{(dex)}   & \colhead{(dex)}
}
\startdata
C(CH)  & 0.16 &   0.10 &   0.21 &  $-0.06$ &   0.10 & 1 &  0.30 \\
O~I  & $-$0.13 & $-$0.07 &  0.02 &  0.00 &  0.08 & 2 &  0.17  \\
Na~I  & $-$0.15 &  0.06 & $-$0.03 &  0.15 &  0.08 & 1&   0.24  \\
Mg~I  & 0.05 &  0.08 &  0.00 &  0.10 &  0.04 &  1 &  0.14  \\
Al~I  &  $-0.17$ &   0.15 &  $-0.23$ &   0.07 &   0.08 & 1 &  0.34 \\
Si~I  &  0.18 & $-$0.05 &  0.03 &  0.01 &  0.08 & 1\tablenotemark{d} & 0.21  \\
K~I   &  0.01 & $-$0.02 &  0.02 &  0.06 &  0.08 & 1 &  0.10  \\
Ca~I  & 0.08 & $-$0.02 &  0.02 &  0.03 &  0.02 & 1 &   0.10  \\
\enddata
\tablenotetext{a}{Uncertainty from errors in \eqw\ taken
as 20\% (0.08~dex)/$\sqrt{Nlines}$).}
\tablenotetext{b}{Ion. code is 1 if with respect to Fe~I,
2 if with respect to Fe~II.}
\tablenotetext{c}{The rms sum of the 5 contributing terms.}
\tablenotetext{d}{The total uncertainty for
[Si/Fe] is the same using either Fe~I or Fe~II,
although the values of the individual terms contributing are different.}
\tablenotetext{e}{Average of [Ti/Fe](Ti~I,Fe~I) and [Ti/Fe](Ti~II,Fe~II).}
\tablenotetext{f}{This is [Fe/H](Fe~I) -- [Fe/H](Fe~II) and vice
versa, given for illustrative purposes only.}
\end{deluxetable}

\clearpage

\begin{deluxetable}{l | rrrr | rrrr | rrrr | rrrr  }
\tablenum{7a}
\tabletypesize{\tiny}
\tablewidth{0pt}
\setlength{\tabcolsep}{.1cm}
\tablecaption{Abundances for the First Four Draco Stars\label{table_abunda}}
\tablehead{\colhead{Species} & 
\colhead{} & \multispan{3}{Draco~361~~~[Fe/H] $-$1.49} &
\colhead{} & \multispan{3}{Draco~3053~~~[Fe/H] $-$1.73} &
\colhead{} & \multispan{3}{Draco~3150~~~[Fe/H] $-$1.89} &
\colhead{} & \multispan{3}{Draco~3157~~~[Fe/H] $-$2.45}   \\
\colhead{} & 
\colhead{[X/Fe]} & \colhead{log$\epsilon(X)$} & \colhead{No.} & 
       \colhead{$\sigma$\tablenotemark{a}} &  
\colhead{[X/Fe]} & \colhead{log$\epsilon(X)$} & \colhead{No.} & 
    \colhead{$\sigma$\tablenotemark{a}} &
\colhead{[X/Fe]} & \colhead{log$\epsilon(X)$} & \colhead{No.} & 
    \colhead{$\sigma$\tablenotemark{a}} &
\colhead{[X/Fe]} & \colhead{log$\epsilon(X)$} & \colhead{No.} & 
    \colhead{$\sigma$\tablenotemark{a}} \\
\colhead{} & \colhead{(dex)} & \colhead{(dex)} & \colhead{Lines} & \colhead{(dex)} &
\colhead{(dex)} & \colhead{(dex)} & \colhead{Lines} & \colhead{(dex)} &
\colhead{(dex)} & \colhead{(dex)} & \colhead{Lines} & \colhead{(dex)} &
\colhead{(dex)} & \colhead{(dex)} & \colhead{Lines} & \colhead{(dex)} 
}
\startdata 
C(CH)\tablenotemark{b}  &  $-$0.54 &   6.56   &    1 &  \nodata   &  $-$0.80 &   6.06   &    1 & \nodata   &  $-$0.64 &   6.06   &    1 & \nodata   &  $-$0.29 &   5.85   &    1 &  \nodata   \\
 O~I    &   0.00 &   7.34   &    2 &    0.04   &   0.32 &   7.42   &    2 &    0.04 
   & \nodata & \nodata   &  \nodata &  \nodata   &   0.40 &   6.78   &    1 &  \nodata   \\   
Na~I   &  $-$0.48 &   4.35   &    3 &    0.25   & 
  $-$0.62 &   3.96   &    4 &    0.28   &  $-$0.47 & 
    3.96   &    2 &    0.08   &   0.01 &   3.88   &    4 &    0.20   \\
Mg~I   &  $-$0.01 &   6.05   &    5 &    0.14   &   0.00 & 
  5.81   &    5 &    0.32   &   0.18 &   5.83   &    5 &   
   0.22   &   0.19 &   5.28   &    4 &    0.06   \\
Al~I   & \nodata & \nodata   &  \nodata &  \nodata   & \nodata & \nodata   &  \nodata &  \nodata   & \nodata & \nodata   &  \nodata &  \nodata   &  $-$0.62 &   3.40   &    1 & \nodata   \\
Si~I   &  $-$0.23 &   5.83   &    4 &    0.23   &  $-$0.01 & 
  5.81   &    5 &    0.13   &   0.08 &   5.74   &    4 &    0.13 
    &   0.26 &   5.36   &    3 &    0.12   \\
K~I    & \nodata & \nodata   &  \nodata &  \nodata   &   0.25 & 
  3.64   &    1 &  \nodata   & \nodata & \nodata   &  \nodata &  \nodata   &   0.10 &   2.77   &    1 & \nodata  \\
Ca~I   &  $-$0.16 &   4.71   &   18 &    0.15   &  $-$0.08 & 
  4.55   &   17 &    0.18   & 0.00 &   4.47   &   18 &    0.14   & 
    0.15 &   4.06   &   17 &    0.17   \\
\enddata
\tablenotetext{a}{$\sigma$ is the dispersion of the abundance ratio 
for each absorption line 
   about the mean for the species.}
\tablenotetext{b}{The 4320~\AA\ region of the G band of CH was used.}
\tablenotetext{c}{[Fe/H](Fe~I) is given instead of [X/Fe].}
\end{deluxetable}

\clearpage

\begin{deluxetable}{l | rrrr | rrrr | rrrr | rrrr  }
\tablenum{7b}
\tabletypesize{\tiny}
\tablewidth{0pt}
\setlength{\tabcolsep}{.1cm}
\tablecaption{Abundances for the Last Four Draco Stars\label{table_abundb}}
\tablehead{\colhead{Species} & 
\colhead{} & \multispan{3}{Draco~19219~~~[Fe/H] $-$3.01} &
\colhead{} & \multispan{3}{Draco~19629~~~[Fe/H] $-$2.65} &
\colhead{} & \multispan{3}{Draco~21456~~~[Fe/H] $-$2.34} &
\colhead{} & \multispan{3}{Draco~XI-2~~~[Fe/H] $-$1.99}   \\
\colhead{} & 
\colhead{[X/Fe]} & \colhead{log$\epsilon(X)$} & \colhead{No.} & 
       \colhead{$\sigma$\tablenotemark{a}} &  
\colhead{[X/Fe]} & \colhead{log$\epsilon(X)$} & \colhead{No.} & 
    \colhead{$\sigma$\tablenotemark{a}} &
\colhead{[X/Fe]} & \colhead{log$\epsilon(X)$} & \colhead{No.} & 
    \colhead{$\sigma$\tablenotemark{a}} &
\colhead{[X/Fe]} & \colhead{log$\epsilon(X)$} & \colhead{No.} & 
    \colhead{$\sigma$\tablenotemark{a}} \\
\colhead{} & \colhead{(dex)} & \colhead{(dex)} & \colhead{Lines} & \colhead{(dex)} &
\colhead{(dex)} & \colhead{(dex)} & \colhead{Lines} & \colhead{(dex)} &
\colhead{(dex)} & \colhead{(dex)} & \colhead{Lines} & \colhead{(dex)} &
\colhead{(dex)} & \colhead{(dex)} & \colhead{Lines} & \colhead{(dex)} 
}
\startdata 
C(CH)\tablenotemark{b}  &   0.32 &   5.90   &    1 &  \nodata   &  $-$0.44 &   5.50   &    1 &  \nodata   &  $-$0.49 &   5.76   &    1 &  \nodata   &  $-$0.84 &   5.76   &    1 &  \nodata   \\
O~I    & \nodata & \nodata   &  \nodata &  \nodata   & \nodata & \nodata   &  \nodata &  
\nodata   &   0.38 &   6.87   &    1 &  \nodata   & 
 $-$0.13\tablenotemark{c} &   6.71\tablenotemark{c}   &    1 &  \nodata   \\ 
Na~I   &  $-$0.30 &   3.02   &    2 &    0.15   & 
   $-$0.29 &   3.38   &    2 &    0.10   &  $-$0.46 & 
     3.52   &    1 &  \nodata   &  $-$1.06 &   3.27   &    1 &  \nodata   \\
Mg~I   &   0.33 &   4.86   &    4 &    0.17   &   0.20 &  
 5.09   &    5 &    0.10   &   0.19 &   5.39   &    4 &    0.25 
   &  $-$0.37 &   5.18   &    5 &    0.31   \\
Si~I   &   0.35 &   4.89   &    1 &  \nodata   &   0.51 &   5.41   &    1 &  \nodata   &   0.51 &   5.71   &    1 &  \nodata   &  $-$0.21 &   5.34   &    1 &  \nodata   \\
K~I    &   0.56 &   2.67   &    1 &  \nodata   &   0.28 &   2.75   &    1 &  \nodata   &   0.36 &   3.13   &    1 &  \nodata   & \nodata & \nodata   &  \nodata &  \nodata   \\
Ca~I   &  $-$0.09 &   3.26   &    6 &    0.18  &  
 0.09 &   3.81   &   13 &    0.16   &
   $-$0.06 &   3.95   &   11 &    0.16   & 
    $-$0.29 &   4.07   &   14 &    0.17   \\
Sc~II  &  $-$0.01 &   0.08   &    4 &    0.24   &  $-$0.08 &   0.38   &    5 &    0.15   &   0.09 &   0.85   &    6 &    0.12   &  $-$0.39 &   0.72   &    7 &    0.21   \\
\enddata
\tablenotetext{a}{$\sigma$ is the dispersion of the abundance ratio 
for each absorption line 
   about the mean for the species.}
\tablenotetext{b}{The 4320~\AA\ region of the G band of CH was used.}
\tablenotetext {c}{Extremely uncertain, could be upper limit.}
\tablenotetext{d}{[Fe/H](Fe~I) is given instead of [X/Fe].}
\end{deluxetable}

\clearpage

\begin{deluxetable}{l rr rr rr}
\tablenum{8}
\tablewidth{0pt}
\tablecaption{Parameters for the Toy Model of Abundance Ratios\tablenotemark{a} 
Applied to the Draco Giants \label{table_fit}}
\tablehead{\colhead{Species [X/Fe]} & 
\colhead{A(X)} & \colhead{B(X)} & \colhead{[Fe/H](A)} &
\colhead{[Fe/H](B)} & \colhead{[Fe/H](low,X)} &  
\colhead{[Fe/H](high,X)}  \\
\colhead{} & \colhead{(dex)} & \colhead{(dex)} & \colhead{(dex)} & \colhead{(dex)}
 & \colhead{(dex)} & \colhead{(dex)} }
\startdata    
${\rm{[Na/Fe]}}$ & 0.07 & $-0.34$ & $-2.96$ & $-1.53$ & $-2.86$ & $-2.29$ \\
${\rm{[Mg/Fe]}}$ & 0.47 & $-0.03$ & $-2.96$ & $-1.53$ & $-2.86$ & $-2.10$ \\
~~thick disk\tablenotemark{b} &
   0.35 & 0.00 & $-1.00$ & 0.00 & $-0.53$ & $-0.07$ \\     
~~thin disk\tablenotemark{b} &
     0.11 & 0.02 & $-$0.70 & 0.05 & $-0.50$ & $-0.10$ \\
${\rm{[Si/Fe]}}$\tablenotemark{c} &
    0.28 & 0.01 & $-2.96$ & $-1.53$ & $-2.31$ & $-2.02$  \\
${\rm{[Ca/Fe]}}$ & $-0.05$ & 0.00 & $-2.96$ & $-1.53$ & $-1.72 $ & $-1.72$ \\
${\rm{[Ti12/Fe]}}$\tablenotemark{d} & 
  0.13 & $-0.04$ & $-2.96$ & $-1.53$ & $-2.39$ & $-2.39$ \\
${\rm{[Cr/Fe]}}$ & $-0.45$ & $-0.07$ & $-2.96$ & $-1.53$ & $-1.82 $ & $-1.63$ \\
\enddata
\tablenotetext{a}{The model and its parameters are described 
in \S\ref{section_toy_model}.}
\tablenotetext{b}{Fits to the Milky Way thin and thick disk sample
of \cite{reddy03} and \cite{reddy06} for [Mg/Fe] vs [Fe/H].}
\tablenotetext{c}{The upper limits for Draco~119 of \cite{fulbright}
for [Si/Fe], [Sr/Fe] and for [Ba/Fe] are treated as detections.}
\tablenotetext{d}{${\rm{[Ti12/Fe]}} ~ = ~{\rm{[Ti ~ from ~ TiI/Fe~ from ~ FeI]}} 
~ + ~ {\rm{[Ti ~ from ~ TiII/Fe ~ from ~ FeII]}})/2$.}
\tablenotetext{e}{[Sr/Fe] is only available for the 8 Draco giants
whose analyses are presented here.}
\end{deluxetable}

\clearpage

\begin{deluxetable}{l rrrrr}
\tablenum{9}
\tablewidth{0pt}
\tablecaption{Abundance Range for Eight Draco Giants \label{table_range}}
\tablehead{\colhead{Species [X/Fe]} & 
\colhead{Nu. stars} & \colhead{Mean [X/Fe]} & \colhead{$\sigma$} &
\colhead{Min.} & \colhead{Max.} \\
\colhead{} & \colhead{(dex)} & \colhead{(dex)} & \colhead{(dex)} & \colhead{(dex)}
 & \colhead{(dex)} }
\startdata     
${\rm{[C(CH)/Fe]}}$\tablenotemark{a} & 8 & $-0.46$ & 0.37 & $-0.84$ & 0.32 \\
${\rm{[O(6300)/Fe]}}$ & 5 & 0.20 & 0.24  & $-0.13$ & 0.40 \\
${\rm{[Na/Fe]}}$ & 8 & $-0.46$ & 0.31 & $-1.06$ & 0.01 \\
${\rm{[Mg/Fe]}}$ & 8 & 0.09 & 0.22 & $-0.37$ & 0.33 \\
${\rm{[Al/Fe]}}$\tablenotemark{b} & 1 & $-0.62$ & \nodata & \nodata & \nodata \\
${\rm{[Si/Fe]}}$ & 8 & 0.16 & 0.30 & $-0.23$ & 0.51 \\
${\rm{[K/Fe]}}$ & 5 & 0.31 & 0.17 & 0.10 & 0.57 \\
${\rm{[Ca/Fe]}}$ & 8 & $-0.05$ & 0.14 & $-0.29$ & 0.15 \\
\enddata
\tablenotetext{a}{The 4320~\AA\ region of the G band of CH was used.}
\tablenotetext{b}{Non-LTE correction of +0.6~dex has been applied for 
[Al/Fe]  from the 3950~\AA\ doublet.}
\tablenotetext{c}{From Ti~I lines.}
\tablenotetext{d}{From Ti~II lines} 
\tablenotetext{e}{${\rm{[Ti12/Fe]}} ~ = ~{\rm{[Ti ~ from ~ TiI/Fe~ from ~ FeI]}} 
~ + ~ {\rm{[Ti ~ from ~ TiII/Fe ~ from ~ FeII]}})/2$.}
\end{deluxetable}

\clearpage

\clearpage

\clearpage

\clearpage

\clearpage

%

\clearpage

\begin{thebibliography}{}

\bibitem[Aaronson(1983)]{aaronson83}
Aaronson, M., 1983, \apj, 266, L11


\bibitem[Adelman-McCarthy et al(2007)]{sdss_dr5}
Adelman-McCarthy, J.~K. et al, 2007, \apjs, 172, 634

\bibitem[Andrievsky \etal(2007)]{andrievsky07}
Andievsky, S.~M., Spite, M., Korotin, S.~A., Spite, F.,
Bonifacio, P., Cayrel, R., Hill, V. \& Fracois, P., 2007
\aap, 464, 1081

\bibitem[Aoki \etal(2009)]{aoki09}
Aoki, W. \etal, 2009, \aap, in press (Astro-ph/0904.4307)


\bibitem[Armandroff, Olszewski \& Pryor(1995)]{armandroff95}
Armandroff, T.~E., Olszewski. E.~W. \& Pryor, C., 1995, \aj, 110, 2131

\bibitem[Arnett(1971)]{arnett}
Arnett, D., 1971, \apj, 166, 153

\bibitem[Asplund \etal(2004)]{asplund04}
Asplund, M., Grevesse, N., Sauval, A.~J., Allende Prieto, C.
\& Kisselman, D., 2004, \aap, 417, 751

\bibitem[Asplund(2005)]{asplund_araa}
Asplund, M. 2005, \araa, 43, 481

\bibitem[Asplund \etal(2005)]{asplund05}
Asplund, M., Grevesse, N., Sauval, A.~J., Allende Prieto, C.
\& Blomme, R., 2005, \aap, 431, 693

\bibitem[Baade \& Swope(1961)]{baade61}
Baade, W. \& Swope, H., 1961, \aj, 66, 300

\bibitem[Battaglia \etal(2008)]{battaglia08}
Battagali, G., Irwin, M., Tolstoy, E., Hill, V., Helmi, A.,
Letarte, B. \& Jablonka, P., 2008, \mnras, 383, 183

\bibitem[Belokurov et al(2006)]{belokurov06}
Belokurov, V. et al, 2006, \apj, 654, 897


\bibitem[Bonanos \etal(2004)]{distance}
Bonanos, A.~Z., Stanke, K.~Z., Szentgyorgyi, A.~H., Sasselov, D.~D.
\& Bakos, G.~A., 2004, \aj, 127, 861


\bibitem[Bonifacio \etal(2009)]{bonifacio09}
Bonifacio, P. \etal, 2009, \aap, in press (Astro-ph/0903.4174)


\bibitem[Busso, Gallino \& Wasserburg(1999)]{busso99}
Busso, M., Gallino, R. \& Wasserburg, G.J., 1999, \araa, 37, 239

\bibitem[Carigi, Hernandez \& Gilmore(2002)]{carigi02}
Carigi, L, Hernandez, X. \& Gilmore, G., 2002, \mnras, 334, 117

\bibitem[Carigi \& Hernandez(2008)]{carigi08}
Carigi, L. \& Hernandez, X., 2008, \mnras, in press

\bibitem[Carpenter(2001)]{carpenter}
Carpenter, J., 2001, \aj, 121, 2851

\bibitem[Carollo et al(2007)]{carollo}
Carollo, D. et al, 2007, Nature, 450, 1020

\bibitem[Cayrel et al(2004)]{cayrel04}
Cayrel, R. et al, 2004, \aap, 416, 1117

\bibitem[Cioni \& Habing(2005)]{cioni}
Cioni, M.~R.~L. \& Habing, H., 2005, \aap, 442, 165

\bibitem[Clayton(2003)]{clayton03}
Clayton, D., ``Handbook of Isotopes in the Cosmos'', 2003,
Cambridge U. Press

\bibitem[Cohen \etal(2002)]{cohen02} Cohen,   J.~G.,  Christlieb,  N., 
Beers,   T.~C.,  Gratton,   R.~G. \& Carretta,   E., 2002, \aj, 124, 470

\bibitem[Cohen \etal(2004)]{cohen04} 
Cohen, J.~G., Christlieb, N.,  McWilliam, A., Shectman, S.,
Thompson, I., Wasserburg, G.~J., Ivans, I., Dehn, Karlsson, T. \& 
Melendez, J., 2004, \apj, 612, 1107

\bibitem[Cohen \etal(2005)]{cohen_cstarfreq}
Cohen, J.~G. et al, 2005, \apjl, 633, L109

\bibitem[Cohen \etal(2006)]{cohen06} 
Cohen, J.~G. \etal, 2006, \aj, 132, 137

\bibitem[Cohen \etal(2008)]{cohen08} 
Cohen, J.~G.,  Christlieb, N.,  McWilliam, A., Shectman, S.,
Thompson, I., Melendez, J., Wisotzki, L. \& Reimers, D.,   2008,
\apj, 672, 320

\bibitem[Cohen, Briley \& Stetson(2005)]{cohen_m15}
Cohen, J.~G., Briley, M.~M. \& Stetson, P.~B., 2005, \aj, 130, 1177

\bibitem[Cohen \& Melendez(2005)]{m3_m13}
Cohen, J.~G. \& Melendez, J., 2005, \aj, 129, 303

\bibitem[Cohen et al(2007)]{cohen07}
Cohen et al 2007, \apjl, 659, L161


\bibitem[Collet, Asplund \& Trampedach(2006)]{collet07}
Collet, R., Asplund, M. \& Trampedach, R., 2007, \aap, 469, 687

\bibitem[Cutri \etal(2003)]{2mass2} Cutri, R.~M. \etal, 2003,
``Explanatory Supplement to the 2MASS All-Sky Data Release,
http:\\www.ipac.caltech.edu/2mass/releases/allsky/doc/explsup.html

\bibitem[Dotter \etal(2008)]{dotter08}
Dotter, A., Chaboyer, B., Jevremovic, D., Kotov, V., Baron, E.
\& Ferguson, J.~W., 2008, \apjs, in press

\bibitem[Faria \etal(2007)]{faria07}
Faria, D., Feltzing, S., Lundstrom, I., Gilmore, G., Wahlgren, G.~M.,
Ardeberg, A. \& Linde, P., , 2007, \aap, 465,357


\bibitem[Francois et al(2007)]{francois}
Francois, P. et al, 2007, \aap, 476, 935

\bibitem[Frebel \etal(2009)]{frebel09}
Frebel, A., Simon, J.~D., Geha, M. \& Willman, B., 2009, \apj,
submitted (Astro-ph/0902.2395)


\bibitem[Fuhrmann, Axer \& Gehren(1995)]{fuhrman_mg}
Fuhrmann, K. Axer, M. \& Gehren, T., 1995, \aap, 301, 492

\bibitem[Fulbright(2002)]{fulbright02}
Fulbright, J.~P., 2000, \aj, 120, 1841

\bibitem[Fulbright, Rich \& Castro(2004)]{fulbright}
Fulbright, J., Rich, R.~M. \& Castro, S., 2004, \apj, 612, 447

\bibitem[Fulbright, McWilliam \& Rich(2007)]{fulbright_bulge}
Fulbright, J.~P., McWilliam, A. \& Rich, R.~M., 2007, \apj, 661, 1152

\bibitem[Geisler et al(2005)]{geisler05}
Geisler, D., Smith, V~V., Wallerstein, G., Gonzalez, G.
\& Charbonnel, C., 2005, \aj, 129, 1482

\bibitem[Geisler et al(2008)]{geisler08}
Geisler, D., Wallerstein, G., Smith, V.~V. \&
Casetti-Dinescu, D.~I., 2008, \pasp, 119, 859

\bibitem[Gratton, Sneden \& Carretta(2004)]{gratton_araa}
Gratton, R., Sneden, C. \& Carretta, E., 2004, \araa, 42, 385

\bibitem[Greggio, Renzini \& Daddi(2008)]{greggio08}
Greggio, L., Renzini, A. \& Daddi, E., 2008, \mnras, 388, 829

\bibitem[Grevesse \& Sauval(1998)]{grevesse98}
Grevesse, N. \& Sauval, A.~J., 1998, Space Science Reviews, 85, 161

\bibitem[Grillmair et al(1998)]{grillmair98}
Grillmair, C.~J. et al, 1998, \aj, 115, 144

\bibitem[Helmi et al(2006)]{helmi06}
Helmi, A. et al, 2006, \apj, 651, L121

\bibitem[Hernandez, Gilmore \& Valls-Gabaud(2000)]{hernandez00}
Hernandez,X.,  Gilmore, G. \& Valls-Gabaud, D., 2000, \mnras, 317, 831

\bibitem[Houdashelt, Bell \& Sweigart(2000)]{houdashelt00}
Houdashelt, M.~L., Bell, R.~A. \& Sweigart, A.~V., 2000, \aj, 119, 1448

\bibitem[Huber \& Herzberg(1979)]{huber79}
Huber, K.~P. \& Herzberg, G., 1979, {\it{Constants of Diatomic Molecules}},
(New York, Van Nostrand)

\bibitem[Iwamoto et al(1999)]{iwamoto}
Iwamoto, K., Brachwitz, F., Nomoto, K., Kishimoto, N., Umeda, H.,
Hix, W.~R. \& Thieleman, F.~K., 1999, \apjs, 125, 439

\bibitem[Johnson(2002)]{johnson02}
Johnson~J. 2002, \apjs, 139, 219

\bibitem[Kirby et al(2008)]{kirby08}
Kirby, E.~N., Simon, J.~D., Geha, M., Guhathakurta, P.
\& Frebel, A., 2008, \apjl, 685, L43

\bibitem[Klypin et al(1999)]{klypin99}
Klypin, A., Kravtsov, A.~V., Valenzuela, O. \& Prada, F., 1999,
\apj, 522, 82

\bibitem[Kobayashi et al(2006)]{kobayashi}
Kobayashi, C., Umeda, H., Nomoto, K., Tominaga, N. \& Ohkubo, T.,
2003, \apj, 653, 1145

\bibitem[Koch et al(2008a)]{koch_carina}
Koch, A., Grebel, E.~K., Gilmore, G.~F., Wyse, R.~F.~G.,
Kleyna, J.~T., Harbeck, D.~R., Wilkinson, M.~I. \& Evans, N.~W.,
\aj, 135, 1580

\bibitem[Koch et al(2008b)]{koch_hercules}
Koch, A., McWilliam, A., Grebel, E.~K., Zucker, D.~B.
\& Belokurov, V., 2008, \apjl, in press

\bibitem[Koposov et al(2008)]{koposov08}
Koposov, S. et al, 2008, \apj, 686, 279

\bibitem[Kurucz(1993)]{kurucz93} Kurucz, R. L., 1993, ATLAS9 Stellar 
Atmosphere Programs and 2 km/s Grid, (Kurucz CD-ROM No. 13)

\bibitem[Lanfranchi \& Matteucci(2004)]{lanfranchi04}
Lanfranchi, G.~A. \& Matteucci, F., 2004, \mnras, 351, 1338

\bibitem[Lanfranchi, Matteucci \& Cescutti(2008)]{lanfranchi08}
Lanfranchi, G.~A., Matteucci, F. \& Cescutti, G., 2008, \aap, 481, 635

\bibitem[Layden \& Sarajedini(2000)]{layden00}
Layden, A.~C. \& Sarajedini, A., 2000, \aj, 119, 1760

\bibitem[Lehnert et al(1992)]{lehnert92}
Lehnert, M.~D., Bell, R.~A., Hesser, J.~E. \& Oke, J.~B., 1992,
\apj, 395, 466

\bibitem[Limongi \& Chieffi(2003)]{limongi03}
Limongi, M. \& Chieffi, A., 2003, \apj, 592, 404

\bibitem[Marcolini et al(2006)]{marcolini06}
Marcolini, A., D'Ercole, A., Brighenti, F. \& Recchi, S.,
2006, \mnras, 371, 643

\bibitem[Marin-Franch et al(2009)]{acs_ages}
Marin-Franch, A. et al 2009, \apj, 694, 1498


\bibitem[Martin, Kobulnicky \& Heckman(2002)]{martin02}
Martin, C.~L., Kobulnicky, H.~A. \& Heckman, T.~M., 2002,
\apj, 574, 663

\bibitem[Mashonkina \& Gehren(2001)]{halo_sr}
Mashonkina, L. \& Gehren, T., 2001, \aap, 376, 232

\bibitem[Mashonkina et al(2003)]{mashonkina}
Mashonkina, L., Gehren, T., Travaglio, C. \& Borkova, T., 2003, 
\aap, 397, 275

\bibitem[Matteucci(2008)]{matteucci08}
Matteucci, F., 2008, in ``Chemical Evolution of the Milky Way
and Its Satellites'', in Saas-Fe Advance Course, ``The Origin
of the Galaxy and the Local Group'', ed. E.~Grebel \& B.~Moore


\bibitem[McWilliam \etal(1995a)]{mcwilliam95a}
McWilliam, A., Preston, G.~W., Sneden, C. \& Shectman, S., 1995, \aj,
109, 2736

\bibitem[McWilliam \& Smecker-Hane(2005)]{mcw_sgr}
McWilliam, A. \& Smecker-Hane, T.~A., 2005, in
``Cosmic Abundances as Records of Stellar Evolution
and Nucleosynthesis'', ed. F.N.Bash \& T.G.Barnes

\bibitem[Melendez et al(2004)]{melendez06}
Melendez, J., Shchukina, N.~G., Vasiljeva, I.~E. \& Ramirez, I., 2006,
\apj, 641, 1082

\bibitem[Mishenina \etal(2002)]{mishenina02}
Mishenina, T.~V., Kovtyukh, V.~V., Soubiran, C., Travaglio, C.
\& Busso, M., 2002, \aap, 396, 189

\bibitem[Monaco et al(2005)]{monaco05}
Monaco, L. et al, 2005, \aap, 441, 141

\bibitem[Munoz et al(2005)]{munoz}
Munoz, R.~R. et al, 2005, \apjl, 631, L137

\bibitem[Nissen \& Schuster(1997)]{nissen97}
Nissen, P.~E. \& Schuster, W.~J.,1997, \aap, 326, 751

\bibitem[Nissen et al(2000)]{nissen00}
Nissen, P.~E. et al 2000, \aap, 353, 722

\bibitem[Nissen et al(2002)]{nissen02}
Nissen, P.~E., Primas, F., Asplund, M. \& Lambert, D.~L., 
2002, \aap, 390, 325

\bibitem[Nissen et al(2008)]{nissen08}
Nissen, P.~E., Ackerman, C., Asplund, M., Fabian, D., Kerber, F.,
Kaufl, H.~U. \& Pettini, M., 2008, \aap, 649, 319

\bibitem[Norris et al(2008)]{norris08}
Norris, J.~E., Gilmore G., Wyse, R.~F.~G., Wilkinson, M.~I.,
Belokurov, V., Wyn Evans, N. \& Zucker, D.~B., 2008, \apjl, 689, L113

\bibitem[Ohkubo, Umeda, Nomoto \& Yoshida(2006)]{ohkubo}
Ohkubo,~T., Umeda,~H., Nomoto,~K. \& Yoshida,~T., 2006,
AIPC, 847, 458

\bibitem[Orban et al(2006)]{orban08}
Orban, C., Gnedin, O.~Y., Weisz, D.~R., Skillman, E.~D.,
Dolphin, A.~E. \& Holtzman, J.~A., 2008, \aj, 686, 1030

\bibitem[Piatek \etal(2001)]{piatek01}
Piatek, S., Pryor, C., Armandroff, T.~E. \& Olszewski, E.~W., 2001, 
\aj, 121, 841

\bibitem[Prantzos(2008)]{prantzos}
Prantzos, N., 2008, in ``Stellar Nucleosynthesis: 50 Years After B2FH'',
2008, ed. C. Charbonnel \& J.~P. Zahn, EAS Publications Series

\bibitem[Prochaska et al(2000)]{prochaska}
Prochaska, J.~X., Naumov, S.~O., Carney, B.~W., McWilliam, A. \&
Wolfe, A.~M., 2000, \aj, 120, 2513

\bibitem[Qian \& Wasserburg (2003)]{qian03}
Qian, Y.~Z. \& Wasserburg, G~J., 2003. \apj, 588, 1099

\bibitem[Qian \& Wasserburg (2007)]{qian07}
Qian, Y.~Z. \& Wasserburg, G~J., 2007, Physics Reports, Review
Section of Physics Letters, 442, 237

\bibitem[Qian \& Wasserburg (2008)]{qian08}
Qian, Y.~Z. \& Wasserburg, G~J., 2008, \apj, 687, 272


\bibitem[Ramirez,  Allende Prieto \& Lambert(2007)]{ramirez07}
Ramirez, I., Allende Prieto, C. \& Lambert, D.~L., 2007, \aap, 465, 271

\bibitem[Reddy et al(2003)]{reddy03}
Reddy, B.~E., Tomkin, J., Lambert, D.~L., \& Allende Prieto, C., 2003,
\mnras, 340, 304

\bibitem[Reddy, Lambert \& Allende Prieto(2006)]{reddy06}
Reddy, B.~E., Lambert, D.~L. \& Allende Prieto, C., 2006, \mnras, 367, 1329

\bibitem[Roederer(2008)]{roederer08}
Roederer, I.~U., 2008, \aj, 137, 272


\bibitem[Sbordone et al(2005)]{sbordone05}
Sbordone, L., Bonifacio, P, Marconi, G., Buonanno, R. \&
Zaggia, S., 2005, \aap, 437, 905

\bibitem[Sbordone et al(2007)]{sbordone07}
Sbordone, L., Bonifacio, P., Buonanno, R., Marconi, G.,
Monaco, L. \& Zaggia, S., 2007, \aap, 465, 815


\bibitem[Schlegel, Finkbeiner \& Davis (1998)]{extinct98}
Schlegel, D., Finkbeiner, D.~P. \& Davis, M., 1998, \apj, 500, 525

\bibitem[Sch\"orck et al(2009)]{hes_mdf}
Sch\"orck, T. et al, 2009, \aap, submitted (Astro-ph/0809.1172)



\bibitem[Schuler et al(2005)]{schuler}
Schuler, S.~C., Hatzes, A.~P., King, J.~R., Kurster, M. \& The, L.-S.,
2005, \apj, 636, 432

\bibitem[Segall \etal(2007)]{megacam}
Segall, M., Ibata, R.~A., Irwin, M.~J., Martin, M.~F. \& Chapman, S.,
2007, \mnras, 375, 831

\bibitem[Shetrone, Bolte \& Stetson(1998)]{shetrone1}
Shetrone, M.~D., Bolte, M. \& Stetson, P.~B., 1998, \aj, 115, 1888

\bibitem[Shetrone, C\^ot\'e \& Sargent(2001)]{shetrone2}
Shetrone, M.~D., C\^ot\'e, P. \& Sargent, W.~L.~W., 2001, \apj, 548, 592
      
\bibitem[Shetrone, C\^ot\'e \& Stetson(2001)]{shetrone_cstar}      
Shetrone, M.~D., C\^ot\'e, P. \& Stetson, P.~B., 2001, \pasp, 113, 1122

\bibitem[Shetrone et al(2003)]{shetrone03}
Shetrone, M.~D. et al, 2003, \aj, 125, 684

\bibitem[Shetrone et al(2009)]{shetrone09}
Shetrone, M.~D., Siegel, M~H., Cook, D.~O. \& Bosler, T.,
2009, \aj, in press

\bibitem[Short \& Hauschildt(2006)]{short06}
Short, C.~I. \& Hauschildt, P.~H., 2006, \apj, 641, 494



\bibitem[Shortridge(1993)]{shortridge93}
Shortridge K. 1993, in {\it{Astronomical Data Analysis Software and
      Systems II}}, A.S.P. Conf. Ser., Vol 52, eds. R.J. Hannisch, 
      R.J.V. Brissenden, \& J. Barnes, 219
      
\bibitem[Siegel et al(2007)]{siegel07}
Siegel, M.~H. et al, 2007, \apjl, 667, L57
      
\bibitem[Simmerer et al(2004)]{simmerer}
Simmerer, J., Sneden, C., Cowan, J.~J., Collier, J.,
Woolf, V.~M. \& Lawler, J.~E.,  2004, \apj, 617, 1091

\bibitem[Skrutskie \etal(1997)]{2mass1}
Skrutskie, M.~F., Schneider, S.E., Stiening, R., Strom, S.E.,
Weinberg, M.D., Beichman, C., Chester, T. \etal, 1997, in {\it{The
Impact of Large Scale Near-IR Sky Surveys}}, ed. 
F.Garzon \etal\ (Dordrecht: Kluwer), p. 187

\bibitem[Smith \& Raggett(1981)]{cagf_old}
Smith, G. \& Raggett, D.~St.~J., 1981, J.Phys.B, 14, 4015


\bibitem[Smith et al.(2002)]{smith02}
Smith, J.~A. \etal, 2002, \aj, 123, 2121

\bibitem[Sneden(1973)]{moog} Sneden, C., 1973, Ph.D. thesis, Univ. 
of Texas

\bibitem[Spite et al(2005)]{spite}
Spite, M. et al, 2005, \aap, 430, 655

\bibitem[Stephens(1999)]{stephens99}
Stephens, A., 1999, \aj, 117, 1771

\bibitem[Takeda et al(2002)]{takeda02}
Takeda, Y., Zhao, G., Chen, Y.~Q., Qui, H.~M. \& Takada-Hidei, M.,
2002, PASJ, 54, 275

\bibitem[Takeda et al(2003)]{takeda_na}
Takeda, Zhao, Takeda-Hidai et al
2003, Chinese Jrl Astron \& Astrophys, 3, 316

\bibitem[Timmes, Woosley \& Weaver(1995)]{timmes}
Timmes,~F.~X., Woosley, S.~E. \& Weaver, T.~A., 1995, \apjs, 98, 617

\bibitem[Tinsley(1973)]{tinsley}
Tinsley, B.~M., 1973, \apj, 186, 35

\bibitem[Tolstoy et al(2003)]{tolstoy03}
Tolstoy, E. et al, 2003, \aj, 125, 707 

\bibitem[Travaglio et al(2003)]{travaglio04}
Travaglio, C., Gallino, R., Arnone, E., Cowan, J.,
Jordan, F. \& Sneden, C., 2004, \apj, 601, 864

\bibitem[Tsujimoto(2006)]{tsujimoto06}
Tsujimoto, T., 2006, \aap, 447, 81

\bibitem[Vogt \etal(1994)]{vogt94} Vogt, S.~E. \etal\, 1994, SPIE, 2198, 362

\bibitem[Wallace, Hinkle \& Livingston(1998)]{wallace98}
Wallace, L., Hinkle, K. \& Livingston, W.~C.,1998,
``An Atlas of the Spectrum of the Solar Photosphere from 13,500
to 28,000 cm$^{-1}$'', N.S.O. Technical Report 98-001,
ftp:\\nsokp.nso.edu/pub/atlas/visatl.

\bibitem[Walker et al(2008)]{walker07}
Walker, M.~G., Mateo, M., Olszewski, E.~W., Gnedein, O.~Y.,
Wang, W., Sen, B. \& Woodroofe, M., 2007., \apjl, 667, L53


\bibitem[Winnick(2003)]{winnick03}
Winnick, R.~A., 2003, PhD thesis,  Yale University,
``Metallicity Distributions in the Draco, Ursa Minor
and Sculptor Dwarf Spheroidal Galaxies''

\bibitem[Woosley \& Weaver(1995)]{woosley}
Woosley, S.~E. \& Weaver, T.~A., 1995, \apj, 101, 181                                                                               

\bibitem[Yi \etal(2002)]{yi01}
Yi, S., Demarque, P., Kim, Y.-C. ,  Lee, Y.-W., Ree, C.
Lejeune, Th. \&  Barnes, S., 2001, \apjs, 136, 417

\bibitem[Yong et al(2003)]{yong}
Yong, D., Lambert, D.~L., Allende Prieto, C. \& Paulson, D.~B.,
2003, \apj. 500. 1357

\bibitem[York \etal(2000)]{york00}
York, D.~G., Adelman, J., Anderson Jr., J.~E. \etal, 2000, \aj, 120, 1579

\bibitem[Young(1999)]{young99}
Young, L.~M., 1999, \aj, 117, 1758

\bibitem[Zinn(1978)]{zinn78}
Zinn, R.~J., 1988, \apj, 225, 790

\end{thebibliography}
\end{document}